\documentclass[%
 aip,
 amsmath,amssymb,
 reprint,
]{revtex4-1}

\usepackage{graphicx}
\usepackage{dcolumn}
\usepackage{bm}

\usepackage[utf8]{inputenc}
\usepackage[T1]{fontenc}
\usepackage{mathptmx}
\usepackage{etoolbox}
\usepackage{placeins}

\usepackage{multirow}
\usepackage{siunitx}
\DeclareSIUnit\angstrom{\text{Å}}
\DeclareSIUnit{\calorie}{cal}
\DeclareSIUnit{\Calorie}{\kilo\calorie}
\DeclareSIUnit{\radiant}{rad}

\DeclareSIUnit{\gauss}{G}
\DeclareSIUnit{\arbitraryunit}{a.u.}

\makeatletter
\def\@email#1#2{%
 \endgroup
 \patchcmd{\titleblock@produce}
  {\frontmatter@RRAPformat}
  {\frontmatter@RRAPformat{\produce@RRAP{*#1\href{mailto:#2}{#2}}}\frontmatter@RRAPformat}
  {}{}
}%
\makeatother
\begin{document}

\preprint{AIP/123-QED}

\title[Optimizing the Structure of Acene Clusters]{Optimizing the Structure of Acene Clusters}
\author{P. Elsässer}
\author{T. Schilling}%
 \email{tanja.schilling@physik.uni-freiburg.de.}
 \affiliation{Institute of Physics, University of Freiburg, 79104 Freiburg (Breisgau), Germany}

\date{1 March 2023}

\begin{abstract}
We present a study of the potential energy surface (PES) of anthracene, tetracene and pentacene clusters with up to 30 molecules. We have applied the basin-hopping Monte Carlo (BHMC) algorithm to clusters of acene molecules in order to find their lowest energy states. The acene molecules are described by the polymer-consistent force field - interface force field (PCFF-IFF). We present the structures with the lowest observed energy, and we discuss the relative stability and accessibility of structures corresponding to local energy minima.
\end{abstract}

\maketitle

\section{Introduction}
Polycyclic aromatic hydrocarbons are a group of molecules that have been heavily investigated in the last few years because their optoelectric properties make them applicable in organic solar cells \cite{walker2011small}. In other areas where semiconductors are needed, like thin film transistors, polycyclic aromatic hydrocarbons have been found to be suitable candidates as well \cite{anthony2006functionalized}.

In this study we concentrate on a subgroup of polycyclic aromatic hydrocarbons in which the aromatic rings form a chain. This group is known as acene molecules. So far the synthetization of acene molecules with up to twelve aromatic rings (dodecacene) has been reported \cite{eisenhut2019dodecacene}. The longest of these molecules have a high reactivity and are thus very unstable, which prevents their use in technical applications. Hence we will focus here on the three shortest molecules: Anthracene, tetracene and pentacene with three, four and five aromatic rings. Clusters of these molecules play an important role in the production of thin films for organic solar cells and thin film transistors \cite{kim2002poly,abthagir2005studies}. They are found as well in interstellar medium where infrared emission bands are associated with them \cite{hudgins1995infrared,roser2014anthracene,mackie2015anharmonic}. Another research area in which polycyclic aromatic hydrocarbons play an important role is environmental and health chemistry or combustion science as parts of soot \cite{morselli1988pah,schuetz2002nucleation,totton2010modelling,menon2020reactivity}.

In 2013, Takeuchi searched for the minimal energy structures of anthracene clusters with up to ten molecules \cite{takeuchi2013structures}. Takeuchi employed a heuristic method combined with geometrical perturbations for geometry optimization of molecular clusters \cite{takeuchi2007novel} to find the global energy minimum. The observed acene structures were compared with those of other acene derivatives. For the di- and trimer, there have also been other studies, both experimental and theoretical\cite{gonzalez2000electronic,gonzalez2000quantum,piuzzi2002spectroscopy,podeszwa2008physical}. Gonzalez and Lim \cite{gonzalez2000electronic,gonzalez2000quantum} analyzed differences between the parallel and orthogonal orientation of molecules in dimers of benzene, naphthalene, and anthracene. Piuzzi et al. \cite{piuzzi2002spectroscopy} investigated anthracene clusters with up to five molecules with respect to their dynamics, spectroscopy and structure. Podeszwa and Szalewicz \cite{podeszwa2008physical} did SAPT(DFT) calculations to find the physical origins of the interactions in naphthalene, anthracene and pyrene dimers. Research on tetramers has been done e.g. by Park et al. \cite{park2007ab} on chains of aromatic rings with different lengths from benzene up to pentacene. On larger clusters, there have been reports by e.g. Mitsui et al. \cite{mitsui2007mass} on electron photoelectron spectroscopy of tetracene cluster anions, and Eilmes \cite{eilmes2012theoretical} on stability and electron detachment energies.

There are many different approaches for the optimization of the chemical structure of clusters \cite{zhang2021global}. We apply the basin-hopping Monte Carlo (BHMC) algorithm to find the global minimum of the potential energy surface (PES) for acene clusters. BHMC is a method, developed by Wales et al. \cite{wales1997global}, which combines local optimization with Monte Carlo steps. It has so far been mainly applied to solids and clusters which consist of single atoms \cite{xu2022geometrical} or small molecules \cite{burnham2019crystal} or the folding of single proteins \cite{roder2018mutation}. In the context of developing coarse-grained models, this method has been used e.g. by Hernández-Rojas et al.\cite{hernandez2016coarse,hernandez2017dynamics}. Examples of applications to larger systems are the work by Totton et al. \cite{totton2010modelling} who applied BHMC to clusters of up to $50$ rigid corone and pyrene molecules and Banerjee et al. \cite{banerjee2021crystal} who searched for the structure of crystal benzene. Here, we apply BHMC to search for the configurations at the global minimum of the PES for large, non rigid molecules which are described by an all atom force field.

\section{Model and Simulations}\label{s:mod_sim}
\subsection{Molecular Modeling}
The interface force field (IFF), which was originally developed by Heinz et al.\cite{heinz2013thermodynamically} to simulate organic and inorganic surfaces, features parameters to model accurately a wide range of molecules. Additionally it can be applied very flexibly and it can be combined with the potentials of other force fields like the polymer-consistent force field (PCFF)\cite{sun1994ab,sun1994force,sun1998compass}, the consistent-valence force field (CVFF) \cite{dauber1988structure} or the CHARMM force field \cite{mackerell1998all} as has been described in Refs.~\cite{heinz2013thermodynamically,pramanik2017carbon}. In this work we use the polymer-consistent force field - interface force field (PCFF-IFF)\cite{heinz2013thermodynamically,pramanik2017carbon,heinzWeb2021}. The potential energy of this force field is given by
\begin{align}
    E_{\text{pot}}&=\sum_{i,j\text{ bonded}} \sum_{n=2}^4 K_{n,ij}(r_{ij}-r_{0,ij})^n \nonumber \\
    &+ \sum_{i,j,k\text{ bonded}}\sum_{n=2}^4 K_{n,ijk}(\theta_{ijk}-\theta_{0,ijk})^n \nonumber\\
    &+ \sum_{i,j,k,l\text{ bonded}}\sum_{n=1}^3 K_{n,ijkl}(1-\cos(n\phi_{ijkl}-\phi_{n,ijkl})) \nonumber \\
    &+ \sum_{i,j,k,l\text{ bonded}}K\left(\dfrac{\chi_{ijkl}+\chi_{kjli}+\chi_{ljik}}{3}-\chi_{0,ijkl}\right) \nonumber \\
    &+ E_{\text{cross}} \nonumber \\
    &+ \sum_{i,j\text{ not bonded}}a_{ij}\dfrac{q_iq_j}{r_{ij}} \nonumber \\
    &+ \sum_{i,j\text{ not bonded}}\epsilon_{ij}\left(2\left(\dfrac{\sigma_{ij}}{r_{ij}}\right)^9-3\left(\dfrac{\sigma_{ij}}{r_{ij}}\right)^6\right). \label{eq:PCFF_potential}
\end{align}
It consists of two main parts: In the first five lines of Eq.~\ref{eq:PCFF_potential} the bonds of the atoms are described, which have contributions by the bond lengths $r$, the angles $\theta$, the torsion angles $\phi$ and the out of plane angles $\chi$.  Additionally there is a term which contains the cross terms between the previous contributions. The last two terms contain a Coulomb potential, which acts onto the partial charges of the molecules, and a 9-6 van der Waals term for the non-bonded interactions.

We observed that the anharmonic contributions to the bond-bond and the angle-angle interactions did not have a significant impact on the structures of the clusters and that they only induced a small shift in the PES. In retrospect, we conclude that these terms could have been omitted in our study.

\subsection{Optimization Method}
We search for the global minimum of the PES of acene clusters with respect to the positions of the atoms in the cluster with the basin hopping (BH) algorithm \cite{wales1997global}. We use the PELE package\cite{PELEsource} which is an adaption for python of the public domain package GMIN \cite{GMINsource}. For the type of problem which we study here, the number of minima depends exponentially on the size of the system, rendering the optimization a computationally expensive problem \cite{doye2002saddle}.

\begin{figure}[tb]
    \centering
    \includegraphics[width=0.99\linewidth]{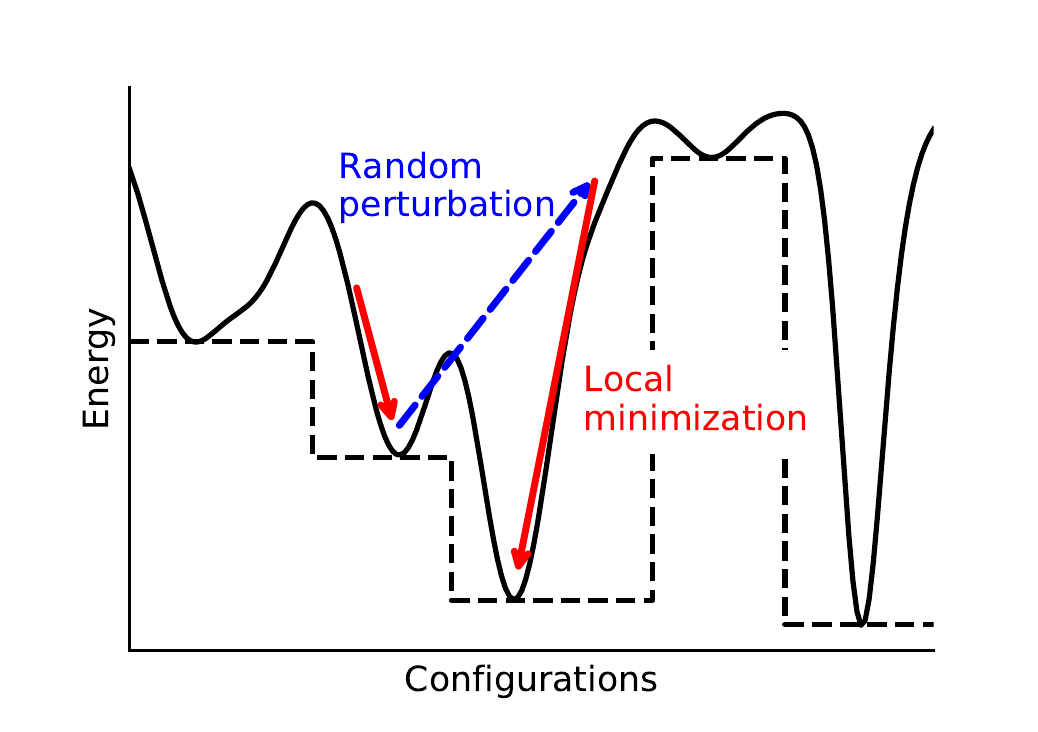}
    \caption{Schematic of the BHMC method. The black continuous line is an exemplary PES. The dashed line is the effective PES which is obtained by the local minimization.}
    \label{fig:basinHopping}
\end{figure}

The principle of this optimization method is illustrated in Fig.~\ref{fig:basinHopping}. The BH algorithm transforms the PES into an effective PES $\tilde{E}(\textbf{x})=\text{min}\{E(\textbf{x})\}$ where $\textbf{x}$ is the $3N$ dimensional vector of the atom positions and $\text{min}\{\}$ is the local minimization of the PES, starting from $\textbf{x}$ \cite{wales1997global}.
The algorithm contains three steps which are repeated between \num{3000} times for dimers and \num{10000} times for $N=30$ clusters:
\begin{itemize}
    \item[$1$] The system is perturbed by random moves.
    \item[$2$] The energy of the new configuration is minimized by a local minimization algorithm.
    \item[$3$] The new state is accepted or rejected, according to the Metropolis criterion.
\end{itemize}
In the following, we will discuss the details of these steps. The most basic movement is to move each atom individually by a small random value and direction. Although, in principle, this is enough to reach every molecular configuration of the cluster, it turned
out that by performing additional movements (translations and rotations) of the whole molecules, the global minima are often found faster. The size of the random movements is not fixed but gets adjusted in each step to approach the global minimum as fast as possible. The same is done with the temperature to get a better acceptance ratio, starting from initial temperatures $k_BT=\num{1},~\num{3}$ and \SI{5}{\Calorie\per\mol}. Sometimes molecules get stuck at one side of the cluster and are not able to find a way to the energetically preferable place on the other side of the cluster. We use an additional movement type to increase the probability of reaching minima with lower energy by moving the molecule with the highest energy by a random angle on a sphere around the cluster.

For the local minimization, we use the quasi-Newtonian \cite{Press1988Numerical} Broyden-Fletcher-Goldfarb-Shanno (BFGS) algorithm \cite{broyden1970convergence,fletcher1970new,goldfarb1970family,shanno1970conditioning}. In this method, the Hessian matrix does not need to be calculated exactly. It is iteratively approximated.

The new, optimized configuration is finally accepted with the probability given by Eq.~\ref{eq:metropolisProb}, the Metropolis criterion.
\begin{align}
    p_{\text{a}}=\text{min}\left(1,\exp\left(-\dfrac{\Delta}{k_BT}\right)\right) \quad ,\label{eq:metropolisProb}
\end{align}

where $\Delta$ is the difference in energy between the old and the new local minimum. In order for a configuration to be interpreted as a "new structure", the energy between a previously found PES minimum and a new one has to be at least \SI{0.001}{\Calorie\per\mol}. To verify that the resulting global minima do not depend on the initial conditions of the optimization processes, we run for each cluster up to four independent simulations.

\subsection{Heat Bath Coupling} \label{sec:heatBathCoupling}
In section~\ref{sec:tempDep}, we discuss the behavior of local minima at temperatures larger than \SI{0}{\kelvin}. We have performed  Metropolis Monte Carlo (MC) simulations \cite{metropolis1953equation}. The starting configurations of these simulations are local minimum configurations of the BH simulations. The aim of these MC simulations is to assess how the free energy landscape is transformed on heating, and in particular which of the minima becomes unstable at which temperature. This method is thus similar to the transition pathway algorithms which were used e.g. by Wales to find crossovers between different states. One way to illustrate such transitions would be by disconnectivity graphs\cite{becker1997topology,wales1998archetypal,evans2003free}.

In the MC simulations, we begin with a first stage of \num{10,000} sweeps, where a sweep consists of as many random translations of the molecules and single atoms or rotations of the molecules as there are atoms in the system. These \num{10,000} sweeps are needed to equilibrate the system at a temperature of $k_{\text{B}}T=\SI{0.02}{\Calorie\per\mol}$. After that we start heating the system. The first \num{50,000} sweeps are again carried out at $k_{\text{B}}T=\SI{0.02}{\Calorie\per\mol}$. Then the temperature is increased by $k_{\text{B}}\Delta T=\SI{0.05}{\Calorie\per\mol}$. This is repeated 14 times such that the final temperature is at $k_{\text{B}}T=\SI{0.72}{\Calorie\per\mol}$ which is sufficient to split the cluster into individual molecules.

At each temperature, we take snapshots from the last \num{25,000} MC sweeps since there the clusters are in equilibrium at the new temperature. We use these snapshots to determine the accessibility of local minima of the PES at a given temperature, starting from some other local energy minimum. In order to do so, we identify the closest minimum by minimizing the energies of these snapshots. For this, we utilize again the BFGS algorithm of the PELE package. The results discussed in Sec.~\ref{sec:tempDep}, have been obtained by running five independent heating simulations for each of the 36 starting configurations.

\section{Results}\label{s:results}
\subsection{Structures of Anthracene, Tetracene and Pentacene Clusters}
In studies of crystalline acene molecules several arrangements of the molecules have been observed \cite{robertson1961crystal}. Depending on the specific arrangement of the aromatic rings and whether there are other chemical groups attached to the molecules, the molecules form e.g.~graphene like layers or have a T-shaped stacking. The perhaps most recognizable structure is the herring bone shape.

\begin{figure*}[tbp]
    \centering
    \includegraphics[width=\textwidth]{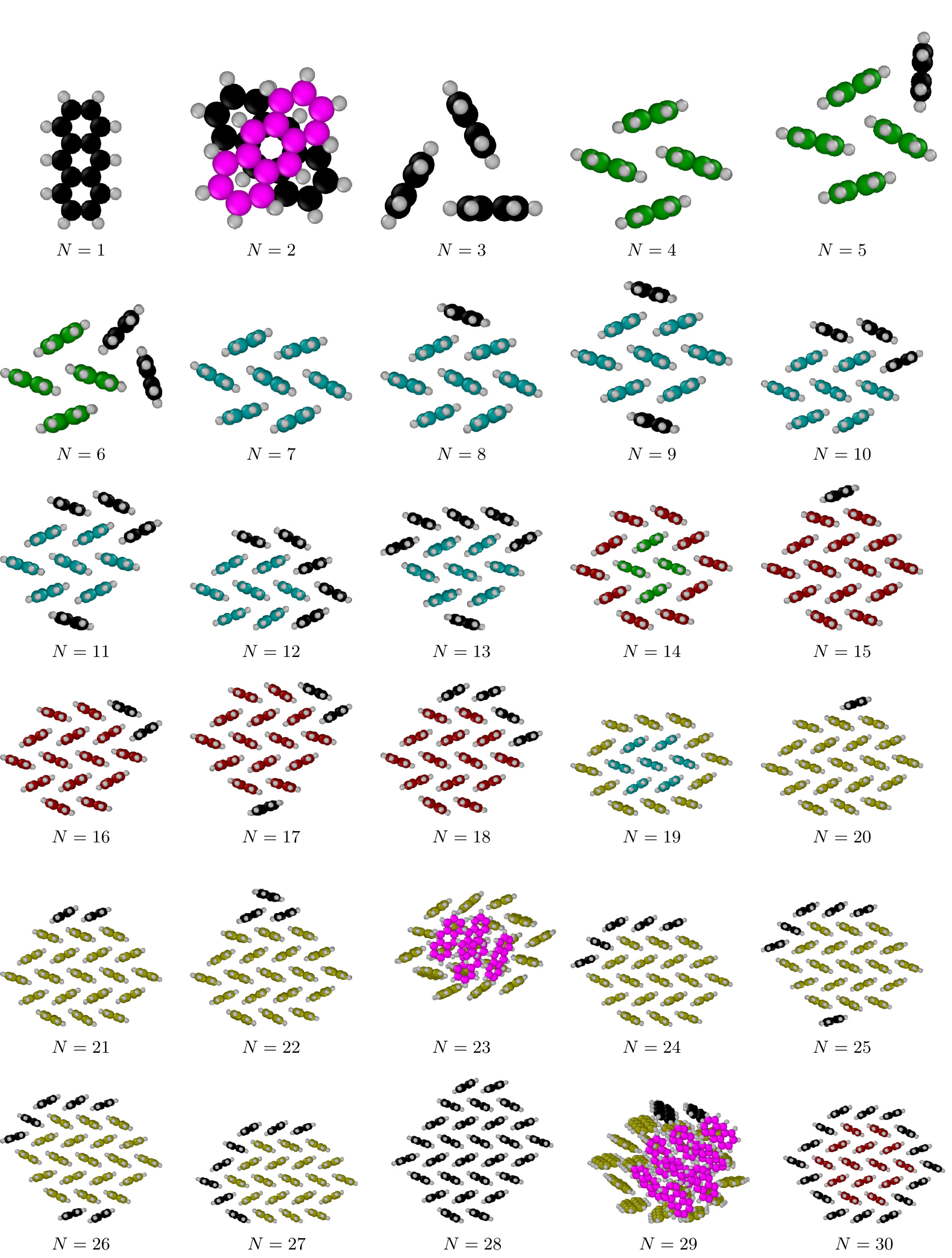}
    \caption{Minimal energy structures of the anthracene clusters. The size of the clusters increases from the top left (one molecule) to the bottom right (30 molecules). We see the clusters with more than three anthracene molecules along the long axis of the molecules. The colors indicate the formation of layers of molecules on top of clusters with high symmetry. The structures are plotted with Ovito\cite{ovito}.}
    \label{fig:3aceneStruct}
\end{figure*}

\begin{figure*}[tbp]
    \centering
    \includegraphics[width=\textwidth]{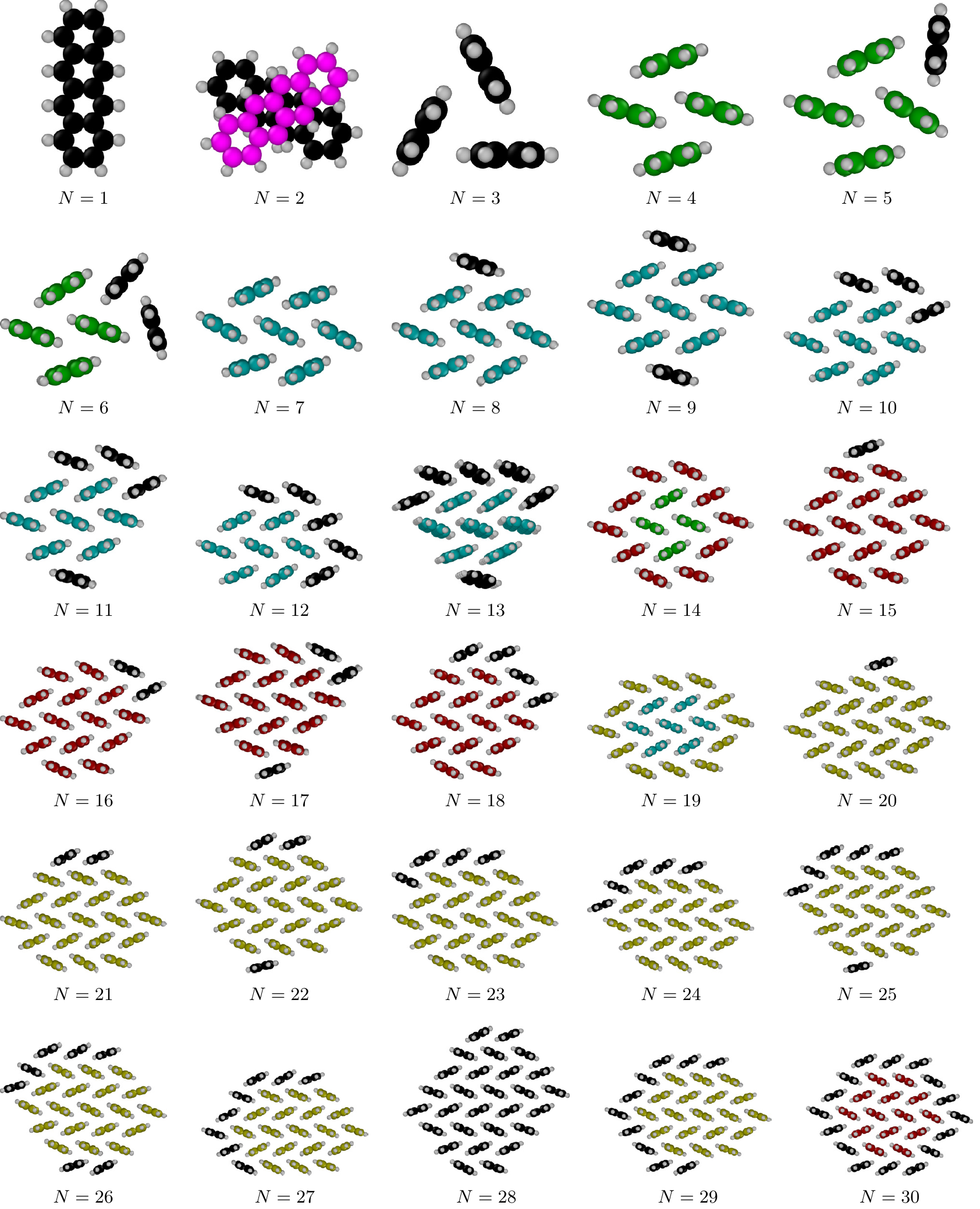}
    \caption{Minimal energy structures of the tetracene clusters. The size of the clusters increases from the top left (one molecule) to the bottom right (30 molecules). We see the clusters with more than three tetracene molecules along the long axis of the molecules. The colors indicate the formation of layers of molecules on top of clusters with high symmetry. The structures are plotted with Ovito\cite{ovito}.}
    \label{fig:4aceneStruct}
\end{figure*}

\begin{figure*}[tbp]
    \centering
    \includegraphics[width=\textwidth]{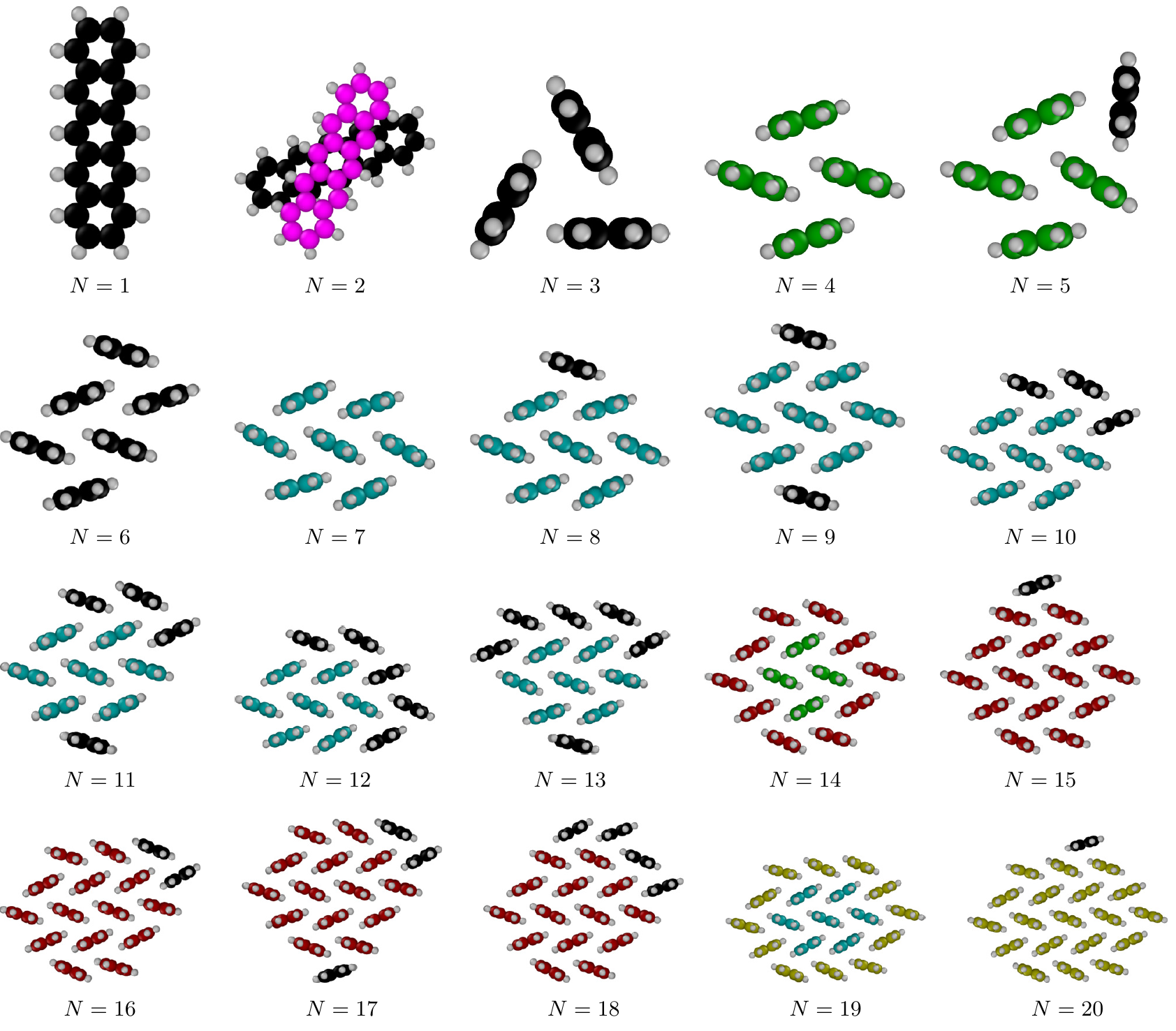}
    \caption{Minimal energy structures of the pentacene clusters. The size of the clusters increases from the top left (one molecule) to the bottom right (20 molecules). We see the clusters with more than three pentacene molecules along the long axis of the molecules. The colors indicate the formation of layers of molecules on top of clusters with high symmetry. The structures are plotted with Ovito\cite{ovito}.}
    \label{fig:5aceneStruct}
\end{figure*}

The structures which we obtained with the BHMC algorithm for the global minima of the PES of anthracene, tetracene and pentacene clusters with up to \num{30} molecules per cluster are presented in Fig.~\ref{fig:3aceneStruct},~\ref{fig:4aceneStruct} and~\ref{fig:5aceneStruct}.


\begin{figure}[tb]
    \centering
    \includegraphics[width=0.45\linewidth]{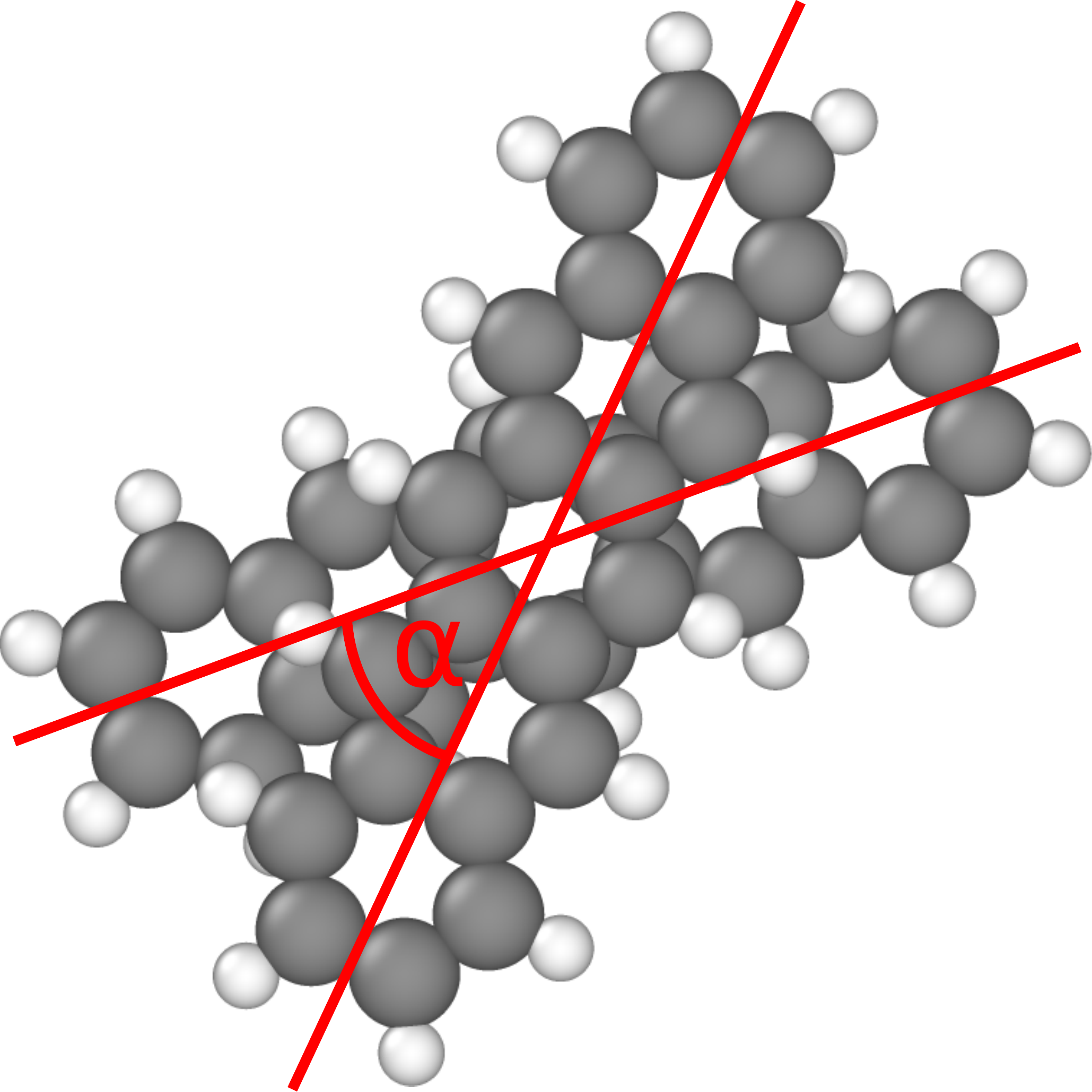}
    \includegraphics[width=0.45\linewidth]{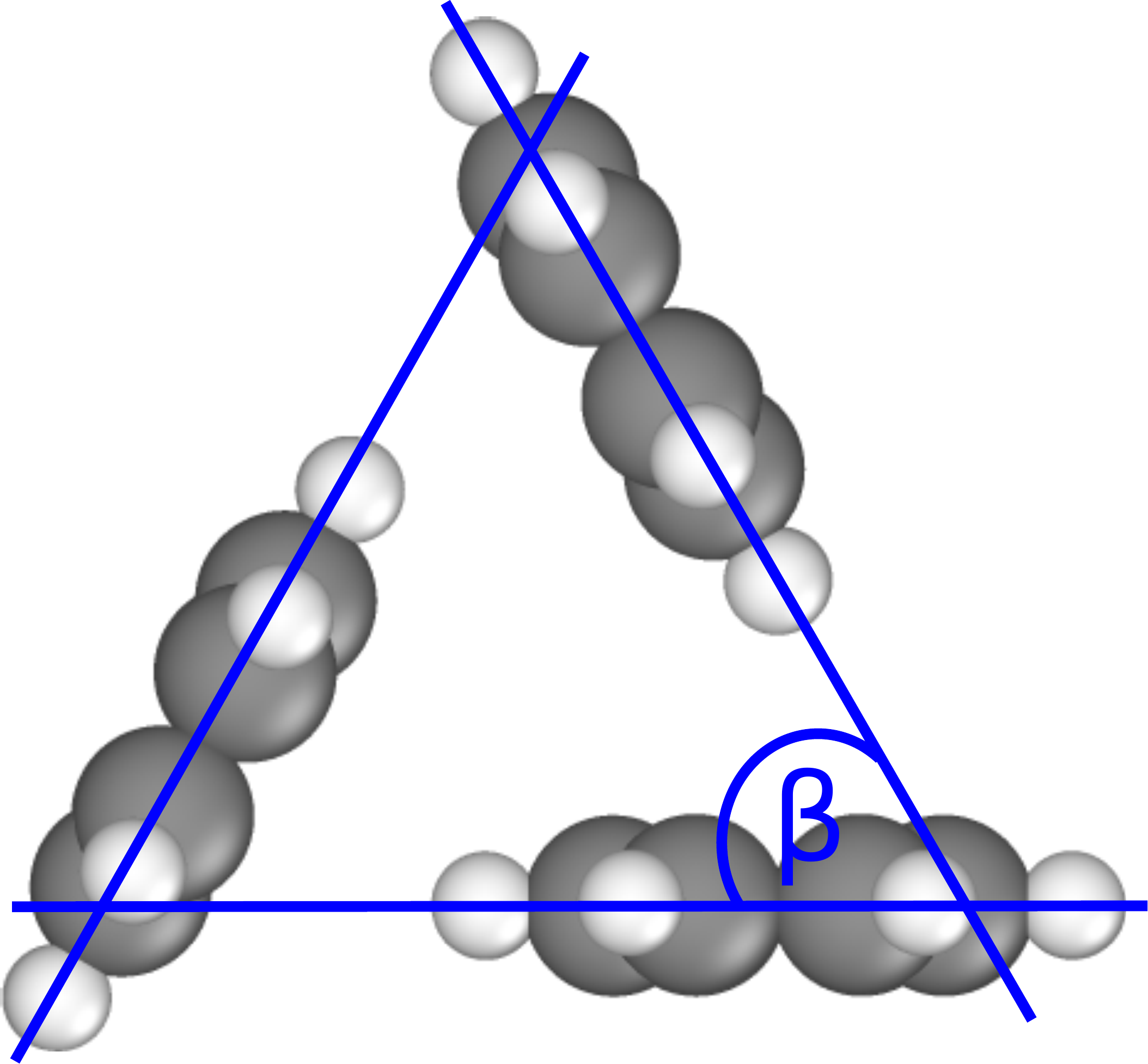}
    \caption{Relevant angles in the di- and trimer of acene clusters, depicted exemplary for pentacene. $\alpha$ is the rotation angle between the long axes of the two molecules in the dimer. $\beta$ is the tilting angle between the wide axes of the molecules.}
    \label{fig:ClusterAngles}
\end{figure}

We will start discussing the structures with the smallest cluster. In the $N=2$ clusters, the molecules of all three types of acene molecules are parallel to each other, forming a double layer. In the anthracene cluster this leads to a nearly orthogonal configuration with an angle $\alpha = \SI{89.6}{\degree}$ between the long axes of the two molecules. (In Fig.~\ref{fig:ClusterAngles}, the angles which we consider here, are illustrated.) This orientation is much more efficient than a parallel orientation of the long axes of the molecules. The energy difference between a parallel configuration, where the hydrogen atoms are above the centers of the aromatic rings, and the orthogonal configuration, is $\Delta E = \SI{0.6114}{\Calorie\per\mol}$. In tetracene, the angle between the two molecules is $\alpha = \SI{65.7}{\degree}$. There is again a large energy difference of $\Delta E =\SI{0.732}{\Calorie\per\mol}$ to a parallel arrangement of the molecules. In the pentacene dimer, the angle between the molecules is even smaller with $\alpha=\SI{44.5}{\degree}$. Nevertheless the energy difference to the structure with two parallel molecules is $\Delta E = \SI{0.533}{\Calorie\per\mol}$, which is of comparable size to the two smaller molecules.

The smallest cluster, in which we found the minimal PES for a structure where the long axes of the molecules are parallel, has three acene molecules. This observation is in agreement with previous experiments and simulations \cite{takeuchi2013structures,piuzzi2002spectroscopy}. These molecules are arranged in a triangular shape with average angles of $\beta = \SI{60.00(2)}{\degree}$ for anthracene, $\beta = \SI{60.0000(14)}{\degree}$ for tetracene and $\beta = \SI{60.0000(2)}{\degree}$ for pentacene as illustrated in Fig.~\ref{fig:ClusterAngles}. The deviations from \SI{60}{\degree} originate from the convergence of the local optimization. These clusters are, together with the tetramer, the basic building blocks of larger clusters.

For the clusters with four molecules, there have been reports of different structures. Takeuchi \cite{takeuchi2013structures} and Piuzzi et al. \cite{piuzzi2002spectroscopy} found for anthracene a structure similar to the trimer with one additional molecule attached to it. This structure corresponds to a local minimum in our optimization. The energy difference between our global minimum and that structure is \SI{0.319}{\Calorie\per\mol}. The structures which we observe are instead the same as those reported by Park et al. \cite{park2007ab} for benzene tetramers.

The configurations depicted for anthracene and tetracene in Fig.~\ref{fig:3aceneStruct} and \ref{fig:4aceneStruct} have the same structures for all clusters with fever than 22 molecules. Pentacene already differs from these structures at smaller cluster sizes. The cluster with 6 pentacene molecules is, in contrast to the clusters of the two smaller acenes, not a combination of the trimer and the tetramer, but instead a combination of the structure of two tetramers. To check whether our algorithm has missed the global minimum in one of these cases, we searched for corresponding configurations in the other clusters with 6 molecules. In all three cases, we found local minima that have the same overall structure as the global minimum in pentacene or anthracene/tetracene. The difference in energy between the two minima for anthracene is $\Delta E= \SI{0.159}{\Calorie\per\mol}$. In the case of tetracene it nearly vanishes with $\Delta E=\SI{0.003}{\Calorie\per\mol}$. In the pentacene cluster, the difference increases again to $\Delta E=\SI{00.167}{\Calorie\per\mol}$. This behavior may hint at a transition between the two structure types at tetracene, since the energy difference is here smaller by two orders of magnitude than that of anthracene and pentacene.

Another cluster size where we have found differences between the structure are the anthracene and tetracene clusters with 22 molecules. The crosssection of the anthracene cluster has a triangular shape with two rounded corners. Such a structure can be found for tetracene as a local minimum with an energy difference $\Delta E=\SI{0.010}{\Calorie\per\mol}$ to the global minimum. The tetracene structure is at its PES ground state more circular since the 'tip' of the triangle is attached at the other side. In anthracene, such a structure has an energy difference of $\Delta E=\SI{0.015}{\Calorie\per\mol}$ to the global minimum.

The smallest cluster which for the global PES minimum has more than one layer of molecules, is the anthracene cluster with 23 molecules. There is one layer which is again very similar to the structure of the cluster with 19 molecules and the remaining particles are attached to it. We observe the same behavior for the cluster with 29 molecules. In this cluster, the first layer resembles that of the cluster with 21 molecules. The additional molecules in both cases form on top of these known layers a three dimensional structure, i.e.~they do not lie flat on the first layer but instead form a distorted herring bone structure. We will revisit this behavior in Sec.~\ref{sec:relStab}.

\subsection{Molecules in the $N=30$ Anthracene Cluster}
\begin{figure}[tb]
    \centering
    \includegraphics[width=0.99\linewidth]{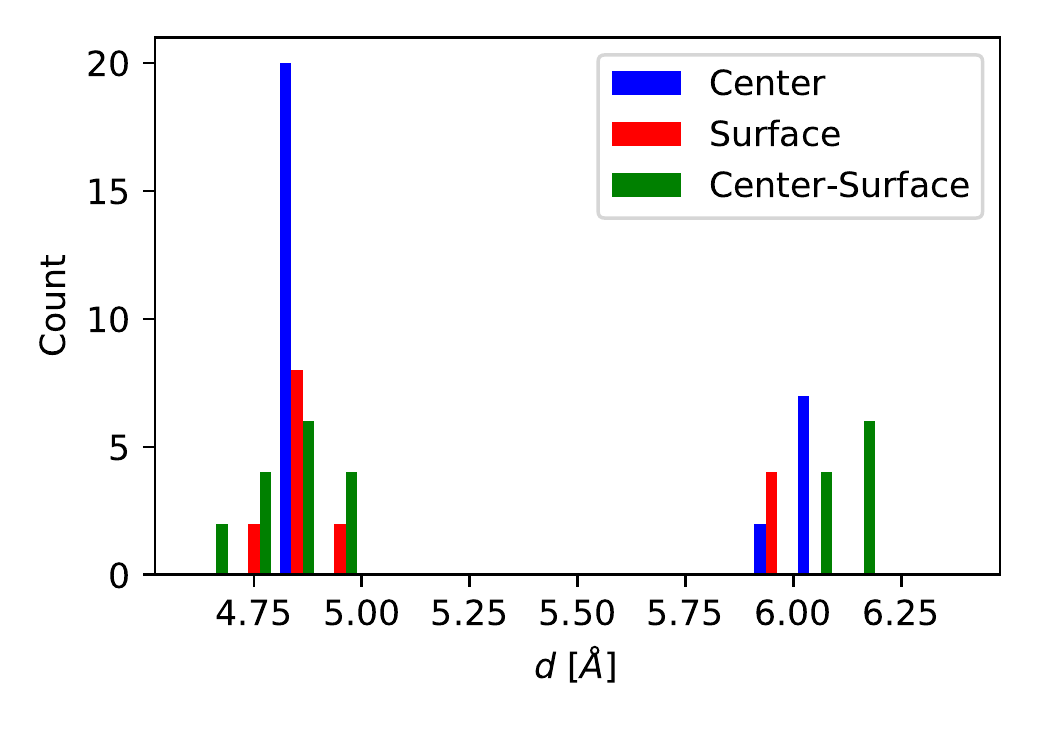}
    \caption{Distances between the center of mass of the molecules in the $N=30$ anthracene cluster. The width of the bins is \SI{0.1}{\angstrom}.}
    \label{fig:3distDev}
\end{figure}

To give an example of the influence which the relative location of a molecule in the cluster has, we will take here a closer look at the distance and angle distributions between the molecules of the anthracene cluster with 30 molecules. In Fig.~\ref{fig:3distDev}, the distribution of the distances between the molecular center of masses at the global minimum of the PES is depicted.
We distinguish here three types of molecular distances: Distances between molecules which have at least six other molecules closer than \SI{7}{\angstrom} are identified as molecules in the center of the cluster. The second group are distances between all other molecules which are thus at the cluster-surface. The third distance type are the distances between the molecules inside of the cluster and those at the surface.

\begin{figure}[tb]
    \centering
    \includegraphics[width=0.99\linewidth]{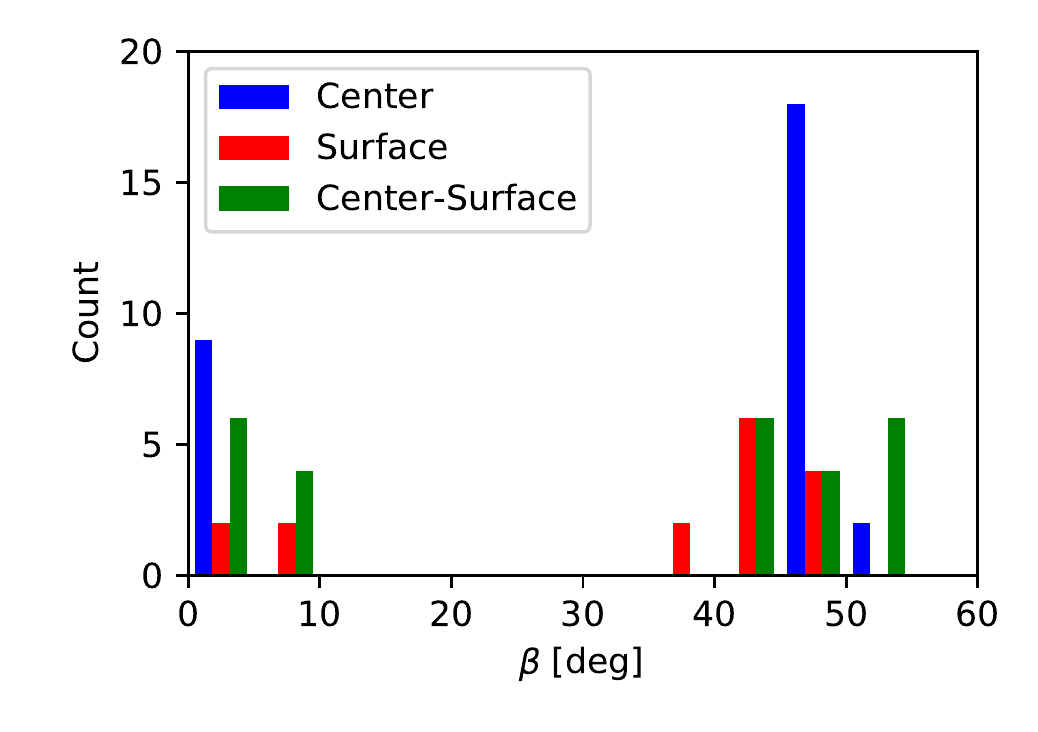}
    \caption{Relative angles $\beta$ between the molecules in the global minimum structure of the $N=30$ anthracene cluster. The width of the bins is \SI{5}{\degree}.}
    \label{fig:3angleDev}
\end{figure}

Independent of the type of the involved molecules, there are two distinct peaks. They are a result of the herring bone structure. The peak around \SI{6}{\angstrom} are the distances between the molecules in the horizontal bands in the cluster depicted in Fig.~\ref{fig:3aceneStruct}. The other peak at about \SI{4.8}{\angstrom} consists of the distances between molecules which are not parallel to each other. In this peak, the distances between molecules in the center of the cluster are much more focused than those including molecules at the surface but we do not observe that the distances are squeezed inside of the cluster. At the other peak, a slight shift of the distances can be observed, between distances which are entirely in the cluster and those between the center- and the surface-molecules. In the distribution of the relative angles $\beta$ (see Fig.~\ref{fig:ClusterAngles}) between the molecules, we observe the same splitting into two peaks due to the herring bone structure of the acene clusters. This is depicted in Fig.~\ref{fig:3angleDev}. In both peaks, around \SI{0}{\degree} and around \SI{45}{\degree}, the angles inside of the cluster have a narrower distribution than the angles which include molecules at the surface.

\subsection{Relative Stability of Acene Clusters} \label{sec:relStab}
\begin{figure}[tb]
    \centering
    \includegraphics[width=0.99\linewidth]{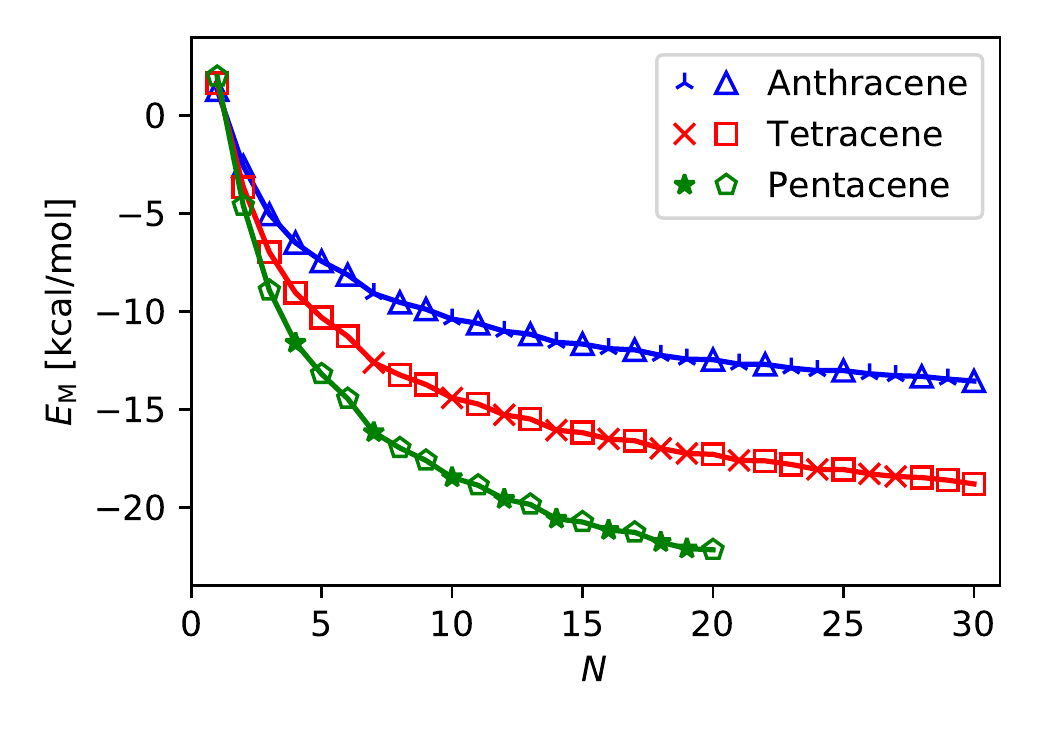}
    \caption{Average energy per molecule $E_{\text{M}}$ as a function of cluster size $N$. The stars and the cross symbols correspond to the stable structures in Fig.~\ref{fig:relStab}.}
    \label{fig:EperN}
\end{figure}

An important property for the stability of clusters is the average potential energy of one molecule $E_{\text{M}}$ which is closely related to the binding energy $E_{\text{B}}$ \cite{bartolomei2017modeling}. It is calculated by
\begin{align}
    E_{\text{M}}=\dfrac{E_{\text{C}}}{N} = -1\cdot E_{\text{B}}
\end{align}
with the total energy $E_{\text{C}}$ of a cluster with $N$ molecules. In Fig.~\ref{fig:EperN}, $E_{\text{M}}$ is depicted for clusters with up to 30 molecules for anthracene and tetracene. For pentacene we present the energies for up to 20 molecules.
We observe here, that after an initial steep drop in energy for the smallest few clusters,
starting at a cluster size of about ten to 15 molecules, the decrease becomes  approximately linear. The size at which this linear regime begins depends on the size of the molecule: for anthracene it starts more closely to the cluster with ten molecules, for tetracene closer to 15 and for pentacene we can not identify such a regime in the analyzed clusters with up to 20 molecules. It is as well worth to note, that the energy per molecule decreases for clusters with larger molecules since larger acene molecules have more atoms with opposing partial charges which allows for stronger binding forces between them.

\begin{figure}[tb]
    \centering
    \includegraphics[width=0.99\linewidth]{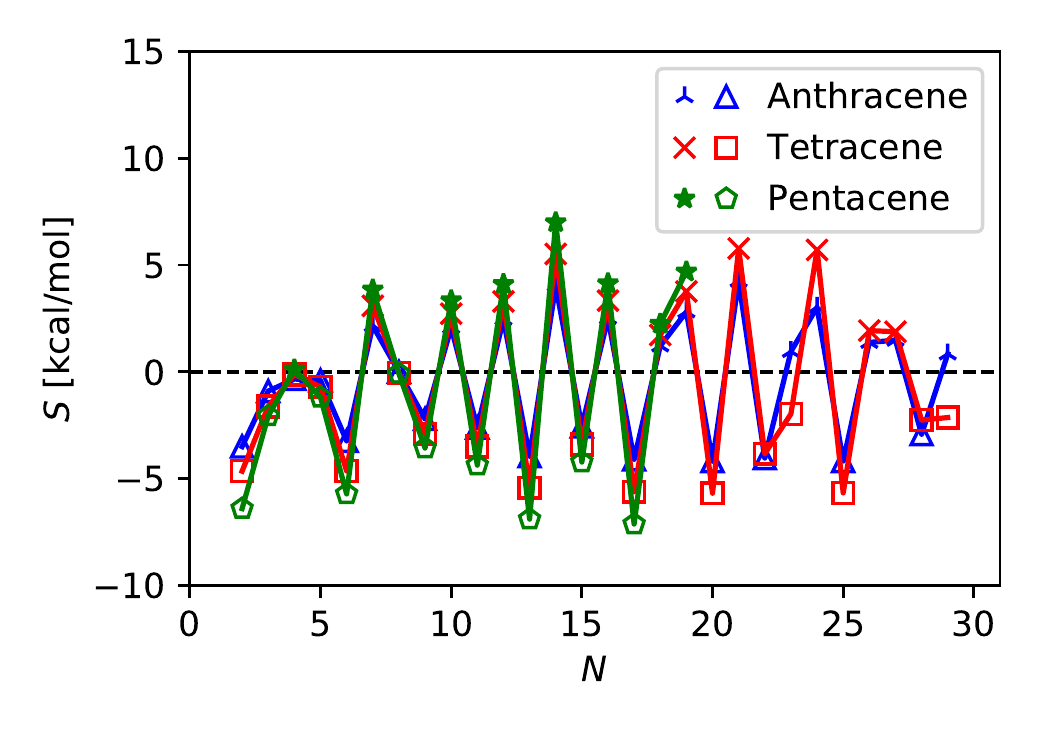}
    \caption{Relative energy difference $S$ of anthracene, tetracene and pentacene clusters of size $N$. The energy difference is calculated according to Eq.~\ref{eq:relStab}. The configurations with a positive relative energy difference are marked by three-pointed stars for anthracene, crosses for tetracene and five-pointed stars for pentacene.}
    \label{fig:relStab}
\end{figure}

Another quantity which helps to understand the stability of clusters is the relative energy difference
\begin{align}
    S(N)=E_{\text{C}}(N-1)+E_{\text{C}}(N+1)-2E_{\text{C}}(N)   \label{eq:relStab}
\end{align}
in which the global minimum of the PES of one cluster $E_{\text{C}}(N)$ of size $N$ is twice subtracted from the sum of the global minima of the PES of the clusters with one molecule more and with one less. This is plotted in Fig.~\ref{fig:relStab}, again for the same clusters as the average energy per molecule in Fig.~\ref{fig:EperN}.
A value larger than zero indicates a cluster which is more stable than its two direct neighbors. We see here directly the oscillation in the relative energy difference for all three molecules which has been reported previously for tetracene anions \cite{mitsui2007mass}, although there are some deviations in the stable cluster sizes. For the smallest clusters ($2\leq N\leq6$), the most stable clusters have four molecules with $S\approx0$. This makes sense since this cluster already has the herring bone shape which we encounter in the larger clusters and as well in crystals.

\begin{figure}[tb]
    \centering
    \includegraphics[width=0.99\linewidth]{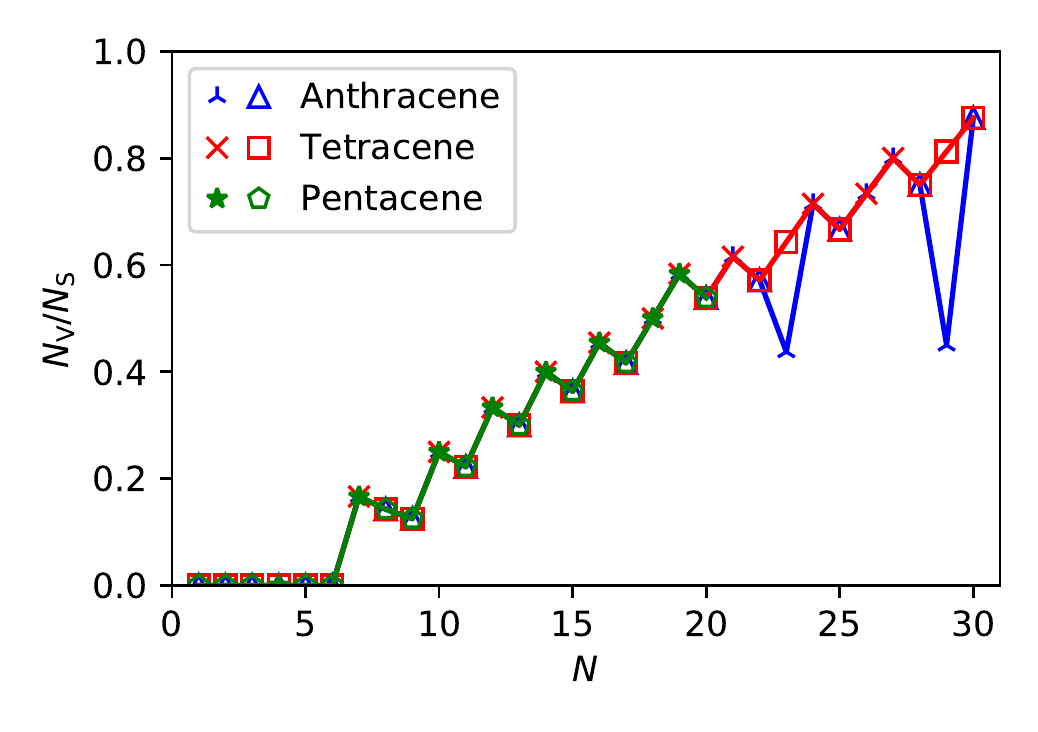}
    \caption{Ratio of the number of molecules in the volume of the cluster $N_{\text{V}}$ (with at least six neighbors) to the number of moleculres at the surface $N_{\text{S}}$. The cutoff distance for nearest neighbors is \SI{7}{\angstrom}. The symbol scheme is the same as in Fig.~\ref{fig:EperN} and \ref{fig:relStab}.}
    \label{fig:NearestNeighbor}
\end{figure}

The behavior of the stability of the larger clusters can be explained by the number of molecules which are surrounded by other molecules. In Fig.~\ref{fig:NearestNeighbor}, the ratio between the number of molecules which are surrounded by at least six other molecules in a distance of \SI{7}{\angstrom} and the number of molecules at the surface of the cluster is depicted. 
The first cluster, larger than $N=6$, with a high relative energy difference is the one with seven molecules, which is as well the first cluster in which we find a molecule that has a complete outer shell. For the next two cluster sizes, the ratio between volume and surface molecules drops again. This results in a decreased stability. $N=10$ has again a higher ratio and as well a higher relative energy difference etc.. A cluster with an especially high relative energy difference is the cluster with 14 molecules which thus corresponds to a magic number. While this does not result in an extremely high ratio in Fig.~\ref{fig:NearestNeighbor}, we see in Fig.~\ref{fig:3aceneStruct} to \ref{fig:5aceneStruct}, that this cluster has again a similar structure as the $N=7$ cluster, but with one additional layer inserted in the middle.

For the clusters $N>22$, the relative energy differences differentiate between anthracene and tetracene. For tetracene, we continue to have a match between the relatively most stable structures and the ratios of the center and surface molecules. In anthracene, there are clusters in which the molecules do not form a monolayer. These lead to a very low ratio in Fig.~\ref{fig:relStab}. Nevertheless these structures have, other than tetracene, a relative energy difference $S>0$.

The influence of the symmetry can be seen in the stability as well. For the smallest clusters of all three molecules, non trivial point groups do not contribute strongly to the overall stability: The clusters with $N=2$ and $N=3$ have D$_2$ respectively C$_{3\text{h}}$ symmetries while a negative relative energy difference is observed. The $N=4$ clusters are the first ones with a C$_{2\text{h}}$ symmetry. The same symmetry is again found for $N=7$, $9$, $14$, $19$ and for anthracene and tetracene for $N=30$. Except for $N=9$ and $N=30$, these cluster sizes have a high relative energy difference. Overall, the most important factor for the relative energy difference is nevertheless the ratio of molecules in the volume to that on the surface.

\subsection{Temperature Dependence of Local Minima of Anthracene Clusters} \label{sec:tempDep}
\begin{figure}[tbp]
    \centering
    \includegraphics[width=\linewidth]{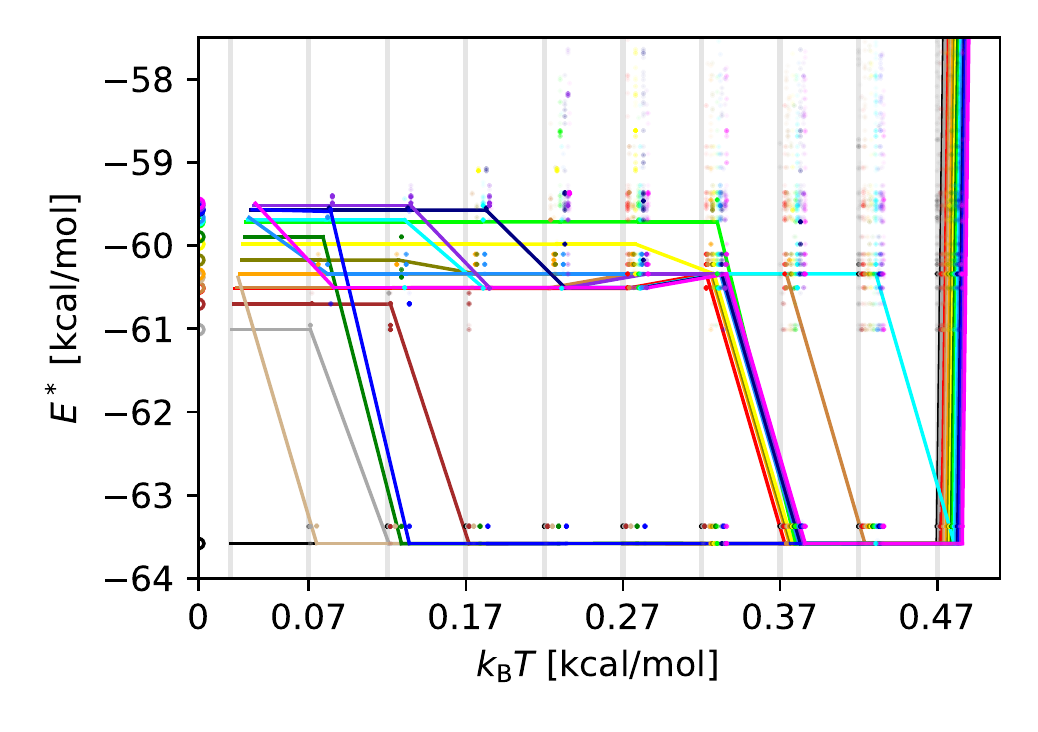}
    \includegraphics[width=\linewidth]{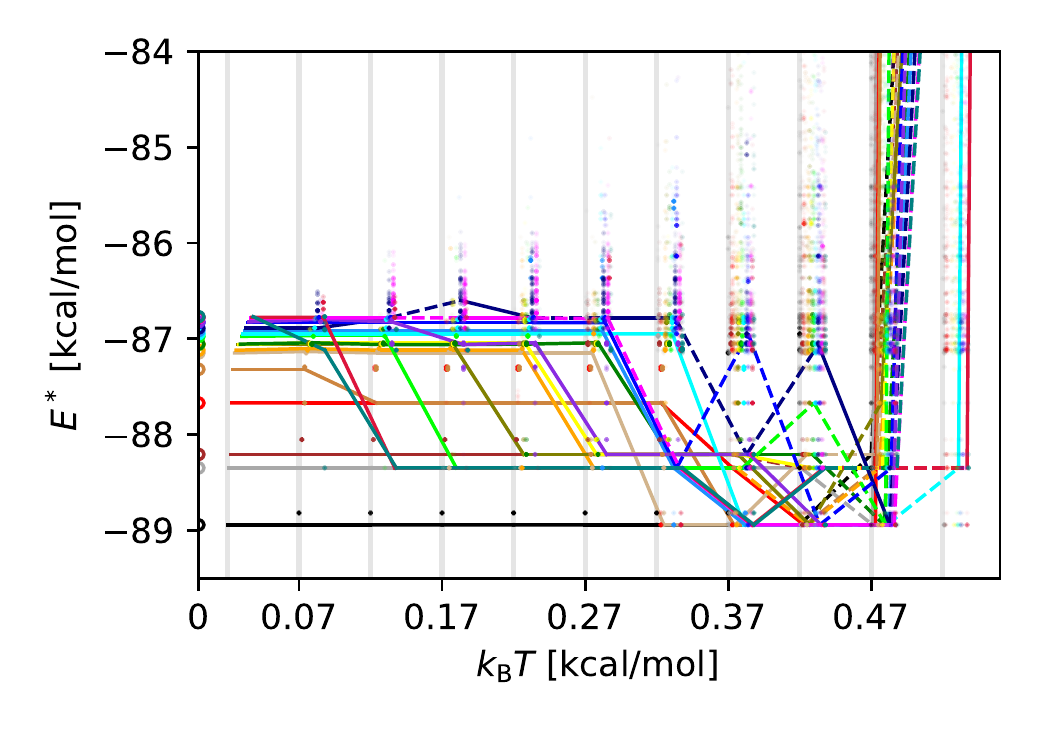}
    \caption{Transition pathways for clusters of seven (top) and nine (bottom) anthracene molecules. The semi-circles on the ordinate indicate the energy of the local minima at $k_{\text{B}}T=\SI{0}{\Calorie\per\mol}$ which are used as a starting configurations. The vertical lines indicate the temperatures at  which the MC simulations are run. The solid lines (color online) indicate the most likely transitions on heating of the clusters. If the transition with the highest probability occurs in less than \SI{15}{\percent} of the simulations, it is drawn as a dashed line. All other observed configurations are depicted by circles where the probability of visiting a state scales with the intensity of the circle. The temperature values are shifted in order to make the data points more easily distinguishable, i.e.~all data points between two horizontal lines belong to the lower one.}
    \label{fig:heating}
\end{figure}

In this section we present an estimate of how the free energy landscape of a cluster gets transformed during heating. All data presented here have been obtained by the MC method which we have explained in Sec.~\ref{sec:heatBathCoupling} with the subsequent optimization of the PES. 
In Fig.~\ref{fig:heating}, the colored dots on the ordinate indicate the potential energy minima in which the simulations were initialized.
On the abscissa we plotted the temperatures at which we relaxed each system in its local free energy minimum.
Horizontal lines indicate that the simulations remained in a given cluster structure on heating, i.e.~the local free energy minimum was stable. If a minimum became unstable and the simulation relaxed into another structure, we indicated the transition by a line to the location of the potential energy minimum corresponding to the new structure. Thus the graph presents "heating pathways".  In some cases simulations initialized on one given structure relaxed into several different minima. Then we plotted only the path which occurred with the highest probability. Other minima which were also reached, but less frequently, are marked by colored dots. At temperatures close to the thermal dissociation of the cluster, the variability of the transition paths increases since a large number of local minima get visited. Paths to minima which are visited with a probability of less than \SI{15}{\percent}, but which are still the most visited configurations, are marked as dashed lines.
These two cluster sizes serve as examples for clusters with a global minimum of high relative stability ($N=7$) and of low relative stability ($N=9$).

As starting configurations, we choose for the cluster with seven molecules the configuration of the lowest \num{17} minima which can be distinguished by visual inspection. For the nine-molecule cluster, we use the lowest \num{19} distinguishable minima. We use for each starting configuration and temperature \num{250} snapshots from five independent simulations.

\begin{figure}[tbp]
    \centering
    \includegraphics[width=\linewidth]{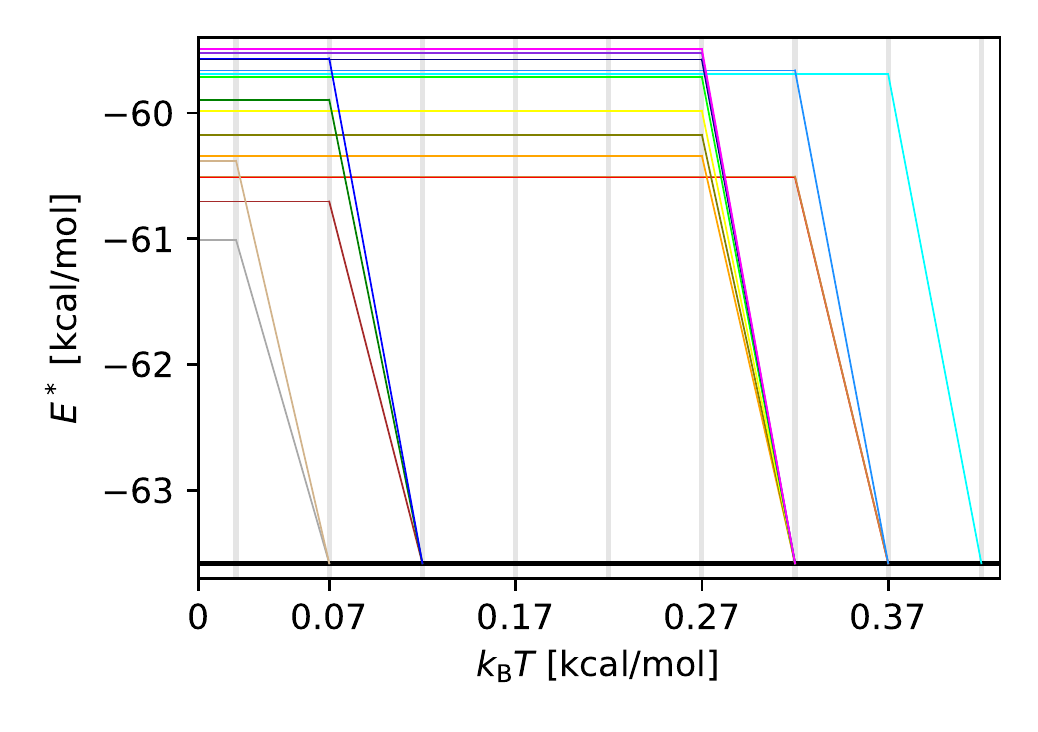}
    \includegraphics[width=\linewidth]{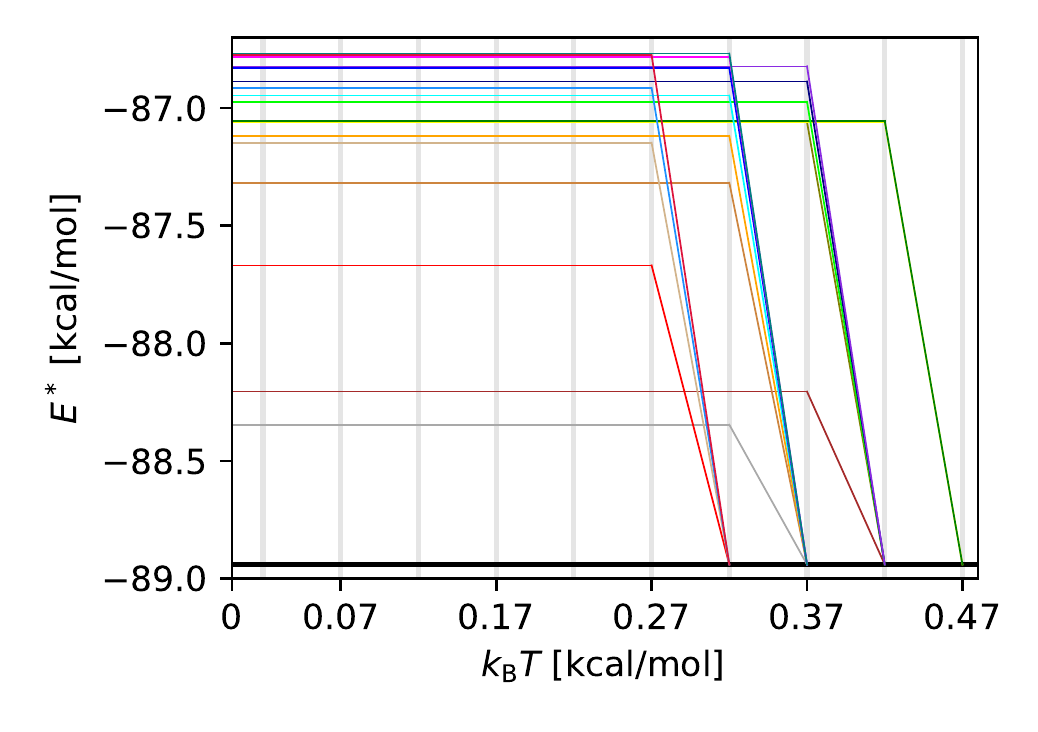}
    \caption{Lowest temperatures at which a transition from a local minimum to the global minimum of anthracene cluster with seven (top) and nine (bottom) molecules has been observed. The local minima which we use here as starting configurations are the same as in Fig.~\ref{fig:heating}. The black horizontal line marks the energy of the global minimum of the PES.}
    \label{fig:heatingMinTrans}
\end{figure}

For the $N=7$-cluster, the most prominent feature is, that there are some starting configurations, whose clusters approach the global minimum at \SI{-63.579}{\Calorie\per\mol} at very low temperatures. Already at an effective temperature of $k_{\text{B}}T=\SI{0.07}{\Calorie\per\mol}$ there is a transition from a local minimum to the global one. In comparison, the $N=9$ cluster has for the first time a transition from a local minimum to the ground state with high probability at $k_{\text{B}}T=\SI{0.37}{\Calorie\per\mol}$. This difference is even more visible if one considers the lowest temperatures at which transitions to the ground state occur. This is depicted in Fig.~\ref{fig:heatingMinTrans}.
We see here that in the $N=7$-cluster about one third of the local minima can reach the global minimum below $k_{\text{B}}T=\SI{0.12}{\Calorie\per\mol}$. All the other clusters reach the global minimum at temperatures above $k_{\text{B}}T=\SI{0.27}{\Calorie\per\mol}$. In Fig.~\ref{fig:heating}, we can see that most local minima which do not cross over directly to the global minimum move to configurations with lower energy where they get stuck and thus a higher temperature is needed in order to transfer them to into the global minimum. We conclude from this, that in experiments already the first assembly of the cluster at a high temperatures ($k_BT>\SI{0.27}{\Calorie\per\mol}$) for some clusters (e.g. $N=9$) has a strong influence on whether the structure with the lowest energy is found after the cooling process. Other clusters, like the one with seven molecules, have lower free energy barriers around the global optimum, so that local minima can still be reached at lower temperatures ($k_BT>\SI{0.02}{\Calorie\per\mol}$).

\section{Conclusion}\label{s:conc}
We have investigated the structure of anthracene, tetracene and pentacene clusters at the global PES minimum as an application of the BHMC optimization algorithm to complex systems with several hundreds to thousands of degrees of freedom. We were able to reproduce most structures of small clusters which have been reported in other publications. For the tetramer we found a structure with similar angles as the crystalline herring bone orientation, instead of a structure close to that of the trimer which was observed previously. On increasing the number of molecules in the cluster, we did not find any difference between the structures of anthracene and tetracene up to a cluster size of \num{21} molecules. For tetracene, we found the first deviation from the other clusters at a size of six molecules. Additionally, we found that there are anthracene clusters with fewer than \num{30} molecules which have a three dimensional cluster structure at the minimum of the PES. The cluster sizes which we identified to have a high relative stability are \num{7}, \num{10}, \num{12}, \num{14}, \num{16}, \num{18} and \num{19} for all three molecules and \num{21} and \num{24} for anthracene and tetracene. For anthracene we found, that \num{23} and \num{29} are additional stable structures, which are the three dimensional clusters.

The accessibility of the global minimum of the PES has been investigated for the cluster with seven molecules which has a high relative stability and the cluster with nine molecules as an example for a cluster with a low relative stability. For the $N=7$-cluster, we found that there are already at very low temperatures transitions from multiple local minima to the global minimum, indicating that for this kind of cluster, the starting preparation is not as crucial as for the $N=9$-cluster. In that cluster, transitions were only observed close to the temperature at which the cluster dissolves.

The table of structures presented here could serve as a guideline when photoelectron spectroscopy experiments of acene clusters are evaluated. Also when organic semiconductor films are produced by means of cluster deposition, the table of stable structures might be useful, as the structure and quality of the film depends on the clusters structures \cite{lee2003studies,kim2019organic}.

\section*{Supplementary Material}
See appendix B for the positions of the atoms in the clusters presented in Fig.~\ref{fig:3aceneStruct}-\ref{fig:5aceneStruct}. The full potential energy (see Eq.~\ref{eq:PCFF_potential}) and the force field parameters for carbon and hydrogen are provided in appendix A.

\section*{Conflicts of interest}
This article has been accepted by The Journal of Chemical Physics. After it is published, it will be found at \url{https://doi.org/10.1063/5.0138961}.

\begin{acknowledgments}
This project has received support by the DFG funded Research Training Group "Dynamics of Controlled Atomic and Molecular Systems" (RTG 2717).

The authors acknowledge support by the state of Baden-Württemberg through bwHPC and the German Research Foundation (DFG) through grant no INST 39/963-1 FUGG (bwForCluster NEMO).
\end{acknowledgments}

\section*{Data Availability Statement}
The data that supports the findings of this study are available in the appendix of this article.

\nocite{*}
\bibliography{aipsamp}

\newpage

\appendix
\section{Interaction Potential for Acene Molecules in PCFF-IFF}

In this supplementary material to 'Optimizing the Structure of Acene Clusters', we give all interaction terms of the potential which we used to determine the structures of the clusters made of anthracene, tetracene and pentacene molecules\footnote{H. Sun et al., “An ab initio cff93 all-atom force field for polycarbonates,” Journal of the American Chemical society 116, 2978–2987 (1994)\\H. Sun, “Force field for computation of conformational energies, structures, and vibrational frequencies of aromatic polyesters,” Journal of Computational Chemistry 15, 752–768 (1994).\\H. Sun, “Compass: an ab initio force-field optimized for condensed-phase applications overview with details on alkane and benzene compounds,” The Journal of Physical Chemistry B 102, 7338–7364 (1998).}. Equation \ref{eq:1} is the whole energy. Equations \ref{eq:2} to \ref{eq:12} are the contributing energies. In equations \ref{eq:2} to \ref{eq:5}, the main parts of the interaction energy between bound atoms is given. The following two equations give the energy between two atoms which have no bond between them. Equation \ref{eq:8} lists all cross terms between different interaction types. These cross terms are then given by equations \ref{eq:9} to \ref{eq:16}. The Parameters\footnote{H. Heinz et al., “Thermodynamically consistent force fields for the assembly of inorganic, organic, and biological nanostructures: the interface force field,” Langmuir 29, 1754–1765 (2013).} for the interaction are given in table \ref{tab:param1} and \ref{tab:param2}.

\begin{align}
    E_{\text{pot}}&= E_{\text{Bond}}+E_{\text{Angle}}+E_{\text{Torsion}}+E_{\text{Plane}}+E_{\text{cross}}\nonumber\\
    &+E_{\text{Coulomb}}+E_{\text{vdW}} \label{eq:1}\\
    E_{\text{Bond}}&=\sum_{i,j\text{ bonded}} \sum_{n=2}^4 K_{n,ij}(r_{ij}-r_{0,ij})^n  \label{eq:2}\\
    E_{\text{Angle}}&= \sum_{i,j,k\text{ bonded}}\sum_{n=2}^4 K_{n,ijk}(\theta_{ijk}-\theta_{0,ijk})^n \label{eq:3}\\
    E_{\text{Torsion}}&= \sum_{i,j,k,l\text{ bonded}}\sum_{n=1}^3 K_{n,ijkl}(1-\cos(n\phi_{ijkl}-\phi_{n,ijkl}))  \label{eq:4}\\
    E_{\text{Plane}}&= \sum_{i,j,k,l\text{ bonded}}K\left(\dfrac{\chi_{ijkl}+\chi_{kjli}+\chi_{ljik}}{3}-\chi_{0,ijkl}\right) \label{eq:5}\\
    E_{\text{Coulomb}}&= \sum_{i,j\text{ not bonded}}a_{ij}\dfrac{q_iq_j}{r_{ij}}  \label{eq:6}\\
    E_{\text{vdW}}&= \sum_{i,j\text{ not bonded}}\epsilon_{ij}\left(2\left(\dfrac{\sigma_{ij}}{r_{ij}}\right)^9-3\left(\dfrac{\sigma_{ij}}{r_{ij}}\right)^6\right) \label{eq:7}
\end{align}
\begin{align}
    E_{\text{cross}} &=E_{\text{BondBond}}+E_{\text{BondAngle}}+E_{\text{AngleAngle}}\nonumber\\
    &+E_{\text{AngleAngleTorsion}}+E_{\text{EndBondTorsion}}\nonumber\\
    &+E_{\text{MiddleBondTorsion}}+E_{\text{BondBond13}}+E_{\text{AngleTorsion}} \label{eq:8}\\
    E_{\text{BondBond}}&= K_{ijk}(r_{ij}-r_1)(r_{jk}-r_2)\label{eq:9}\\
    E_{\text{BondAngle}}&= K_1(r_{ij}-r_1)(\theta_{ijk}-\theta_0)+K_2(r_{jk}-r_1)(\theta_{ijk}-\theta_0)\label{eq:10}\\
    E_{\text{AngleAngle}}&= K_1 (\theta_{ijk} - \theta_1) (\theta_{kjl} - \theta_3) + K_2 (\theta_{ijk} - \theta_1) (\theta_{ijl} - \theta_2)\nonumber\\ 
    &+ K_3 (\theta_{ijl} - \theta_2) (\theta_{kjl} - \theta_3)\label{eq:11}\\
    E_{\text{AngleAngleTorsion}}&=K (\theta_{ijk} - \theta_1) (\theta_{jkl} - \theta_2) \cos (\phi_{ijkl})\label{eq:12}\\
    E_{\text{EndBondTorsion}}&=(r_{ij} - r_1) \sum_{n=1}^3 K_{n,\text{left}} \cos (n\phi_{ijkl})\nonumber\\ 
    &+ (r_{kl} - r_3) \sum_{n=1}^3 K_{n,\text{right}} \cos (n\phi_{ijkl})\label{eq:13}\\
    E_{\text{MiddleBondTorsion}}&=(r_{jk} - r_2) \sum_{n=1}^3  K_n \cos (n\phi_{ijkl})\label{eq:14}\\
    E_{\text{BondBond13}}&=N (r_{ij} - r_1) (r_{kl} - r_3)\label{eq:15}\\
    E_{\text{AngleTorsion}}&=(\theta_{ijk} - \theta_1) \sum_{n=1}^3 K_{n,\text{left}} \cos (n\phi_{ijkl})\nonumber\\ 
    &+ (\theta_{jkl} - \theta_2) \sum_{n=1}^3 K_{n,\text{right}} \cos (n\phi_{ijkl})\label{eq:16}
\end{align}

\begin{table}[htb]
    \centering
\begin{tabular}{c|c|c c c}
        & & C & H &\\
        \multirow{2}{*}{van der Waals} &$\epsilon$ & \SI{0.06400}{} & \SI{0.02000}{} &\\
        &$\sigma$ & \SI{4.0100}{} & \SI{2.9950}{} &\\
        \hline
        \hline
        & & C-C & C-H \\
        Charge & $q$ & \SI{0}{} & $\pm$\SI{0.1268}{} \\
        \hline
        \multirow{4}{*}{Bond} &$r_0$ & \SI{1.4170}{} & \SI{1.0982}{}\\
        &$K_2$ & \SI{470.8361}{} & \SI{372.8251}{}\\
        &$K_3$ & \SI{-627.6179}{} & \SI{-803.4526}{}\\
        &$K_4$ & \SI{1327.6345}{} & \SI{894.3173}{}\\
        \hline
        \hline
        & & C-C-C & C-C-H \\
        \multirow{4}{*}{Angle} &$\theta_0$ & \SI{118.9000}{} & \SI{117.9400}{}\\
        &$K_2$ & \SI{61.0226}{} & \SI{35.1558}{}\\
        &$K_3$ & \SI{-34.9931}{} & \SI{-12.4682}{}\\
        &$K_4$ & \SI{0.0000}{} & \SI{0.0000}{}\\
        \hline
        Bond Bond & $K$ & \SI{68.2856}{} & \SI{1.0795}{}\\
        \hline
        \multirow{2}{*}{Bond Angle}&$K_1$ & \SI{28.8708}{} & \SI{20.0033}{} \\
        &$K_2$ & \SI{28.8708}{} & \SI{24.2183}{} \\
        \hline
        \hline
        & & C-C-C-C & C-C-C-H \\
        \multirow{2}{*}{Plane}&$K$ & \SI{7.1794}{} & \SI{4.8912}{}\\
        &$\chi_0$ & \SI{0.0000}{} & \SI{0.0000}{}\\
        \hline
        \multirow{3}{*}{Angle Angle} & $K_1$ & \SI{0.0}{} & \SI{0.0}{} \\
        & $K_2$ & \SI{0.0}{} & \SI{0.0}{} \\
        & $K_3$ & \SI{0.0}{} & \SI{0.0}{} \\
            \end{tabular}
    \caption{Parameters of the polymer-consistent force field - interface force field (PCFF-IFF) which are needed to describe acene molecules. The original data can be found at https://bionanostructures.com/interface-md/. The units of the prefactors $K$ are such, that the energies are in \si{\Calorie\per\mol}. The distances are in Angstrom and the angles are given in degree.}
    \label{tab:param1}
\end{table}

\begin{table}[htb]
    \centering
    \begin{tabular}{c|c|c c c}
        & & C-C-C-C & C-C-C-H & H-C-C-H \\
        \multirow{6}{*}{Torsion}&$K_2$ & \SI{8.3667}{} & \SI{0.0000}{} & \SI{0.0000}{}\\
        &$\phi_2$ & \SI{0.0}{} & \SI{0.0}{} & \SI{0.0}{}\\
        &$K_3$ & \SI{1.1932}{} & \SI{3.9661}{} & \SI{1.8769}{}\\
        &$\phi_3$ & \SI{0.0}{} & \SI{0.0}{} & \SI{0.0}{}\\
        &$K_4$ & \SI{0.0000}{} & \SI{0.0000}{} & \SI{0.0000}{}\\
        &$\phi_4$ & \SI{0.0}{} & \SI{0.0}{} & \SI{0.0}{}\\
        \hline
        Angle Angle Torsion & $K$ & \SI{0.0000}{} & \SI{-4.8141}{} & \SI{0.3598}{}\\
        \hline
        \multirow{6}{*}{End Bond Torsion}&$K_{1,\text{left}}$ & \SI{-0.1185}{} & \SI{0.0000}{} & \SI{0.0000}{}\\
        &$K_{2,\text{left}}$ & \SI{6.3204}{} & \SI{-6.8958}{} & \SI{-0.6890}{}\\
        &$K_{3,\text{left}}$ & \SI{0.0000}{} & \SI{0.0000}{} & \SI{0.0000}{}\\
        &$K_{1,\text{right}}$ & \SI{-0.1185}{} & \SI{0.0000}{} & \SI{0.0000}{}\\
        &$K_{2,\text{right}}$ & \SI{6.3204}{} & \SI{-0.4669}{} & \SI{-0.6890}{}\\
        &$K_{3,\text{right}}$ & \SI{0.0000}{} & \SI{0.0000}{} & \SI{0.0000}{}\\
        \hline
        \multirow{3}{*}{Middle Bond Torsion}&$K_1$ & \SI{27.5989}{} & \SI{0.0000}{} & \SI{0.0000}{} \\ 
        &$K_2$ & \SI{-2.3120}{} & \SI{-1.1521}{} & \SI{4.8228}{} \\
        &$K_3$ & \SI{0.0000}{} & \SI{0.0000}{} & \SI{0.0000}{} \\
        \hline
        Bond Bond 13 & $K$ & \SI{53.0000}{} & \SI{-6.2741}{} & \SI{-1.7077}{} \\
        \hline
        \multirow{6}{*}{Angle Torsion}&$K_{1,\text{left}}$ & \SI{1.9767}{} & \SI{0.0000}{} & \SI{0.0000}{} \\
        &$K_{2,\text{left}}$ & \SI{1.0239}{} & \SI{2.5014}{} & \SI{2.4501}{} \\
        &$K_{3,\text{left}}$ & \SI{0.0000}{} & \SI{0.0000}{} & \SI{0.0000}{} \\
        &$K_{1,\text{right}}$ & \SI{1.9767}{} & \SI{0.0000}{} & \SI{0.0000}{} \\
        &$K_{2,\text{right}}$ & \SI{1.0239}{} & \SI{2.7147}{} & \SI{2.4501}{} \\
        &$K_{3,\text{right}}$ & \SI{0.0000}{} & \SI{0.0000}{} & \SI{0.0000}{} \\
        \end{tabular}
    \caption{Continue table \ref{tab:param1}.}
    \label{tab:param2}
\end{table}

\FloatBarrier
\section{Cluster Structures}
\subsection{Anthracene Cluster}

\begin{table}[p]
\begin{tabular}{c|c c c}
type & $x$ & $y$ & $z$\\ 
\hline
\hline\hline
C14H10\\ 
\hline\hline
C & 3.4141 & 2.13 & 1.1891\\ 
C & 3.5709 & 3.4481 & 0.698\\ 
C & 4.5513 & 1.3173 & 1.4086\\ 
C & 2.1201 & 1.6258 & 1.4602\\ 
C & 4.8647 & 3.9524 & 0.4268\\ 
C & 2.4336 & 4.2608 & 0.4785\\
\end{tabular}
\caption{}
\end{table}
\begin{table}[p]
\begin{tabular}{c|c c c}
type & $x$ & $y$ & $z$\\ 
\hline
H & 30.6693 & 5.0174 & -251.1885\\ 
H & 29.9329 & 6.133 & -253.2383\\ 
H & 35.4126 & 5.1345 & -252.8286\\ 
H & 34.6765 & 6.2506 & -254.8782\\ 
H & 30.4648 & 7.1665 & -255.4661\\ 
H & 32.8378 & 7.2253 & -256.2864\\ 
H & 31.4088 & 3.926 & -249.1272\\
\end{tabular}
\caption{}
\end{table}
\clearpage

\begin{table}[p]
\begin{tabular}{c|c c c}
type & $x$ & $y$ & $z$\\ 
\hline
C & 16.3176 & 16.9524 & 16.4788\\ 
C & 18.7504 & 13.4778 & 16.6137\\ 
C & 14.6187 & 15.8191 & 17.8309\\ 
C & 19.1917 & 14.6254 & 15.9132\\ 
C & 15.0619 & 16.9647 & 17.129\\ 
C & 19.5609 & 12.3185 & 16.6506\\ 
C & 20.4409 & 14.6132 & 15.2505\\
\end{tabular}
\caption{}
\end{table}
\begin{table}[p]
\begin{tabular}{c|c c c}
type & $x$ & $y$ & $z$\\ 
\hline
H & 25.9356 & 20.0701 & 10.5921\\ 
H & 23.9407 & 23.0636 & 7.0905\\ 
H & 26.8947 & 20.0972 & 8.2726\\ 
H & 25.895 & 21.5953 & 6.5181\\ 
C & 17.5752 & 18.6404 & 8.5204\\ 
C & 17.0103 & 19.4839 & 7.5347\\ 
C & 18.6808 & 17.8189 & 8.1993\\
\end{tabular}
\caption{}
\end{table}
\clearpage

\begin{table}[p]
\begin{tabular}{c|c c c}
type & $x$ & $y$ & $z$\\ 
\hline
C & 15.9511 & 25.6111 & 21.6067\\ 
C & 17.0569 & 25.5338 & 23.7919\\ 
C & 16.1749 & 26.1698 & 22.887\\ 
H & 18.825 & 22.0894 & 22.4653\\ 
H & 19.243 & 20.326 & 20.8218\\ 
H & 15.689 & 24.3396 & 19.2555\\ 
H & 16.1095 & 22.5739 & 17.6089\\
\end{tabular}
\caption{}
\end{table}
\begin{table}[p]
\begin{tabular}{c|c c c}
type & $x$ & $y$ & $z$\\ 
\hline
C & 14.6303 & 19.9552 & 20.0303\\ 
C & 11.5894 & 22.0472 & 20.649\\ 
C & 12.0174 & 20.2828 & 18.9993\\ 
C & 13.3387 & 22.6781 & 22.2444\\ 
C & 14.1931 & 19.1561 & 18.9489\\ 
C & 12.0308 & 22.8452 & 21.7311\\ 
C & 12.8859 & 19.3196 & 18.4335\\
\end{tabular}
\caption{}
\end{table}
\clearpage

\begin{table}[p]
\begin{tabular}{c|c c c}
type & $x$ & $y$ & $z$\\ 
\hline
H & 18.5927 & 13.8503 & 13.1943\\ 
H & 18.3393 & 12.0196 & 14.8989\\ 
H & 11.7786 & 16.022 & 12.6479\\ 
H & 11.2499 & 12.3764 & 16.0589\\ 
H & 9.5129 & 15.5179 & 13.6003\\ 
H & 9.2471 & 13.6908 & 15.3069\\ 
C & 13.402 & 18.3255 & 11.1434\\
\end{tabular}
\caption{}
\end{table}
\begin{table}[p]
\begin{tabular}{c|c c c}
type & $x$ & $y$ & $z$\\ 
\hline
C & 17.0907 & 21.5206 & 13.9272\\ 
C & 17.4974 & 22.8334 & 16.3995\\ 
C & 18.3942 & 21.6247 & 14.4664\\ 
C & 18.598 & 22.2826 & 15.7022\\ 
H & 14.528 & 21.45 & 13.1315\\ 
H & 12.1231 & 21.8683 & 13.3059\\ 
H & 15.2488 & 23.7754 & 17.523\\
\end{tabular}
\caption{}
\end{table}
\clearpage

\begin{table}[p]
\begin{tabular}{c|c c c}
type & $x$ & $y$ & $z$\\ 
\hline
C & 13.9634 & 14.682 & 10.8538\\ 
C & 14.4316 & 12.4059 & 10.0817\\ 
C & 12.1904 & 13.3566 & 9.805\\ 
C & 15.3013 & 14.8086 & 11.2937\\ 
C & 13.0602 & 15.7573 & 11.0164\\ 
C & 15.7695 & 12.5325 & 10.5217\\ 
C & 11.2891 & 14.4355 & 9.9609\\
\end{tabular}
\caption{}
\end{table}
\begin{table}[p]
\begin{tabular}{c|c c c}
type & $x$ & $y$ & $z$\\ 
\hline
H & 11.2876 & 9.2343 & 8.4467\\ 
H & 9.24 & 10.9067 & 11.9022\\ 
H & 9.2655 & 10.3297 & 9.4543\\ 
H & 15.777 & 8.6128 & 13.7978\\ 
H & 15.8181 & 7.4451 & 8.9151\\ 
H & 17.814 & 7.5564 & 12.7775\\ 
H & 17.8333 & 6.9721 & 10.3317\\
\end{tabular}
\caption{}
\end{table}
\clearpage

\begin{table}[p]
\begin{tabular}{c|c c c}
type & $x$ & $y$ & $z$\\ 
\hline
C & 12.0981 & 19.682 & 15.0513\\ 
C & 17.07 & 15.5341 & 14.8096\\ 
C & 17.9536 & 17.9598 & 15.9631\\ 
C & 18.4163 & 15.6712 & 15.2185\\ 
C & 18.8585 & 16.8848 & 15.7954\\ 
H & 14.4796 & 15.5238 & 14.1202\\ 
H & 12.2359 & 16.4496 & 13.871\\
\end{tabular}
\caption{}
\end{table}
\begin{table}[p]
\begin{tabular}{c|c c c}
type & $x$ & $y$ & $z$\\ 
\hline
C & 14.2171 & 9.3241 & 21.3586\\ 
C & 15.2077 & 8.8003 & 22.2226\\ 
C & 14.5709 & 10.2737 & 20.3712\\ 
C & 12.8723 & 8.9047 & 21.4895\\ 
C & 16.5513 & 9.2238 & 22.0974\\ 
C & 14.8531 & 7.8586 & 23.2162\\ 
C & 15.9146 & 10.699 & 20.2476\\
\end{tabular}
\caption{}
\end{table}
\clearpage

\begin{table}[p]
\begin{tabular}{c|c c c}
type & $x$ & $y$ & $z$\\ 
\hline
H & 9.7731 & 18.6694 & 23.5513\\ 
H & 7.9727 & 17.3723 & 24.6038\\ 
H & 8.8713 & 14.7236 & 27.9369\\ 
H & 7.2093 & 15.7228 & 26.336\\ 
H & 14.8753 & 17.9345 & 25.6984\\ 
H & 11.5792 & 19.9617 & 22.5014\\ 
H & 15.6566 & 19.5778 & 23.9735\\
\end{tabular}
\caption{}
\end{table}
\begin{table}[p]
\begin{tabular}{c|c c c}
type & $x$ & $y$ & $z$\\ 
\hline
C & 14.4193 & 19.0213 & 28.5884\\ 
C & 10.8547 & 16.4357 & 30.7425\\ 
C & 16.5957 & 17.98 & 28.1636\\ 
C & 14.8807 & 20.2213 & 27.9974\\ 
C & 17.0572 & 19.1785 & 27.5724\\ 
C & 16.1993 & 20.3002 & 27.4902\\ 
H & 15.4825 & 15.828 & 29.3105\\
\end{tabular}
\caption{}
\end{table}
\clearpage

\begin{table}[p]
\begin{tabular}{c|c c c}
type & $x$ & $y$ & $z$\\ 
\hline
H & 15.6663 & 8.4364 & 31.7019\\ 
C & 20.0885 & 17.8801 & 30.5438\\ 
C & 20.4575 & 16.6105 & 31.0474\\ 
C & 18.722 & 18.2317 & 30.4456\\ 
C & 21.0871 & 18.7935 & 30.1324\\ 
C & 19.4603 & 15.6924 & 31.4492\\ 
C & 21.8235 & 16.2572 & 31.1426\\
\end{tabular}
\caption{}
\end{table}
\begin{table}[p]
\begin{tabular}{c|c c c}
type & $x$ & $y$ & $z$\\ 
\hline
H & 15.3168 & 15.7657 & 21.1375\\ 
H & 18.4385 & 11.6081 & 19.0979\\ 
H & 16.0547 & 11.1178 & 19.3898\\ 
H & 13.3854 & 14.1706 & 20.9649\\ 
H & 13.7553 & 11.8432 & 20.0878\\ 
H & 20.0762 & 16.7417 & 20.5715\\ 
H & 20.8248 & 12.0951 & 18.8243\\
\end{tabular}
\caption{}
\end{table}
\clearpage

\begin{table}[p]
\begin{tabular}{c|c c c}
type & $x$ & $y$ & $z$\\ 
\hline
C & 23.4482 & 15.087 & 34.2497\\ 
C & 18.8918 & 12.7961 & 34.3413\\ 
C & 23.6567 & 13.7778 & 33.7558\\ 
C & 17.3818 & 14.5181 & 35.2117\\ 
C & 17.8032 & 11.9005 & 34.2265\\ 
C & 16.2929 & 13.6225 & 35.0945\\ 
C & 16.5035 & 12.3136 & 34.6009\\
\end{tabular}
\caption{}
\end{table}
\begin{table}[p]
\begin{tabular}{c|c c c}
type & $x$ & $y$ & $z$\\ 
\hline
H & 15.6701 & 5.3595 & 27.2579\\ 
\hline\hline
C140H100\\ 
\hline\hline
C & 15.1093 & 24.1642 & 3.2208\\ 
C & 15.6555 & 25.1094 & 4.1211\\ 
C & 15.823 & 22.9808 & 2.9184\\ 
C & 13.8479 & 24.4016 & 2.6263\\ 
C & 16.9153 & 24.8713 & 4.7182\\
\end{tabular}
\caption{}
\end{table}
\clearpage

\begin{table}[p]
\begin{tabular}{c|c c c}
type & $x$ & $y$ & $z$\\ 
\hline
H & 10.3302 & 18.7671 & 14.4995\\ 
H & 8.3604 & 20.2066 & 14.2426\\ 
H & 13.3178 & 22.7219 & 13.7029\\ 
H & 11.3571 & 24.1536 & 13.4459\\ 
H & 7.3629 & 22.4698 & 13.8153\\ 
H & 8.8633 & 24.4484 & 13.4155\\ 
H & 12.3036 & 17.3294 & 14.7417\\
\end{tabular}
\caption{}
\end{table}
\begin{table}[p]
\begin{tabular}{c|c c c}
type & $x$ & $y$ & $z$\\ 
\hline
C & 11.0818 & 16.747 & 5.4541\\ 
C & 7.0839 & 19.531 & 4.9462\\ 
C & 10.2673 & 15.6091 & 5.6625\\ 
C & 6.2659 & 18.3964 & 5.158\\ 
C & 12.4897 & 16.6236 & 5.4786\\ 
C & 10.8646 & 14.3473 & 5.8944\\ 
C & 13.0871 & 15.3635 & 5.7106\\
\end{tabular}
\caption{}
\end{table}
\clearpage

\begin{table}[p]
\begin{tabular}{c|c c c}
type & $x$ & $y$ & $z$\\ 
\hline
H & 20.432 & 22.1234 & 8.9812\\ 
H & 21.9051 & 24.1276 & 8.6059\\ 
C & 9.4039 & 18.9295 & 9.2881\\ 
C & 9.989 & 19.9042 & 10.1302\\ 
C & 10.0994 & 17.7308 & 9.0047\\ 
C & 8.1267 & 19.1586 & 8.725\\ 
C & 11.2688 & 19.6806 & 10.6876\\
\end{tabular}
\caption{}
\end{table}
\begin{table}[p]
\begin{tabular}{c|c c c}
type & $x$ & $y$ & $z$\\ 
\hline
C & 14.4228 & 11.0269 & 13.9472\\ 
H & 9.8512 & 9.6897 & 15.0299\\ 
H & 7.5727 & 10.536 & 14.7511\\ 
H & 11.6451 & 13.774 & 12.7293\\ 
H & 9.3682 & 14.6199 & 12.4558\\ 
H & 5.9946 & 12.3097 & 13.9493\\ 
H & 6.894 & 14.3621 & 12.7961\\
\end{tabular}
\caption{}
\end{table}
\clearpage

\begin{table}[p]
\begin{tabular}{c|c c c}
type & $x$ & $y$ & $z$\\ 
\hline
C & 9.7211 & 13.7473 & 9.2011\\ 
C & 13.7272 & 14.3548 & 10.454\\ 
C & 9.1706 & 16.0475 & 9.8364\\ 
C & 13.3335 & 13.1017 & 9.9291\\ 
C & 8.7786 & 14.7964 & 9.305\\ 
C & 15.0627 & 14.5565 & 10.8713\\ 
C & 14.2755 & 12.0525 & 9.8227\\
\end{tabular}
\caption{}
\end{table}
\begin{table}[p]
\begin{tabular}{c|c c c}
type & $x$ & $y$ & $z$\\ 
\hline
H & 6.3984 & 7.3245 & 9.4118\\ 
H & 5.5248 & 11.5324 & 10.0925\\ 
H & 4.7611 & 9.2268 & 9.4378\\ 
C & 12.2379 & 14.6971 & 2.8362\\ 
C & 11.7382 & 13.4346 & 2.4378\\ 
C & 13.5985 & 14.8381 & 3.1956\\ 
C & 11.3751 & 15.8166 & 2.8822\\
\end{tabular}
\caption{}
\end{table}
\clearpage

\begin{table}[p]
\begin{tabular}{c|c c c}
type & $x$ & $y$ & $z$\\ 
\hline
C & 10.9861 & 4.5859 & 15.015\\ 
H & 8.7029 & 8.4267 & 13.0483\\ 
H & 6.4288 & 9.2589 & 12.7452\\ 
H & 6.9115 & 4.37 & 15.4029\\ 
H & 4.6234 & 5.2085 & 15.0989\\ 
H & 3.9534 & 8.9958 & 13.0645\\ 
H & 3.0492 & 6.9672 & 14.2434\\
\end{tabular}
\caption{}
\end{table}
\begin{table}[p]
\begin{tabular}{c|c c c}
type & $x$ & $y$ & $z$\\ 
\hline
C & 10.2848 & 14.659 & -2.3833\\ 
C & 8.7375 & 16.159 & -1.2181\\ 
C & 12.4537 & 14.1173 & -1.3802\\ 
C & 9.3565 & 17.1154 & 0.9507\\ 
C & 11.5207 & 13.973 & -2.4343\\ 
C & 8.4241 & 16.9712 & -0.1034\\ 
C & 13.6882 & 13.4288 & -1.4301\\
\end{tabular}
\caption{}
\end{table}
\clearpage

\begin{table}[p]
\begin{tabular}{c|c c c}
type & $x$ & $y$ & $z$\\ 
\hline
H & 5.9706 & 7.7423 & -3.9179\\ 
H & 9.0424 & 7.7503 & 0.0505\\ 
H & 8.0837 & 6.3873 & -3.9655\\ 
H & 9.6242 & 6.3936 & -1.9741\\ 
C & 11.7896 & 5.9136 & 10.035\\ 
C & 10.8212 & 5.7706 & 9.0136\\ 
C & 13.0121 & 5.2065 & 9.9535\\
\end{tabular}
\caption{}
\end{table}
\begin{table}[p]
\begin{tabular}{c|c c c}
type & $x$ & $y$ & $z$\\ 
\hline
C & 12.3594 & 15.1982 & 5.4775\\ 
C & 11.3622 & 14.9601 & 4.5028\\ 
H & 15.9083 & 13.221 & 4.9463\\ 
H & 17.4125 & 11.6801 & 3.7755\\ 
H & 12.3631 & 12.3793 & 1.4914\\ 
H & 13.8606 & 10.8409 & 0.3276\\ 
H & 17.8581 & 10.0991 & 1.8787\\
\end{tabular}
\caption{}
\end{table}
\clearpage

\begin{table}[p]
\begin{tabular}{c|c c c}
type & $x$ & $y$ & $z$\\ 
\hline
C & 9.9482 & 11.5671 & 0.0432\\ 
C & 8.4653 & 13.1238 & 1.2179\\ 
C & 10.8536 & 9.6191 & 1.2182\\ 
C & 7.9022 & 12.7434 & 3.5709\\ 
C & 10.7442 & 10.3986 & 0.0432\\ 
C & 7.7877 & 13.5202 & 2.3942\\ 
C & 11.6446 & 8.4476 & 1.2162\\
\end{tabular}
\caption{}
\end{table}
\begin{table}[p]
\begin{tabular}{c|c c c}
type & $x$ & $y$ & $z$\\ 
\hline
H & 17.1505 & 6.7183 & 7.5471\\ 
H & 17.9513 & 4.3079 & 11.8776\\ 
H & 18.6533 & 4.709 & 7.6021\\ 
H & 19.0545 & 3.5016 & 9.7711\\ 
C & 7.7733 & 7.7346 & 4.12\\ 
C & 8.7031 & 7.8137 & 5.1828\\ 
C & 8.054 & 6.9279 & 2.993\\
\end{tabular}
\caption{}
\end{table}
\clearpage

\begin{table}[p]
\begin{tabular}{c|c c c}
type & $x$ & $y$ & $z$\\ 
\hline
C & 5.5146 & 2.3028 & -54.3147\\ 
C & 5.6102 & 4.0633 & -52.6136\\ 
C & 6.1848 & 2.9323 & -53.2398\\ 
H & 2.0032 & 5.309 & -54.0963\\ 
H & 0.075 & 5.1596 & -55.597\\ 
H & 4.0396 & 1.2914 & -56.3166\\ 
H & 2.1117 & 1.1439 & -57.8173\\
\end{tabular}
\caption{}
\end{table}
\begin{table}[p]
\begin{tabular}{c|c c c}
type & $x$ & $y$ & $z$\\ 
\hline
C & -1.8824 & 7.0621 & -51.7553\\ 
C & 0.9973 & 4.7159 & -51.3042\\ 
C & -0.9326 & 4.6156 & -52.8072\\ 
C & 1.2492 & 6.5976 & -49.7563\\ 
C & -2.6158 & 6.3934 & -52.7636\\ 
C & 1.7218 & 5.3727 & -50.2832\\ 
C & -2.1408 & 5.1692 & -53.2896\\
\end{tabular}
\caption{}
\end{table}
\clearpage

\begin{table}[p]
\begin{tabular}{c|c c c}
type & $x$ & $y$ & $z$\\ 
\hline
H & 1.2364 & 9.5114 & -42.7302\\ 
H & -2.7879 & 6.4638 & -48.9441\\ 
H & -4.8138 & 10.364 & -46.5213\\ 
H & -4.9996 & 7.3805 & -49.6849\\ 
H & -6.0156 & 9.3392 & -48.4727\\ 
C & -2.7923 & -0.4119 & -48.113\\ 
C & -2.2309 & -1.5244 & -48.7822\\
\end{tabular}
\caption{}
\end{table}
\begin{table}[p]
\begin{tabular}{c|c c c}
type & $x$ & $y$ & $z$\\ 
\hline
C & 6.518 & -1.5109 & -55.6504\\ 
C & 6.1389 & -1.4064 & -53.2315\\ 
C & 7.0271 & -1.4406 & -54.3319\\ 
H & 2.1502 & -1.521 & -54.1273\\ 
H & 0.2459 & -1.6203 & -55.6499\\ 
H & 5.2938 & -1.6448 & -58.0405\\ 
H & 3.377 & -1.7449 & -59.5724\\
\end{tabular}
\caption{}
\end{table}
\clearpage

\begin{table}[p]
\begin{tabular}{c|c c c}
type & $x$ & $y$ & $z$\\ 
\hline
C & -1.7664 & 6.7651 & -42.0842\\ 
C & -5.4361 & 7.1824 & -41.4763\\ 
C & -3.5359 & 7.2202 & -39.9252\\ 
C & -5.0525 & 6.8154 & -43.868\\ 
C & -1.2585 & 6.8963 & -40.7712\\ 
C & -5.9409 & 7.0461 & -42.7913\\ 
C & -2.1437 & 7.1234 & -39.6913\\
\end{tabular}
\caption{}
\end{table}
\begin{table}[p]
\begin{tabular}{c|c c c}
type & $x$ & $y$ & $z$\\ 
\hline
H & 17.0866 & 19.0034 & 14.2432\\ 
H & 17.1282 & 21.4473 & 14.854\\ 
H & 24.1206 & 17.6044 & 14.1724\\ 
H & 24.2147 & 22.4692 & 15.3936\\ 
H & 26.319 & 18.7563 & 14.5525\\ 
H & 26.3634 & 21.2001 & 15.167\\ 
C & 19.6544 & 13.6852 & 8.4966\\
\end{tabular}
\caption{}
\end{table}
\clearpage

\begin{table}[p]
\begin{tabular}{c|c c c}
type & $x$ & $y$ & $z$\\ 
\hline
C & 24.3671 & 17.4261 & 7.0453\\ 
C & 24.4167 & 20.168 & 7.7289\\ 
C & 25.6038 & 18.0767 & 7.265\\ 
C & 25.6284 & 19.4492 & 7.607\\ 
H & 21.8985 & 16.4294 & 6.6968\\ 
H & 19.4537 & 16.4951 & 6.627\\ 
H & 21.9903 & 21.297 & 7.917\\
\end{tabular}
\caption{}
\end{table}
\begin{table}[p]
\begin{tabular}{c|c c c}
type & $x$ & $y$ & $z$\\ 
\hline
C & 23.0691 & 27.229 & 10.2954\\ 
C & 20.6223 & 27.2552 & 10.2716\\ 
C & 23.1113 & 29.6872 & 8.8962\\ 
C & 20.658 & 29.7143 & 8.8766\\ 
C & 24.3039 & 27.8278 & 9.957\\ 
C & 19.4059 & 27.8807 & 9.9139\\ 
C & 24.3261 & 29.0567 & 9.2561\\
\end{tabular}
\caption{}
\end{table}
\clearpage

\begin{table}[p]
\begin{tabular}{c|c c c}
type & $x$ & $y$ & $z$\\ 
\hline
H & 17.7133 & 23.5806 & 9.755\\ 
H & 17.6486 & 21.496 & 11.1701\\ 
H & 24.8125 & 24.4392 & 9.1326\\ 
H & 24.6753 & 20.2865 & 11.9462\\ 
H & 26.9523 & 23.3462 & 9.8471\\ 
H & 26.8861 & 21.2601 & 11.259\\ 
C & 20.1254 & 15.9848 & 11.8909\\
\end{tabular}
\caption{}
\end{table}
\begin{table}[p]
\begin{tabular}{c|c c c}
type & $x$ & $y$ & $z$\\ 
\hline
C & 23.3545 & 15.4165 & 16.1221\\ 
C & 23.2916 & 12.6812 & 15.4101\\ 
C & 24.5634 & 14.6986 & 15.9715\\ 
C & 24.532 & 13.3299 & 15.6149\\ 
H & 20.9333 & 16.5494 & 16.3445\\ 
H & 18.4932 & 16.6259 & 16.2805\\ 
H & 20.8161 & 11.694 & 15.0724\\
\end{tabular}
\caption{}
\end{table}
\clearpage

\begin{table}[p]
\begin{tabular}{c|c c c}
type & $x$ & $y$ & $z$\\ 
\hline
C & 3.2394 & 10.47 & 3.0962\\ 
C & 4.2265 & 9.3702 & 5.0498\\ 
C & 1.7974 & 9.626 & 4.8886\\ 
C & 4.5319 & 10.6721 & 2.557\\ 
C & 2.0981 & 10.923 & 2.3936\\ 
C & 5.5176 & 9.5684 & 4.5094\\ 
C & 0.6562 & 10.0748 & 4.1846\\
\end{tabular}
\caption{}
\end{table}
\begin{table}[p]
\begin{tabular}{c|c c c}
type & $x$ & $y$ & $z$\\ 
\hline
H & 2.826 & 20.9999 & 2.4616\\ 
H & 0.2043 & 19.6112 & 5.6444\\ 
H & 0.5172 & 20.7107 & 3.4048\\ 
H & 7.0268 & 18.1949 & 7.3238\\ 
H & 7.6851 & 20.4013 & 2.864\\ 
H & 9.3395 & 18.4712 & 6.3964\\ 
H & 9.6714 & 19.5784 & 4.1615\\
\end{tabular}
\caption{}
\end{table}
\clearpage

\begin{table}[p]
\begin{tabular}{c|c c c}
type & $x$ & $y$ & $z$\\ 
\hline
C & 1.195 & 15.7483 & 8.1898\\ 
C & 7.0236 & 14.2632 & 10.6012\\ 
C & 7.3418 & 15.3065 & 7.9944\\ 
C & 8.3173 & 14.3876 & 10.0443\\ 
C & 8.4769 & 14.9097 & 8.7392\\ 
H & 4.4757 & 14.1464 & 11.4339\\ 
H & 2.0554 & 14.4346 & 11.2552\\
\end{tabular}
\caption{}
\end{table}
\begin{table}[p]
\begin{tabular}{c|c c c}
type & $x$ & $y$ & $z$\\ 
\hline
C & 3.559 & 11.1396 & 12.5945\\ 
C & 4.5368 & 10.2707 & 14.6647\\ 
C & 2.1116 & 10.5702 & 14.4863\\ 
C & 4.8489 & 11.2303 & 12.0228\\ 
C & 2.4242 & 11.5251 & 11.8448\\ 
C & 5.8267 & 10.3597 & 14.0922\\ 
C & 0.9758 & 10.9515 & 13.7344\\
\end{tabular}
\caption{}
\end{table}
\clearpage

\begin{table}[p]
\begin{tabular}{c|c c c}
type & $x$ & $y$ & $z$\\ 
\hline
H & 2.6573 & 21.3914 & 11.9909\\ 
H & 0.1174 & 20.4376 & 15.3899\\ 
H & 0.3727 & 21.2034 & 13.0079\\ 
H & 7.0114 & 19.3645 & 17.1018\\ 
H & 7.5105 & 20.8785 & 12.3401\\ 
H & 9.2918 & 19.4998 & 16.063\\ 
H & 9.5432 & 20.2596 & 13.6786\\
\end{tabular}
\caption{}
\end{table}
\begin{table}[p]
\begin{tabular}{c|c c c}
type & $x$ & $y$ & $z$\\ 
\hline
C & 1.5562 & 16.6698 & 17.7314\\ 
C & 7.4308 & 15.1417 & 20.0134\\ 
C & 7.7026 & 16.3099 & 17.4526\\ 
C & 8.7145 & 15.2964 & 19.4389\\ 
C & 8.8507 & 15.8807 & 18.1577\\ 
H & 4.8896 & 14.9523 & 20.8665\\ 
H & 2.4545 & 15.2121 & 20.7194\\
\end{tabular}
\caption{}
\end{table}
\clearpage

\begin{table}[p]
\begin{tabular}{c|c c c}
type & $x$ & $y$ & $z$\\ 
\hline
C & 5.8159 & 4.1128 & 18.954\\ 
C & 6.0076 & 5.1399 & 17.9997\\ 
C & 6.9305 & 3.5401 & 19.61\\ 
C & 4.5094 & 3.6629 & 19.256\\ 
C & 7.3138 & 5.5939 & 17.7018\\ 
C & 4.8926 & 5.7144 & 17.3462\\ 
C & 8.2359 & 3.9988 & 19.317\\
\end{tabular}
\caption{}
\end{table}
\begin{table}[p]
\begin{tabular}{c|c c c}
type & $x$ & $y$ & $z$\\ 
\hline
H & 6.0323 & 16.5317 & 19.9055\\ 
H & 3.6032 & 16.6542 & 19.6652\\ 
H & 1.2807 & 13.0189 & 19.1072\\ 
H & 1.3887 & 15.5243 & 19.3488\\ 
H & 8.2551 & 11.4171 & 19.6754\\ 
H & 8.4611 & 16.4068 & 20.1542\\ 
H & 10.4682 & 12.5467 & 20.0001\\
\end{tabular}
\caption{}
\end{table}
\clearpage

\begin{table}[p]
\begin{tabular}{c|c c c}
type & $x$ & $y$ & $z$\\ 
\hline
C & 6.6707 & 24.3857 & 20.0372\\ 
C & 1.8124 & 24.6487 & 19.435\\ 
C & 7.6251 & 22.6728 & 21.5057\\ 
C & 7.972 & 24.8615 & 19.7497\\ 
C & 8.9255 & 23.1476 & 21.2191\\ 
C & 9.0992 & 24.2428 & 20.3408\\ 
H & 5.0682 & 21.9675 & 21.8921\\
\end{tabular}
\caption{}
\end{table}
\begin{table}[p]
\begin{tabular}{c|c c c}
type & $x$ & $y$ & $z$\\ 
\hline
C & 4.9681 & 6.0893 & 26.2097\\ 
C & 5.1335 & 7.2358 & 25.3975\\ 
C & 6.0992 & 5.4595 & 26.7806\\ 
C & 3.6722 & 5.571 & 26.4426\\ 
C & 6.4274 & 7.7502 & 25.1539\\ 
C & 4.0053 & 7.8633 & 24.8219\\ 
C & 7.3938 & 5.9777 & 26.541\\
\end{tabular}
\caption{}
\end{table}
\clearpage

\begin{table}[p]
\begin{tabular}{c|c c c}
type & $x$ & $y$ & $z$\\ 
\hline
H & 5.4587 & 18.9265 & 27.0168\\ 
H & 3.0267 & 19.0162 & 26.7649\\ 
H & 0.7542 & 15.3443 & 26.2416\\ 
H & 0.8284 & 17.8545 & 26.453\\ 
H & 7.7441 & 13.8373 & 26.8405\\ 
H & 7.8913 & 18.8347 & 27.2613\\ 
H & 9.9447 & 14.9981 & 27.1414\\
\end{tabular}
\caption{}
\end{table}
\begin{table}[p]
\begin{tabular}{c|c c c}
type & $x$ & $y$ & $z$\\ 
\hline
C & 6.5479 & 15.3126 & 30.914\\ 
C & 1.6825 & 15.3957 & 30.3433\\ 
C & 7.5875 & 13.6345 & 32.3638\\ 
C & 7.8246 & 15.8351 & 30.6038\\ 
C & 8.8647 & 14.1593 & 32.0553\\ 
C & 8.9836 & 15.26 & 31.1745\\ 
H & 5.0577 & 12.8246 & 32.7698\\
\end{tabular}
\caption{}
\end{table}
\clearpage

\begin{table}[p]
\begin{tabular}{c|c c c}
type & $x$ & $y$ & $z$\\ 
\hline
H & 11.4635 & 10.7329 & 16.6697\\ 
C & 5.8397 & 4.1369 & 18.9706\\ 
C & 6.0355 & 5.1683 & 18.0218\\ 
C & 6.9512 & 3.5631 & 19.6307\\ 
C & 4.5321 & 3.6837 & 19.2628\\ 
C & 7.3427 & 5.6255 & 17.7336\\ 
C & 4.9235 & 5.7439 & 17.3642\\
\end{tabular}
\caption{}
\end{table}
\begin{table}[p]
\begin{tabular}{c|c c c}
type & $x$ & $y$ & $z$\\ 
\hline
H & 3.3825 & 11.6801 & 19.2211\\ 
H & 6.0263 & 16.5393 & 19.9727\\ 
H & 3.5988 & 16.6677 & 19.7208\\ 
H & 1.2713 & 13.0393 & 19.139\\ 
H & 1.3835 & 15.5437 & 19.3888\\ 
H & 8.2389 & 11.4206 & 19.7411\\ 
H & 8.4535 & 16.4084 & 20.2337\\
\end{tabular}
\caption{}
\end{table}
\clearpage

\begin{table}[p]
\begin{tabular}{c|c c c}
type & $x$ & $y$ & $z$\\ 
\hline
C & 1.6362 & 23.5712 & 20.2871\\ 
C & 6.6639 & 24.3908 & 20.0043\\ 
C & 1.8061 & 24.6607 & 19.4009\\ 
C & 7.6151 & 22.6791 & 21.4762\\ 
C & 7.9661 & 24.8638 & 19.7161\\ 
C & 8.9164 & 23.1512 & 21.1891\\ 
C & 9.0921 & 24.2446 & 20.3088\\
\end{tabular}
\caption{}
\end{table}
\begin{table}[p]
\begin{tabular}{c|c c c}
type & $x$ & $y$ & $z$\\ 
\hline
H & 10.5679 & 13.2172 & 24.0365\\ 
C & 4.9513 & 6.0619 & 26.2195\\ 
C & 5.1199 & 7.2177 & 25.4212\\ 
C & 6.0796 & 5.4286 & 26.7921\\ 
C & 3.655 & 5.5379 & 26.4368\\ 
C & 6.4143 & 7.7378 & 25.193\\ 
C & 3.9946 & 7.8486 & 24.8439\\
\end{tabular}
\caption{}
\end{table}
\clearpage

\begin{table}[p]
\begin{tabular}{c|c c c}
type & $x$ & $y$ & $z$\\ 
\hline
H & 2.7844 & 14.0375 & 26.3613\\ 
H & 5.3531 & 18.941 & 27.0819\\ 
H & 2.9241 & 19.0322 & 26.8131\\ 
H & 0.6555 & 15.3639 & 26.2457\\ 
H & 0.728 & 17.8725 & 26.4745\\ 
H & 7.6401 & 13.8522 & 26.902\\ 
H & 7.7819 & 18.8478 & 27.349\\
\end{tabular}
\caption{}
\end{table}
\begin{table}[p]
\begin{tabular}{c|c c c}
type & $x$ & $y$ & $z$\\ 
\hline
C & 1.4921 & 14.1419 & 31.1515\\ 
C & 6.4781 & 15.1985 & 30.9697\\ 
C & 1.6226 & 15.2845 & 30.3277\\ 
C & 7.5034 & 13.467 & 32.3659\\ 
C & 7.7561 & 15.7429 & 30.7063\\ 
C & 8.7817 & 14.0155 & 32.1067\\ 
C & 8.9083 & 15.1535 & 31.2759\\
\end{tabular}
\caption{}
\end{table}
\clearpage

\begin{table}[p]
\begin{tabular}{c|c c c}
type & $x$ & $y$ & $z$\\ 
\hline
H & 9.7074 & 15.4979 & 16.6547\\ 
H & 11.1093 & 13.4415 & 17.0173\\ 
C & 4.6336 & 15.1486 & 11.4355\\ 
C & 5.6665 & 14.4928 & 10.7259\\ 
C & 4.9517 & 16.1376 & 12.3948\\ 
C & 3.2835 & 14.8106 & 11.1874\\ 
C & 7.0171 & 14.8275 & 10.9756\\
\end{tabular}
\caption{}
\end{table}
\begin{table}[p]
\begin{tabular}{c|c c c}
type & $x$ & $y$ & $z$\\ 
\hline
H & 14.3781 & 13.7428 & 3.6285\\ 
H & 12.6814 & 12.4401 & 2.4253\\ 
H & 10.7465 & 16.3594 & 5.902\\ 
H & 9.0607 & 15.0664 & 4.701\\ 
H & 10.2708 & 11.9415 & 1.93\\ 
H & 8.4558 & 13.2601 & 3.0733\\ 
H & 16.0693 & 15.0452 & 4.8382\\
\end{tabular}
\caption{}
\end{table}
\clearpage

\begin{table}[p]
\begin{tabular}{c|c c c}
type & $x$ & $y$ & $z$\\ 
\hline
C & 5.0557 & 20.3417 & 9.3536\\ 
C & 1.7425 & 17.7501 & 6.8412\\ 
C & 5.7999 & 19.162 & 9.5878\\ 
C & 2.4848 & 16.5694 & 7.0788\\ 
C & 5.5145 & 21.5789 & 9.862\\ 
C & 7.0007 & 19.2216 & 10.3312\\ 
C & 6.713 & 21.6375 & 10.6109\\
\end{tabular}
\caption{}
\end{table}
\begin{table}[p]
\begin{tabular}{c|c c c}
type & $x$ & $y$ & $z$\\ 
\hline
H & 13.7083 & 22.2643 & 3.963\\ 
H & 15.0288 & 20.1748 & 4.4422\\ 
C & 9.5892 & 21.5262 & -1.7854\\ 
C & 8.5375 & 22.2166 & -1.1383\\ 
C & 10.9343 & 21.8536 & -1.4929\\ 
C & 9.293 & 20.502 & -2.7156\\ 
C & 8.8303 & 23.2329 & -0.2003\\
\end{tabular}
\caption{}
\end{table}
\clearpage

\begin{table}[p]
\begin{tabular}{c|c c c}
type & $x$ & $y$ & $z$\\ 
\hline
H & 1.4446 & 23.9083 & 7.3316\\ 
H & -0.2262 & 22.6009 & 6.1146\\ 
H & 5.1005 & 21.63 & 4.7545\\ 
H & 3.4299 & 20.3245 & 3.539\\ 
H & -0.8029 & 20.8617 & 4.4038\\ 
H & 1.0311 & 19.7161 & 3.1102\\ 
H & 3.1173 & 25.2116 & 8.5499\\
\end{tabular}
\caption{}
\end{table}
\begin{table}[p]
\begin{tabular}{c|c c c}
type & $x$ & $y$ & $z$\\ 
\hline
C & 3.0647 & 26.4267 & 3.9612\\ 
C & -0.2721 & 23.8352 & 1.4823\\ 
C & 2.2965 & 27.6 & 3.7749\\ 
C & -1.0431 & 25.0065 & 1.2951\\ 
C & 4.2801 & 26.4841 & 4.6804\\ 
C & 2.7476 & 28.8303 & 4.3085\\ 
C & 4.7319 & 27.713 & 5.2136\\
\end{tabular}
\caption{}
\end{table}
\clearpage

\begin{table}[p]
\begin{tabular}{c|c c c}
type & $x$ & $y$ & $z$\\ 
\hline
H & 12.8054 & 14.6168 & 13.8825\\ 
H & 14.6529 & 13.4699 & 12.6137\\ 
C & -2.6813 & 26.5579 & 6.442\\ 
C & -1.6597 & 25.8225 & 5.7959\\ 
C & -2.3435 & 27.5706 & 7.3711\\ 
C & -4.0389 & 26.2775 & 6.1581\\ 
C & -0.3027 & 26.0977 & 6.0797\\
\end{tabular}
\caption{}
\end{table}
\begin{table}[p]
\begin{tabular}{c|c c c}
type & $x$ & $y$ & $z$\\ 
\hline
C & -7.0521 & 8.916 & -16.4031\\ 
H & -3.2609 & 7.8287 & -13.5263\\ 
H & -1.0935 & 6.7189 & -13.6797\\ 
H & -4.0826 & 6.9112 & -18.3937\\ 
H & -1.8976 & 5.801 & -18.5494\\ 
H & 0.6429 & 5.5095 & -15.029\\ 
H & 0.2403 & 5.0499 & -17.4682\\
\end{tabular}
\caption{}
\end{table}
\clearpage

\begin{table}[p]
\begin{tabular}{c|c c c}
type & $x$ & $y$ & $z$\\ 
\hline
C & 3.245 & 19.7139 & 4.9734\\ 
C & -0.37 & 20.9972 & 3.1562\\ 
C & 3.9796 & 18.7675 & 2.8356\\ 
C & -0.1394 & 21.256 & 4.528\\ 
C & 4.2152 & 19.0264 & 4.2062\\ 
C & -1.5731 & 21.4224 & 2.5473\\ 
C & -1.1163 & 21.9395 & 5.2902\\
\end{tabular}
\caption{}
\end{table}
\begin{table}[p]
\begin{tabular}{c|c c c}
type & $x$ & $y$ & $z$\\ 
\hline
H & -2.3128 & 19.1208 & 4.8402\\ 
H & -5.7852 & 17.2633 & 2.9899\\ 
H & -4.754 & 19.173 & 4.2659\\ 
C & -0.2384 & 13.5284 & -5.4563\\ 
C & -0.406 & 13.3277 & -6.8462\\ 
C & -1.2518 & 14.171 & -4.7087\\ 
C & 0.9419 & 13.0855 & -4.8158\\
\end{tabular}
\caption{}
\end{table}
\clearpage

\begin{table}[p]
\begin{tabular}{c|c c c}
type & $x$ & $y$ & $z$\\ 
\hline
C & -1.6681 & 23.877 & -0.2513\\ 
H & 0.2925 & 19.8716 & -2.2365\\ 
H & 2.4562 & 18.7487 & -2.3567\\ 
H & 2.3247 & 23.4931 & 0.5844\\ 
H & 4.5011 & 22.3654 & 0.458\\ 
H & 4.8893 & 18.6442 & -1.7665\\ 
H & 5.9174 & 20.4587 & -0.3558\\
\end{tabular}
\caption{}
\end{table}
\begin{table}[p]
\begin{tabular}{c|c c c}
type & $x$ & $y$ & $z$\\ 
\hline
C & -2.5419 & 4.1151 & -12.2831\\ 
C & -4.9477 & 7.3254 & -10.901\\ 
C & -0.5981 & 5.0865 & -11.1498\\ 
C & -5.5219 & 6.3052 & -11.6952\\ 
C & -1.1688 & 4.0655 & -11.9454\\ 
C & -5.7446 & 8.4008 & -10.4467\\ 
C & -6.8949 & 6.3625 & -12.0321\\
\end{tabular}
\caption{}
\end{table}
\clearpage

\begin{table}[p]
\begin{tabular}{c|c c c}
type & $x$ & $y$ & $z$\\ 
\hline
H & -6.4281 & 8.7178 & -7.5378\\ 
H & -6.8191 & 12.5902 & -5.5884\\ 
H & -7.8527 & 10.6671 & -6.8471\\ 
C & 2.3561 & 19.2631 & -9.8432\\ 
C & 2.2331 & 19.1722 & -11.2498\\ 
C & 1.3223 & 19.8528 & -9.0798\\ 
C & 3.5097 & 18.7592 & -9.1992\\
\end{tabular}
\caption{}
\end{table}
\begin{table}[p]
\begin{tabular}{c|c c c}
type & $x$ & $y$ & $z$\\ 
\hline
C & -3.6338 & 18.9818 & -0.3103\\ 
H & -0.7304 & 16.9321 & -3.6601\\ 
H & 1.4542 & 15.8421 & -3.7803\\ 
H & -0.1371 & 17.5708 & 1.2828\\ 
H & 2.042 & 16.4676 & 1.1627\\ 
H & 3.5468 & 15.0345 & -2.6626\\ 
H & 3.844 & 15.3466 & -0.1793\\
\end{tabular}
\caption{}
\end{table}
\clearpage

\begin{table}[p]
\begin{tabular}{c|c c c}
type & $x$ & $y$ & $z$\\ 
\hline
C & 7.0585 & 5.5121 & 17.8185\\ 
C & 4.6298 & 5.6148 & 17.5158\\ 
C & 8.0292 & 3.8876 & 19.3725\\ 
C & 3.1724 & 4.088 & 18.7583\\ 
C & 8.1918 & 4.9361 & 18.4373\\ 
C & 3.3347 & 5.138 & 17.8242\\ 
C & 9.1624 & 3.3146 & 19.9936\\
\end{tabular}
\caption{}
\end{table}
\begin{table}[p]
\begin{tabular}{c|c c c}
type & $x$ & $y$ & $z$\\ 
\hline
H & 9.6871 & -8.8241 & 24.5573\\ 
H & 10.2846 & -5.2388 & 21.0983\\ 
H & 11.9911 & -8.0316 & 23.9687\\ 
H & 12.2938 & -6.2297 & 22.2333\\ 
C & 6.4089 & -13.1793 & 24.7582\\ 
C & 6.5885 & -12.0854 & 23.8792\\ 
C & 7.5339 & -13.801 & 25.3503\\
\end{tabular}
\caption{}
\end{table}
\clearpage

\begin{table}[p]
\begin{tabular}{c|c c c}
type & $x$ & $y$ & $z$\\ 
\hline
C & 10.3464 & 1.8877 & 23.5618\\ 
C & 10.3825 & 3.2879 & 23.7599\\ 
H & 6.6505 & 0.2508 & 22.9241\\ 
H & 4.2227 & 0.3485 & 22.6593\\ 
H & 6.7745 & 5.2194 & 23.6182\\ 
H & 4.3451 & 5.3167 & 23.3426\\ 
H & 2.0892 & 1.6718 & 22.6037\\
\end{tabular}
\caption{}
\end{table}
\begin{table}[p]
\begin{tabular}{c|c c c}
type & $x$ & $y$ & $z$\\ 
\hline
C & 7.2177 & 7.7064 & 14.4344\\ 
C & 4.782 & 7.7996 & 14.1752\\ 
C & 8.3793 & 5.5534 & 14.3046\\ 
C & 3.5116 & 5.7413 & 13.7845\\ 
C & 8.4163 & 6.9565 & 14.4812\\ 
C & 3.5464 & 7.1442 & 13.9639\\ 
C & 9.5762 & 4.8028 & 14.3556\\
\end{tabular}
\caption{}
\end{table}
\clearpage

\begin{table}[p]
\begin{tabular}{c|c c c}
type & $x$ & $y$ & $z$\\ 
\hline
H & 10.1989 & -7.8171 & 17.9657\\ 
H & 10.3453 & -2.8356 & 18.543\\ 
H & 12.3958 & -6.6702 & 18.34\\ 
H & 12.4723 & -4.168 & 18.6301\\ 
C & 6.8806 & -11.5863 & 19.955\\ 
C & 6.9661 & -10.1742 & 19.965\\ 
C & 8.0545 & -12.3656 & 20.0844\\
\end{tabular}
\caption{}
\end{table}
\begin{table}[p]
\begin{tabular}{c|c c c}
type & $x$ & $y$ & $z$\\ 
\hline
C & 9.9691 & 0.5365 & 28.9022\\ 
C & 10.1365 & 1.4837 & 27.8647\\ 
H & 6.1096 & -0.5207 & 29.7752\\ 
H & 3.6804 & -0.4349 & 29.4553\\ 
H & 6.7066 & 2.8442 & 26.0961\\ 
H & 4.2854 & 2.9331 & 25.783\\ 
H & 1.677 & 0.4727 & 28.2431\\
\end{tabular}
\caption{}
\end{table}
\clearpage

\begin{table}[p]
\begin{tabular}{c|c c c}
type & $x$ & $y$ & $z$\\ 
\hline
C & 8.2892 & -2.4953 & 12.4087\\ 
C & 5.8645 & -2.3965 & 12.0646\\ 
C & 9.2294 & -4.0722 & 14.0306\\ 
C & 4.3835 & -3.8665 & 13.3498\\ 
C & 9.4106 & -3.0512 & 13.0679\\ 
C & 4.5629 & -2.8522 & 12.3799\\ 
C & 10.3493 & -4.6285 & 14.69\\
\end{tabular}
\caption{}
\end{table}
\begin{table}[p]
\begin{tabular}{c|c c c}
type & $x$ & $y$ & $z$\\ 
\hline
H & 11.018 & -10.1466 & 11.5128\\ 
H & 11.0688 & -5.1365 & 11.2039\\ 
H & 13.1938 & -8.9044 & 11.664\\ 
H & 13.2196 & -6.3944 & 11.5115\\ 
\hline\hline
C294H210\\ 
\hline\hline
C & 5.4785 & 9.8513 & 17.243\\ 
C & 5.6558 & 10.9732 & 16.3991\\
\end{tabular}
\caption{}
\end{table}
\clearpage

\begin{table}[p]
\begin{tabular}{c|c c c}
type & $x$ & $y$ & $z$\\ 
\hline
C & 9.8331 & -0.3154 & 19.9631\\ 
C & 10.8092 & -1.9276 & 21.5271\\ 
C & 10.9684 & -0.8791 & 20.5908\\ 
H & 6.9648 & -3.1278 & 22.2555\\ 
H & 4.541 & -3.033 & 21.9508\\ 
H & 7.5272 & 0.5845 & 18.9243\\ 
H & 5.1039 & 0.6717 & 18.614\\
\end{tabular}
\caption{}
\end{table}
\begin{table}[p]
\begin{tabular}{c|c c c}
type & $x$ & $y$ & $z$\\ 
\hline
C & 3.8257 & 7.1372 & 20.8954\\ 
C & 6.366 & 9.8384 & 21.4041\\ 
C & 3.9356 & 9.9558 & 21.0979\\ 
C & 7.493 & 7.6634 & 21.4042\\ 
C & 2.6383 & 7.8984 & 20.7962\\ 
C & 7.5516 & 9.0731 & 21.5073\\ 
C & 2.6935 & 9.3084 & 20.8973\\
\end{tabular}
\caption{}
\end{table}
\clearpage

\begin{table}[p]
\begin{tabular}{c|c c c}
type & $x$ & $y$ & $z$\\ 
\hline
H & 2.6101 & -1.3853 & 24.811\\ 
H & 9.4836 & -5.4807 & 25.118\\ 
H & 9.6785 & -0.496 & 25.6583\\ 
H & 11.6946 & -4.3493 & 25.4542\\ 
H & 11.7961 & -1.8464 & 25.7236\\ 
C & 6.0709 & -9.2603 & 26.928\\ 
C & 6.1209 & -7.8528 & 27.0613\\
\end{tabular}
\caption{}
\end{table}
\begin{table}[p]
\begin{tabular}{c|c c c}
type & $x$ & $y$ & $z$\\ 
\hline
C & 9.9312 & 1.7801 & 16.5142\\ 
C & 11.0915 & -0.3747 & 16.4222\\ 
C & 11.13 & 1.0316 & 16.5706\\ 
H & 7.394 & -2.026 & 15.8297\\ 
H & 4.9668 & -1.9319 & 15.5565\\ 
H & 7.527 & 2.9641 & 16.3569\\ 
H & 5.0989 & 3.0563 & 16.0861\\
\end{tabular}
\caption{}
\end{table}
\clearpage

\begin{table}[p]
\begin{tabular}{c|c c c}
type & $x$ & $y$ & $z$\\ 
\hline
C & 3.3493 & 5.7749 & 26.5513\\ 
C & 6.1348 & 7.7244 & 24.9861\\ 
C & 3.714 & 7.8098 & 24.6207\\ 
C & 7.073 & 6.1486 & 26.6115\\ 
C & 2.231 & 6.3185 & 25.8769\\ 
C & 7.2551 & 7.1661 & 25.6448\\ 
C & 2.4135 & 7.3358 & 24.9109\\
\end{tabular}
\caption{}
\end{table}
\begin{table}[p]
\begin{tabular}{c|c c c}
type & $x$ & $y$ & $z$\\ 
\hline
H & 2.5288 & -3.3082 & 28.234\\ 
H & 9.1794 & -6.4774 & 31.3955\\ 
H & 9.6584 & -2.5723 & 28.2787\\ 
H & 11.4599 & -5.5785 & 30.8563\\ 
H & 11.7016 & -3.6221 & 29.294\\ 
C & 6.2809 & 1.7345 & 11.9652\\ 
C & 6.4242 & 2.8461 & 11.1018\\
\end{tabular}
\caption{}
\end{table}
\clearpage

\begin{table}[p]
\begin{tabular}{c|c c c}
type & $x$ & $y$ & $z$\\ 
\hline
C & 10.9154 & -8.4347 & 14.8444\\ 
C & 11.9252 & -10.0138 & 16.4209\\ 
C & 12.0626 & -8.9746 & 15.4708\\ 
H & 8.102 & -11.2657 & 17.1839\\ 
H & 5.6706 & -11.1997 & 16.8992\\ 
H & 8.5922 & -7.5745 & 13.8158\\ 
H & 6.1667 & -7.5108 & 13.5316\\
\end{tabular}
\caption{}
\end{table}
\begin{table}[p]
\begin{tabular}{c|c c c}
type & $x$ & $y$ & $z$\\ 
\hline
C & 11.4205 & 6.9775 & 18.6388\\ 
C & 9.0419 & 6.8133 & 19.1894\\ 
C & 10.9942 & 9.5558 & 17.5558\\ 
C & 8.6102 & 9.3913 & 18.1116\\ 
C & 12.5041 & 7.7017 & 18.0917\\ 
C & 7.7436 & 7.3728 & 19.198\\ 
C & 12.2918 & 8.991 & 17.5489\\
\end{tabular}
\caption{}
\end{table}
\clearpage

\begin{table}[p]
\begin{tabular}{c|c c c}
type & $x$ & $y$ & $z$\\ 
\hline
H & 6.3041 & -5.1916 & 20.6328\\ 
H & 5.9247 & -2.9979 & 19.4501\\ 
H & 13.3743 & -5.9277 & 19.6738\\ 
H & 12.622 & -1.57 & 17.303\\ 
H & 15.3061 & -4.7421 & 18.5915\\ 
H & 14.9252 & -2.5533 & 17.4013\\ 
C & 9.1798 & -10.4903 & 20.3377\\
\end{tabular}
\caption{}
\end{table}
\begin{table}[p]
\begin{tabular}{c|c c c}
type & $x$ & $y$ & $z$\\ 
\hline
C & 14.3843 & 4.1889 & 20.8423\\ 
C & 14.4135 & 6.8889 & 21.683\\ 
C & 15.587 & 4.9306 & 20.7914\\ 
C & 15.6013 & 6.2814 & 21.2118\\ 
H & 11.9856 & 3.0066 & 21.0321\\ 
H & 9.6041 & 2.8659 & 21.5474\\ 
H & 12.0301 & 7.7986 & 22.5282\\
\end{tabular}
\caption{}
\end{table}
\clearpage

\begin{table}[p]
\begin{tabular}{c|c c c}
type & $x$ & $y$ & $z$\\ 
\hline
C & 10.8674 & -13.8109 & 23.3432\\ 
C & 8.4653 & -13.9277 & 23.8169\\ 
C & 10.9067 & -11.1215 & 24.2194\\ 
C & 8.508 & -11.2407 & 24.6963\\ 
C & 12.0766 & -13.077 & 23.3197\\ 
C & 7.2738 & -13.3119 & 24.2665\\ 
C & 12.0953 & -11.7318 & 23.7571\\
\end{tabular}
\caption{}
\end{table}
\begin{table}[p]
\begin{tabular}{c|c c c}
type & $x$ & $y$ & $z$\\ 
\hline
H & 5.4437 & -2.5605 & 15.5839\\ 
H & 5.5008 & -0.1891 & 16.4362\\ 
H & 12.3217 & -3.3933 & 13.7285\\ 
H & 12.4404 & 1.3312 & 15.4152\\ 
H & 14.4828 & -2.1139 & 13.7158\\ 
H & 14.5394 & 0.2593 & 14.5626\\ 
C & 8.4179 & -7.9147 & 15.7226\\
\end{tabular}
\caption{}
\end{table}
\clearpage

\begin{table}[p]
\begin{tabular}{c|c c c}
type & $x$ & $y$ & $z$\\ 
\hline
C & 15.3656 & 1.583 & 25.9805\\ 
C & 14.8957 & 4.0038 & 24.5949\\ 
C & 16.4334 & 2.2436 & 25.329\\ 
C & 16.1982 & 3.454 & 24.6354\\ 
H & 13.1725 & 0.5227 & 27.11\\ 
H & 10.7951 & 0.3889 & 27.6796\\ 
H & 12.3392 & 4.817 & 24.6429\\
\end{tabular}
\caption{}
\end{table}
\begin{table}[p]
\begin{tabular}{c|c c c}
type & $x$ & $y$ & $z$\\ 
\hline
C & 9.7228 & 0.3204 & 11.7849\\ 
C & 7.3416 & 0.2135 & 12.3399\\ 
C & 9.3633 & 2.9228 & 10.736\\ 
C & 6.978 & 2.8163 & 11.2975\\ 
C & 10.8246 & 1.0232 & 11.2462\\ 
C & 6.0601 & 0.81 & 12.3642\\ 
C & 10.6453 & 2.3242 & 10.72\\
\end{tabular}
\caption{}
\end{table}
\clearpage

\begin{table}[p]
\begin{tabular}{c|c c c}
type & $x$ & $y$ & $z$\\ 
\hline
H & 4.3399 & -11.9394 & 14.0595\\ 
H & 4.014 & -9.7384 & 12.881\\ 
H & 11.4097 & -12.8083 & 13.1461\\ 
H & 10.7537 & -8.4109 & 10.8174\\ 
H & 13.3743 & -11.6324 & 12.114\\ 
H & 13.0432 & -9.4259 & 10.9457\\ 
C & 6.4314 & -5.8603 & 5.2392\\
\end{tabular}
\caption{}
\end{table}
\begin{table}[p]
\begin{tabular}{c|c c c}
type & $x$ & $y$ & $z$\\ 
\hline
C & 10.3978 & -9.2193 & 7.8266\\ 
C & 10.4352 & -6.463 & 8.4557\\ 
C & 11.6104 & -8.4935 & 7.7545\\ 
C & 11.6289 & -7.1144 & 8.0689\\ 
H & 7.9765 & -10.3622 & 8.0595\\ 
H & 5.5716 & -10.4308 & 8.5368\\ 
H & 8.0538 & -5.4693 & 9.1817\\
\end{tabular}
\caption{}
\end{table}
\clearpage

\begin{table}[p]
\begin{tabular}{c|c c c}
type & $x$ & $y$ & $z$\\ 
\hline
C & 14.7109 & 7.0875 & -58.4627\\ 
C & 13.7627 & 8.0985 & -60.4806\\ 
C & 16.1514 & 8.3442 & -59.994\\ 
C & 13.4369 & 6.5771 & -58.1206\\ 
C & 15.8224 & 6.8365 & -57.6259\\ 
C & 12.4875 & 7.5961 & -60.1331\\ 
C & 17.2629 & 8.0951 & -59.1559\\
\end{tabular}
\caption{}
\end{table}
\begin{table}[p]
\begin{tabular}{c|c c c}
type & $x$ & $y$ & $z$\\ 
\hline
H & 10.6402 & 10.2384 & -41.9416\\ 
H & 13.8568 & 9.6007 & -44.8038\\ 
H & 13.0939 & 10.1601 & -42.4717\\ 
H & 7.5238 & 8.653 & -48.0207\\ 
H & 5.9663 & 9.7929 & -43.3871\\ 
H & 5.0681 & 8.7406 & -47.5122\\ 
H & 4.2875 & 9.3126 & -45.1907\\
\end{tabular}
\caption{}
\end{table}
\clearpage

\begin{table}[p]
\begin{tabular}{c|c c c}
type & $x$ & $y$ & $z$\\ 
\hline
C & 14.5927 & 6.4553 & -46.278\\ 
C & 9.1286 & 6.6356 & -49.7522\\ 
C & 8.6018 & 5.3692 & -47.2792\\ 
C & 7.8255 & 6.1756 & -49.4578\\ 
C & 7.5605 & 5.5419 & -48.2214\\ 
H & 11.6671 & 7.4296 & -50.0624\\ 
H & 14.0005 & 7.7388 & -49.417\\
\end{tabular}
\caption{}
\end{table}
\begin{table}[p]
\begin{tabular}{c|c c c}
type & $x$ & $y$ & $z$\\ 
\hline
C & 15.4308 & 12.3896 & -66.9567\\ 
C & 14.6313 & 11.7608 & -69.1859\\ 
C & 17.0376 & 11.9754 & -68.7577\\ 
C & 14.0962 & 12.4472 & -66.4953\\ 
C & 16.4959 & 12.6785 & -66.0726\\ 
C & 13.2948 & 11.8219 & -68.7242\\ 
C & 18.1034 & 12.2587 & -67.871\\
\end{tabular}
\caption{}
\end{table}
\clearpage

\begin{table}[p]
\begin{tabular}{c|c c c}
type & $x$ & $y$ & $z$\\ 
\hline
H & 14.9567 & 9.8079 & -51.9133\\ 
H & 15.7299 & 11.265 & -47.8875\\ 
H & 16.58 & 10.4746 & -50.1238\\ 
H & 8.6554 & 10.7156 & -48.98\\ 
H & 10.353 & 9.1592 & -53.433\\ 
H & 7.0317 & 9.9781 & -50.7469\\ 
H & 7.8852 & 9.162 & -52.9646\\
\end{tabular}
\caption{}
\end{table}
\begin{table}[p]
\begin{tabular}{c|c c c}
type & $x$ & $y$ & $z$\\ 
\hline
C & 16.8964 & 7.0672 & -51.9793\\ 
C & 11.3068 & 7.2916 & -55.2542\\ 
C & 10.9075 & 5.884 & -52.8366\\ 
C & 10.0322 & 6.761 & -54.9478\\ 
C & 9.832 & 6.0595 & -53.7366\\ 
H & 13.8165 & 8.1734 & -55.6051\\ 
H & 16.1712 & 8.4854 & -55.0309\\
\end{tabular}
\caption{}
\end{table}
\clearpage

\begin{table}[p]
\begin{tabular}{c|c c c}
type & $x$ & $y$ & $z$\\ 
\hline
C & 14.6186 & 14.8067 & -57.6046\\ 
C & 13.9367 & 15.8228 & -59.723\\ 
C & 16.2613 & 16.0062 & -58.9716\\ 
C & 13.2995 & 14.3374 & -57.4039\\ 
C & 15.6197 & 14.5361 & -56.6446\\ 
C & 12.6172 & 15.359 & -59.5189\\ 
C & 17.2632 & 15.7307 & -58.0114\\
\end{tabular}
\caption{}
\end{table}
\begin{table}[p]
\begin{tabular}{c|c c c}
type & $x$ & $y$ & $z$\\ 
\hline
H & 1.7274 & 7.8753 & -55.9784\\ 
H & 2.9105 & 12.0562 & -56.2125\\ 
H & 1.4739 & 10.2503 & -55.2104\\ 
H & 6.8699 & 8.1527 & -60.7349\\ 
H & 3.9891 & 4.5363 & -58.7804\\ 
H & 7.1856 & 5.7673 & -61.4634\\ 
H & 5.7437 & 3.9577 & -60.4821\\
\end{tabular}
\caption{}
\end{table}
\clearpage

\begin{table}[p]
\begin{tabular}{c|c c c}
type & $x$ & $y$ & $z$\\ 
\hline
C & 7.994 & 10.7512 & 16.5001\\ 
C & 3.1372 & 10.939 & 15.8713\\ 
C & 8.9605 & 9.1464 & 18.0781\\ 
C & 9.2916 & 11.2108 & 16.1749\\ 
C & 10.2579 & 9.6065 & 17.7547\\ 
C & 10.4237 & 10.6392 & 16.802\\ 
H & 6.4062 & 8.4521 & 18.5119\\
\end{tabular}
\caption{}
\end{table}
\begin{table}[p]
\begin{tabular}{c|c c c}
type & $x$ & $y$ & $z$\\ 
\hline
C & 6.2677 & -7.3938 & 22.9026\\ 
C & 6.4397 & -6.3802 & 21.9315\\ 
C & 7.396 & -7.9603 & 23.539\\ 
C & 4.9676 & -7.8386 & 23.2364\\ 
C & 7.7388 & -5.9322 & 21.5986\\ 
C & 5.3118 & -5.8182 & 21.2911\\ 
C & 8.6953 & -7.514 & 23.2044\\
\end{tabular}
\caption{}
\end{table}
\clearpage

\begin{table}[p]
\begin{tabular}{c|c c c}
type & $x$ & $y$ & $z$\\ 
\hline
H & 6.3728 & 10.9787 & 21.4246\\ 
H & 3.9385 & 11.0832 & 21.1364\\ 
H & 1.6457 & 7.4124 & 20.687\\ 
H & 1.7355 & 9.923 & 20.8205\\ 
H & 8.6056 & 5.8607 & 21.4356\\ 
H & 8.8082 & 10.8678 & 21.7071\\ 
H & 10.8131 & 7.0045 & 21.7477\\
\end{tabular}
\caption{}
\end{table}
\begin{table}[p]
\begin{tabular}{c|c c c}
type & $x$ & $y$ & $z$\\ 
\hline
C & 8.5434 & -8.0003 & 27.3432\\ 
C & 3.6796 & -7.7669 & 26.8054\\ 
C & 9.6882 & -10.1628 & 27.2367\\ 
C & 9.7845 & -7.3537 & 27.5451\\ 
C & 10.9305 & -9.5153 & 27.4329\\ 
C & 10.9786 & -8.11 & 27.5887\\ 
H & 7.2154 & -11.1395 & 26.8594\\
\end{tabular}
\caption{}
\end{table}
\clearpage

\begin{table}[p]
\begin{tabular}{c|c c c}
type & $x$ & $y$ & $z$\\ 
\hline
C & 6.2196 & -0.2061 & 15.9645\\ 
C & 6.2597 & 1.1976 & 16.1337\\ 
C & 7.416 & -0.9577 & 16.0149\\ 
C & 4.9833 & -0.8562 & 15.7441\\ 
C & 7.4951 & 1.8485 & 16.3555\\ 
C & 5.0629 & 1.9483 & 16.0851\\ 
C & 8.6513 & -0.3068 & 16.2378\\
\end{tabular}
\caption{}
\end{table}
\begin{table}[p]
\begin{tabular}{c|c c c}
type & $x$ & $y$ & $z$\\ 
\hline
H & 8.2911 & -8.4566 & 20.1478\\ 
H & 5.8707 & -8.3105 & 19.8731\\ 
H & 3.4885 & -11.9326 & 19.5376\\ 
H & 3.6406 & -9.4201 & 19.6022\\ 
H & 10.4405 & -13.6092 & 20.2962\\ 
H & 10.7106 & -8.5985 & 20.4286\\ 
H & 12.6664 & -12.482 & 20.5692\\
\end{tabular}
\caption{}
\end{table}
\clearpage

\begin{table}[p]
\begin{tabular}{c|c c c}
type & $x$ & $y$ & $z$\\ 
\hline
C & 8.2726 & -3.9608 & 29.3203\\ 
C & 3.4099 & -3.7937 & 28.7458\\ 
C & 9.2834 & -5.6666 & 30.7579\\ 
C & 9.5578 & -3.4568 & 29.015\\ 
C & 10.5692 & -5.1604 & 30.4542\\ 
C & 10.7068 & -4.055 & 29.582\\ 
H & 6.7401 & -6.4381 & 31.1559\\
\end{tabular}
\caption{}
\end{table}
\begin{table}[p]
\begin{tabular}{c|c c c}
type & $x$ & $y$ & $z$\\ 
\hline
C & 6.852 & -4.0322 & 13.8274\\ 
C & 7.0136 & -2.9735 & 12.9036\\ 
C & 7.9848 & -4.6042 & 14.4507\\ 
C & 5.5581 & -4.5173 & 14.1271\\ 
C & 8.3072 & -2.4882 & 12.6024\\ 
C & 5.8806 & -2.4043 & 12.2784\\ 
C & 9.2779 & -4.1167 & 14.152\\
\end{tabular}
\caption{}
\end{table}
\clearpage

\begin{table}[p]
\begin{tabular}{c|c c c}
type & $x$ & $y$ & $z$\\ 
\hline
H & 7.8274 & 6.1789 & 7.4326\\ 
H & 5.394 & 6.3123 & 7.2093\\ 
H & 3.0726 & 2.704 & 6.4978\\ 
H & 3.179 & 5.1914 & 6.8561\\ 
H & 10.0533 & 1.0765 & 6.9569\\ 
H & 10.2575 & 6.0419 & 7.6797\\ 
H & 12.2639 & 2.1933 & 7.3608\\
\end{tabular}
\caption{}
\end{table}
\begin{table}[p]
\begin{tabular}{c|c c c}
type & $x$ & $y$ & $z$\\ 
\hline
C & -5.2141 & 7.4072 & 17.1788\\ 
C & -0.5499 & 6.4226 & 18.9928\\ 
C & -5.4057 & 6.7815 & 18.4332\\ 
C & 0.9477 & 7.2788 & 17.2531\\ 
C & 0.5705 & 6.032 & 19.7627\\ 
C & 2.068 & 6.8905 & 18.0239\\ 
C & 1.8791 & 6.2682 & 19.2802\\
\end{tabular}
\caption{}
\end{table}
\clearpage

\begin{table}[p]
\begin{tabular}{c|c c c}
type & $x$ & $y$ & $z$\\ 
\hline
H & 3.6658 & 7.8021 & 3.71\\ 
C & -2.6713 & 10.9688 & 9.4631\\ 
C & -2.5363 & 12.3341 & 9.12\\ 
C & -1.524 & 10.2007 & 9.7678\\ 
C & -3.9536 & 10.3742 & 9.5031\\ 
C & -1.255 & 12.9304 & 9.0818\\ 
C & -3.6842 & 13.103 & 8.8211\\
\end{tabular}
\caption{}
\end{table}
\begin{table}[p]
\begin{tabular}{c|c c c}
type & $x$ & $y$ & $z$\\ 
\hline
H & -3.9708 & 5.7229 & 10.5303\\ 
H & -0.956 & 7.9856 & 6.4101\\ 
H & -3.3832 & 8.1287 & 6.1646\\ 
H & -5.9862 & 6.446 & 9.2176\\ 
H & -5.6895 & 7.6526 & 7.0271\\ 
H & 0.8942 & 5.4588 & 11.0343\\ 
H & 1.4702 & 7.851 & 6.6592\\
\end{tabular}
\caption{}
\end{table}
\clearpage

\begin{table}[p]
\begin{tabular}{c|c c c}
type & $x$ & $y$ & $z$\\ 
\hline
C & -5.1804 & 21.5301 & 5.6203\\ 
C & -0.5275 & 20.5664 & 7.47\\ 
C & -5.3787 & 20.9118 & 6.8772\\ 
C & 0.9832 & 21.3985 & 5.7314\\ 
C & 0.5863 & 20.1748 & 8.2473\\ 
C & 2.0976 & 21.0099 & 6.5108\\ 
C & 1.8988 & 20.3981 & 7.7708\\
\end{tabular}
\caption{}
\end{table}
\begin{table}[p]
\begin{tabular}{c|c c c}
type & $x$ & $y$ & $z$\\ 
\hline
H & 2.6961 & 10.0134 & 19.8753\\ 
C & -3.1257 & 11.4098 & 24.32\\ 
C & -2.9245 & 12.0904 & 23.0962\\ 
C & -2.0125 & 10.9687 & 25.0738\\ 
C & -4.4402 & 11.1648 & 24.7827\\ 
C & -1.613 & 12.3304 & 22.6269\\ 
C & -4.0353 & 12.5231 & 22.3367\\
\end{tabular}
\caption{}
\end{table}
\clearpage

\begin{table}[p]
\begin{tabular}{c|c c c}
type & $x$ & $y$ & $z$\\ 
\hline
H & -4.3813 & 15.2413 & 12.3642\\ 
H & -1.289 & 17.4282 & 8.2638\\ 
H & -3.7122 & 17.5848 & 7.9791\\ 
H & -6.3726 & 15.9493 & 11.0077\\ 
H & -6.0332 & 17.1267 & 8.8047\\ 
H & 0.4642 & 14.9328 & 12.9387\\ 
H & 1.1337 & 17.2699 & 8.5499\\
\end{tabular}
\caption{}
\end{table}
\begin{table}[p]
\begin{tabular}{c|c c c}
type & $x$ & $y$ & $z$\\ 
\hline
C & -5.8244 & 20.8416 & 16.5197\\ 
C & -0.7772 & 21.2136 & 15.8839\\ 
C & -5.6311 & 21.4888 & 15.2767\\ 
C & 0.1479 & 20.1722 & 17.897\\ 
C & 0.5325 & 21.4626 & 15.4126\\ 
C & 1.458 & 20.4259 & 17.4291\\ 
C & 1.6507 & 21.0715 & 16.1852\\
\end{tabular}
\caption{}
\end{table}
\clearpage

\begin{table}[p]
\begin{tabular}{c|c c c}
type & $x$ & $y$ & $z$\\ 
\hline
H & 2.8599 & 26.3881 & 7.7502\\ 
C & -2.7954 & 25.5222 & 3.3738\\ 
C & -2.582 & 26.243 & 2.1751\\ 
C & -1.6923 & 25.0619 & 4.1281\\ 
C & -4.1118 & 25.2603 & 3.8184\\ 
C & -1.2633 & 26.5049 & 1.7331\\ 
C & -3.6887 & 26.699 & 1.4204\\
\end{tabular}
\caption{}
\end{table}
\begin{table}[p]
\begin{tabular}{c|c c c}
type & $x$ & $y$ & $z$\\ 
\hline
H & -4.2108 & 23.7729 & 20.6286\\ 
H & -2.2529 & 18.6456 & 21.6677\\ 
H & -4.6744 & 18.8352 & 21.4121\\ 
H & -6.4827 & 22.7083 & 20.5961\\ 
H & -6.7174 & 20.2345 & 20.9896\\ 
H & 0.653 & 23.3818 & 21.1029\\ 
H & 0.171 & 18.4496 & 21.9011\\
\end{tabular}
\caption{}
\end{table}
\clearpage

\begin{table}[p]
\begin{tabular}{c|c c c}
type & $x$ & $y$ & $z$\\ 
\hline
C & -4.915 & 11.4566 & 4.6588\\ 
C & 0.1394 & 11.7427 & 4.0369\\ 
C & -4.7085 & 12.0411 & 3.387\\ 
C & 1.0368 & 10.7858 & 6.1054\\ 
C & 1.4562 & 11.9721 & 3.5731\\ 
C & 2.3529 & 11.0188 & 5.6446\\ 
C & 2.5629 & 11.6133 & 4.3782\\
\end{tabular}
\caption{}
\end{table}
\begin{table}[p]
\begin{tabular}{c|c c c}
type & $x$ & $y$ & $z$\\ 
\hline
H & -5.6398 & -1.6163 & -44.2642\\ 
H & -6.6582 & -0.2212 & -42.4351\\ 
C & 2.2308 & -4.3591 & -56.1214\\ 
C & 1.7174 & -3.6183 & -55.0313\\ 
C & 3.4324 & -3.9541 & -56.7466\\ 
C & 1.5414 & -5.5034 & -56.5848\\ 
C & 2.4054 & -2.4737 & -54.5669\\
\end{tabular}
\caption{}
\end{table}
\clearpage

\begin{table}[p]
\begin{tabular}{c|c c c}
type & $x$ & $y$ & $z$\\ 
\hline
H & -6.3983 & 5.3903 & -58.9266\\ 
H & -4.525 & 6.9506 & -59.079\\ 
H & -8.2699 & 8.0111 & -55.0746\\ 
H & -6.3917 & 9.5749 & -55.2289\\ 
H & -3.3212 & 8.9922 & -58.2596\\ 
H & -4.2562 & 10.3102 & -56.328\\ 
H & -8.2648 & 3.8236 & -58.7682\\
\end{tabular}
\caption{}
\end{table}
\begin{table}[p]
\begin{tabular}{c|c c c}
type & $x$ & $y$ & $z$\\ 
\hline
C & -5.5409 & 3.25 & -50.7303\\ 
C & -1.755 & 6.3397 & -51.0847\\ 
C & -6.046 & 3.9945 & -49.639\\ 
C & -2.2571 & 7.0818 & -49.9899\\ 
C & -6.2356 & 2.1067 & -51.188\\ 
C & -7.2451 & 3.5934 & -49.0067\\ 
C & -7.4324 & 1.7023 & -50.5519\\
\end{tabular}
\caption{}
\end{table}
\clearpage

\begin{table}[p]
\begin{tabular}{c|c c c}
type & $x$ & $y$ & $z$\\ 
\hline
H & -10.9085 & 3.7129 & -61.3073\\ 
H & -9.3098 & 1.7887 & -61.5695\\ 
C & 1.1236 & 0.33 & -61.8628\\ 
C & 1.6213 & -0.3821 & -62.9785\\ 
C & -0.0682 & -0.0924 & -61.2305\\ 
C & 1.8186 & 1.4629 & -61.3802\\ 
C & 0.9239 & -1.5119 & -63.4649\\
\end{tabular}
\caption{}
\end{table}
\begin{table}[p]
\begin{tabular}{c|c c c}
type & $x$ & $y$ & $z$\\ 
\hline
H & 3.7745 & -4.3917 & -51.0406\\ 
H & 5.6643 & -2.8444 & -51.1642\\ 
H & 0.6064 & -0.5029 & -50.8566\\ 
H & 2.496 & 1.042 & -50.9859\\ 
H & 6.559 & -0.5037 & -51.22\\ 
H & 4.9696 & 1.4502 & -51.1311\\ 
H & 1.8833 & -5.9371 & -50.9171\\
\end{tabular}
\caption{}
\end{table}
\clearpage

\begin{table}[p]
\begin{tabular}{c|c c c}
type & $x$ & $y$ & $z$\\ 
\hline
C & -4.0737 & 2.1171 & -64.5276\\ 
C & -0.2613 & 5.1913 & -64.7199\\ 
C & -3.1834 & 1.0202 & -64.4506\\ 
C & 0.6271 & 4.093 & -64.646\\ 
C & -5.4704 & 1.895 & -64.504\\ 
C & -3.689 & -0.2962 & -64.3471\\ 
C & -5.9762 & 0.5779 & -64.4\\
\end{tabular}
\caption{}
\end{table}
\begin{table}[p]
\begin{tabular}{c|c c c}
type & $x$ & $y$ & $z$\\ 
\hline
H & -3.7972 & -3.158 & -41.4034\\ 
H & -2.1887 & -5.0915 & -41.3816\\ 
C & 6.6368 & -4.8737 & -53.785\\ 
C & 7.511 & -5.9701 & -53.9737\\ 
C & 5.2491 & -5.0942 & -53.632\\ 
C & 7.1493 & -3.5563 & -53.7551\\ 
C & 6.9951 & -7.2872 & -54.0117\\
\end{tabular}
\caption{}
\end{table}
\clearpage

\begin{table}[p]
\begin{tabular}{c|c c c}
type & $x$ & $y$ & $z$\\ 
\hline
H & -1.8083 & 1.8376 & -57.5392\\ 
H & 0.0873 & 3.3759 & -57.668\\ 
H & 1.3472 & -2.0624 & -57.7038\\ 
H & 3.2421 & -0.5229 & -57.8265\\ 
H & 2.5637 & 3.7745 & -57.8015\\ 
H & 4.1459 & 1.8142 & -57.8815\\ 
H & -3.7028 & 0.2976 & -57.4098\\
\end{tabular}
\caption{}
\end{table}
\begin{table}[p]
\begin{tabular}{c|c c c}
type & $x$ & $y$ & $z$\\ 
\hline
C & -9.4086 & 7.1524 & -52.0654\\ 
C & -5.5734 & 10.1934 & -52.3408\\ 
C & -8.5398 & 6.0466 & -52.2213\\ 
C & -4.708 & 9.0855 & -52.4987\\ 
C & -10.799 & 6.9403 & -51.9127\\ 
C & -9.0599 & 4.7319 & -52.2232\\ 
C & -11.3199 & 5.6248 & -51.921\\
\end{tabular}%
\caption{}
\end{table}

\FloatBarrier
\subsection{Tetracene Cluster}
\clearpage

\begin{table}[p]
%
\caption{}
\end{table}

\FloatBarrier
\subsection{Pentacene Cluster}
\clearpage

\begin{table}[p]
\begin{tabular}{c|c c c}
type & $x$ & $y$ & $z$\\ 
\hline
\hline\hline
C22H14\\ 
\hline\hline
C & 3.811 & 1.2512 & 5.3018\\ 
C & 3.3333 & 2.1367 & 4.3064\\ 
C & 2.9103 & 0.6815 & 6.2325\\ 
C & 5.1889 & 0.9363 & 5.3657\\ 
C & 1.9553 & 2.4516 & 4.2426\\ 
C & 4.2339 & 2.7064 & 3.3757\\
\end{tabular}
\caption{}
\end{table}
\begin{table}[p]
\begin{tabular}{c|c c c}
type & $x$ & $y$ & $z$\\ 
\hline
H & 2.7836 & 1.216 & -143.7532\\ 
H & 5.0044 & 0.7408 & -144.8256\\ 
H & 2.8961 & -0.5682 & -151.5612\\ 
H & -1.5488 & 0.3881 & -149.432\\ 
H & 1.7919 & -0.8516 & -153.7323\\ 
H & -2.6492 & 0.1158 & -151.5992\\ 
H & 0.6901 & -1.089 & -155.9128\\
\end{tabular}
\caption{}
\end{table}
\clearpage

\begin{table}[p]
\begin{tabular}{c|c c c}
type & $x$ & $y$ & $z$\\ 
\hline
C & 12.3902 & 0.6083 & 2.843\\ 
C & 14.0511 & 0.9882 & 4.6041\\ 
C & 13.4315 & 0.4065 & 1.9064\\ 
C & 15.0919 & 0.7884 & 3.6679\\ 
C & 14.782 & 0.4968 & 2.3187\\ 
H & 4.7487 & 1.7349 & 7.7545\\ 
H & 7.1309 & 1.5928 & 7.227\\
\end{tabular}
\caption{}
\end{table}
\begin{table}[p]
\begin{tabular}{c|c c c}
type & $x$ & $y$ & $z$\\ 
\hline
C & 8.7939 & 3.915 & 8.5083\\ 
C & 10.2116 & 2.2723 & 9.6471\\ 
C & 9.9137 & 4.3754 & 7.7781\\ 
C & 11.3322 & 2.7328 & 8.916\\ 
C & 11.1824 & 3.7849 & 7.9813\\ 
C & 12.6018 & 2.1406 & 9.1158\\ 
C & 12.3007 & 4.2441 & 7.2479\\
\end{tabular}
\caption{}
\end{table}
\clearpage

\begin{table}[p]
\begin{tabular}{c|c c c}
type & $x$ & $y$ & $z$\\ 
\hline
C & 3.1499 & 6.706 & 5.9508\\ 
C & 2.8199 & 7.9488 & 5.3606\\ 
C & 2.1551 & 5.9516 & 6.6134\\ 
C & 4.4742 & 6.2165 & 5.8769\\ 
C & 1.4925 & 8.4338 & 5.4327\\ 
C & 3.8178 & 8.7035 & 4.6997\\ 
C & 0.4964 & 7.6762 & 6.0932\\
\end{tabular}
\caption{}
\end{table}
\begin{table}[p]
\begin{tabular}{c|c c c}
type & $x$ & $y$ & $z$\\ 
\hline
H & 8.0887 & 1.2207 & 10.1769\\ 
H & 6.9351 & 5.6442 & 8.1005\\ 
H & 10.4162 & 1.4861 & 9.4461\\ 
H & 9.2522 & 5.9016 & 7.3593\\ 
H & 12.7446 & 1.757 & 8.7142\\ 
H & 11.5656 & 6.1649 & 6.623\\ 
H & 14.5119 & 3.1022 & 7.542\\
\end{tabular}
\caption{}
\end{table}
\clearpage

\begin{table}[p]
\begin{tabular}{c|c c c}
type & $x$ & $y$ & $z$\\ 
\hline
C & 14.0523 & 6.7777 & 3.2544\\ 
C & 15.811 & 5.3288 & 4.1513\\ 
C & 15.4236 & 6.498 & 3.4556\\ 
H & 5.6787 & 1.1271 & 6.3332\\ 
H & 8.0437 & 1.7221 & 6.0463\\ 
H & 4.3413 & 5.296 & 3.8784\\ 
H & 6.692 & 5.8874 & 3.5907\\
\end{tabular}
\caption{}
\end{table}
\begin{table}[p]
\begin{tabular}{c|c c c}
type & $x$ & $y$ & $z$\\ 
\hline
C & 7.4316 & 11.0066 & 1.9766\\ 
C & 9.1549 & 9.9434 & 3.3565\\ 
C & 8.7918 & 10.8531 & 2.3349\\ 
C & 10.5126 & 9.7914 & 3.7187\\ 
C & 9.7889 & 11.6106 & 1.6758\\ 
C & 11.509 & 10.5457 & 3.0581\\ 
C & 11.1484 & 11.4563 & 2.0368\\
\end{tabular}
\caption{}
\end{table}
\clearpage

\begin{table}[p]
\begin{tabular}{c|c c c}
type & $x$ & $y$ & $z$\\ 
\hline
C & 5.3017 & 12.9227 & 8.7985\\ 
C & 2.8633 & 11.3161 & 6.4572\\ 
C & 5.209 & 11.946 & 6.1451\\ 
C & 1.7132 & 11.2399 & 7.2764\\ 
C & 1.7579 & 11.7276 & 8.6037\\ 
C & 6.4534 & 12.991 & 7.9791\\ 
C & 6.4063 & 12.5017 & 6.652\\
\end{tabular}
\caption{}
\end{table}
\begin{table}[p]
\begin{tabular}{c|c c c}
type & $x$ & $y$ & $z$\\ 
\hline
H & 9.8898 & 3.1457 & 6.357\\ 
H & 8.5307 & 4.7458 & 1.796\\ 
H & 12.2435 & 3.3302 & 5.7208\\ 
H & 10.8905 & 4.9299 & 1.1586\\ 
H & 14.0012 & 3.8923 & 4.0171\\ 
H & 13.3224 & 4.6924 & 1.7322\\ 
C & 3.2707 & 7.7089 & 5.5619\\
\end{tabular}
\caption{}
\end{table}
\clearpage

\begin{table}[p]
\begin{tabular}{c|c c c}
type & $x$ & $y$ & $z$\\ 
\hline
H & 2.4738 & 12.2891 & 7.6533\\ 
H & 4.8165 & 12.491 & 6.9946\\ 
H & 1.0473 & 13.7525 & 3.0689\\ 
H & 3.403 & 13.9522 & 2.4039\\ 
H & -0.6818 & 13.2096 & 4.8071\\ 
H & 0.0352 & 12.4756 & 7.1041\\ 
H & 7.1643 & 12.6949 & 6.3396\\
\end{tabular}
\caption{}
\end{table}
\begin{table}[p]
\begin{tabular}{c|c c c}
type & $x$ & $y$ & $z$\\ 
\hline
C & 9.9373 & 9.2964 & 9.5434\\ 
C & 9.1256 & 10.1438 & 6.9694\\ 
C & 10.9082 & 9.6043 & 8.5614\\ 
C & 10.5011 & 10.0288 & 7.2746\\ 
C & 12.2855 & 9.4793 & 8.8604\\ 
C & 11.47 & 10.3288 & 6.29\\ 
C & 13.2545 & 9.7781 & 7.8735\\
\end{tabular}
\caption{}
\end{table}
\clearpage

\begin{table}[p]
\begin{tabular}{c|c c c}
type & $x$ & $y$ & $z$\\ 
\hline
C & 6.9342 & 2.5075 & 2.2293\\ 
C & 3.2149 & 2.9462 & 2.1988\\ 
C & 3.1124 & 4.3571 & 2.2005\\ 
C & 7.9976 & 4.7148 & 2.2369\\ 
C & 8.1039 & 3.3038 & 2.2361\\ 
C & 9.1647 & 5.5122 & 2.2418\\ 
C & 9.3796 & 2.6914 & 2.2402\\
\end{tabular}
\caption{}
\end{table}
\begin{table}[p]
\begin{tabular}{c|c c c}
type & $x$ & $y$ & $z$\\ 
\hline
H & 14.1839 & 7.1808 & -4.6241\\ 
H & 16.5085 & 5.0944 & -1.5958\\ 
H & 16.4034 & 6.6005 & -3.6073\\ 
C & 6.5171 & 9.9981 & -4.295\\ 
C & 6.4489 & 11.0029 & -5.2886\\ 
C & 5.3293 & 9.4195 & -3.7918\\ 
C & 7.7722 & 9.5795 & -3.7972\\
\end{tabular}
\caption{}
\end{table}
\clearpage

\begin{table}[p]
\begin{tabular}{c|c c c}
type & $x$ & $y$ & $z$\\ 
\hline
H & 6.8718 & 6.4926 & 6.095\\ 
H & 2.2014 & 7.0656 & 5.176\\ 
H & 2.0942 & 8.8508 & 3.408\\ 
H & 9.0887 & 10.2294 & 2.5841\\ 
H & 9.3194 & 6.6641 & 6.1108\\ 
H & 11.5291 & 10.3994 & 2.5935\\ 
H & 11.7668 & 6.8336 & 6.119\\
\end{tabular}
\caption{}
\end{table}
\begin{table}[p]
\begin{tabular}{c|c c c}
type & $x$ & $y$ & $z$\\ 
\hline
C & 13.7969 & 14.6732 & -2.651\\ 
C & 14.8646 & 16.8777 & -2.6388\\ 
C & 15.0693 & 14.0569 & -2.6416\\ 
C & 16.1377 & 16.2603 & -2.6273\\ 
C & 16.2402 & 14.8493 & -2.6289\\ 
H & 4.9942 & 17.2252 & -2.6838\\ 
H & 7.4427 & 17.4195 & -2.6764\\
\end{tabular}
\caption{}
\end{table}
\clearpage

\begin{table}[p]
\begin{tabular}{c|c c c}
type & $x$ & $y$ & $z$\\ 
\hline
C & 5.431 & 3.0394 & 6.0317\\ 
C & 5.0668 & 3.6477 & 4.8069\\ 
C & 6.7959 & 2.8451 & 6.3462\\ 
C & 6.0684 & 4.061 & 3.8973\\ 
C & 7.7973 & 3.2583 & 5.437\\ 
C & 7.4339 & 3.8651 & 4.2112\\ 
C & 9.1621 & 3.07 & 5.7552\\
\end{tabular}
\caption{}
\end{table}
\begin{table}[p]
\begin{tabular}{c|c c c}
type & $x$ & $y$ & $z$\\ 
\hline
C & 2.9591 & 12.4751 & 7.409\\ 
C & 2.5577 & 13.07 & 6.1898\\ 
C & 1.9866 & 12.0493 & 8.3421\\ 
C & 4.333 & 12.3075 & 7.6967\\ 
C & 1.1815 & 13.2394 & 5.907\\ 
C & 3.5332 & 13.494 & 5.2565\\ 
C & 0.2086 & 12.8147 & 6.8428\\
\end{tabular}
\caption{}
\end{table}
\clearpage

\begin{table}[p]
\begin{tabular}{c|c c c}
type & $x$ & $y$ & $z$\\ 
\hline
H & 7.4796 & 1.8745 & 10.942\\ 
H & 6.4018 & 6.2371 & 8.7031\\ 
H & 9.8398 & 2.1459 & 10.331\\ 
H & 8.7525 & 6.5009 & 8.0824\\ 
H & 12.2019 & 2.428 & 9.7284\\ 
H & 11.1025 & 6.7758 & 7.4757\\ 
H & 14.0215 & 3.7692 & 8.6347\\
\end{tabular}
\caption{}
\end{table}
\begin{table}[p]
\begin{tabular}{c|c c c}
type & $x$ & $y$ & $z$\\ 
\hline
C & 13.2463 & 13.4934 & 9.1736\\ 
C & 12.964 & 14.7609 & 8.6122\\ 
H & 3.0192 & 10.7275 & 12.6773\\ 
H & 5.3708 & 10.996 & 12.0706\\ 
H & 1.9916 & 15.2295 & 10.7125\\ 
H & 4.3511 & 15.4969 & 10.0954\\ 
H & 0.1553 & 13.8629 & 11.7435\\
\end{tabular}
\caption{}
\end{table}
\clearpage

\begin{table}[p]
\begin{tabular}{c|c c c}
type & $x$ & $y$ & $z$\\ 
\hline
C & 8.9613 & 9.1169 & 14.7938\\ 
C & 8.6994 & 10.445 & 14.3811\\ 
C & 10.2679 & 8.587 & 14.6839\\ 
C & 9.7451 & 11.2434 & 13.8616\\ 
C & 11.3128 & 9.3842 & 14.1614\\ 
C & 11.0515 & 10.7126 & 13.75\\ 
C & 12.6186 & 8.8528 & 14.0481\\
\end{tabular}
\caption{}
\end{table}
\begin{table}[p]
\begin{tabular}{c|c c c}
type & $x$ & $y$ & $z$\\ 
\hline
C & 4.0716 & 7.0712 & 6.6578\\ 
C & 1.154 & 9.3764 & 6.2438\\ 
C & 3.5039 & 9.6301 & 5.5948\\ 
C & 0.1212 & 8.6103 & 6.8335\\ 
C & 0.4063 & 7.3307 & 7.3652\\ 
C & 5.1036 & 7.8356 & 6.0668\\ 
C & 4.8203 & 9.1162 & 5.5359\\
\end{tabular}
\caption{}
\end{table}
\clearpage

\begin{table}[p]
\begin{tabular}{c|c c c}
type & $x$ & $y$ & $z$\\ 
\hline
H & 7.7951 & 14.7488 & 3.1314\\ 
H & 11.5724 & 12.8881 & 6.8001\\ 
H & 10.1474 & 14.9995 & 2.4744\\ 
H & 13.3017 & 13.6334 & 5.1416\\ 
H & 12.5883 & 14.6906 & 2.9747\\ 
C & 3.4832 & 3.3102 & 11.3081\\ 
C & 3.1758 & 4.5486 & 10.6968\\
\end{tabular}
\caption{}
\end{table}
\begin{table}[p]
\begin{tabular}{c|c c c}
type & $x$ & $y$ & $z$\\ 
\hline
H & 5.3687 & 8.0108 & 13.1968\\ 
H & 1.5807 & 9.3522 & 9.3213\\ 
H & 3.922 & 9.6048 & 8.6621\\ 
H & -0.1435 & 8.7474 & 11.0439\\ 
H & 0.5852 & 7.9462 & 13.3188\\ 
H & 7.7133 & 8.2795 & 12.5438\\ 
H & 6.2663 & 9.8728 & 8.0087\\
\end{tabular}
\caption{}
\end{table}
\clearpage

\begin{table}[p]
\begin{tabular}{c|c c c}
type & $x$ & $y$ & $z$\\ 
\hline
C & 9.7466 & 5.6021 & 12.7415\\ 
C & 11.5344 & 4.8283 & 14.2277\\ 
C & 11.1228 & 5.4344 & 13.0174\\ 
C & 12.913 & 4.6591 & 14.4986\\ 
C & 12.0876 & 5.8715 & 12.0815\\ 
C & 13.8781 & 5.0945 & 13.5598\\ 
C & 13.465 & 5.7014 & 12.3506\\
\end{tabular}
\caption{}
\end{table}
\begin{table}[p]
\begin{tabular}{c|c c c}
type & $x$ & $y$ & $z$\\ 
\hline
C & 1.9886 & 13.5001 & 17.3478\\ 
C & 6.7081 & 13.9886 & 16.1183\\ 
C & 6.3265 & 14.5981 & 14.8994\\ 
C & 8.0783 & 13.7988 & 16.4146\\ 
C & 7.3143 & 15.0193 & 13.9792\\ 
C & 9.0665 & 14.2206 & 15.4943\\ 
C & 8.684 & 14.8307 & 14.276\\
\end{tabular}
\caption{}
\end{table}
\clearpage

\begin{table}[p]
\begin{tabular}{c|c c c}
type & $x$ & $y$ & $z$\\ 
\hline
H & 6.2163 & 1.5352 & -70.9347\\ 
H & 6.7968 & 3.8152 & -71.8227\\ 
C & 3.0772 & 8.945 & -60.9855\\ 
C & 2.0811 & 8.0386 & -61.4177\\ 
C & 2.7806 & 9.8919 & -59.9789\\ 
C & 4.3703 & 8.8998 & -61.555\\ 
C & 0.7893 & 8.0828 & -60.8437\\
\end{tabular}
\caption{}
\end{table}
\begin{table}[p]
\begin{tabular}{c|c c c}
type & $x$ & $y$ & $z$\\ 
\hline
H & 0.0281 & 14.9166 & -61.9043\\ 
H & -0.7486 & 12.617 & -61.2457\\ 
H & 3.5016 & 8.6103 & -65.4157\\ 
H & 5.0613 & 13.1942 & -66.7409\\ 
H & 5.1134 & 7.6126 & -66.9619\\ 
H & 6.6764 & 12.1938 & -68.2925\\ 
H & 6.7355 & 6.6129 & -68.492\\
\end{tabular}
\caption{}
\end{table}
\clearpage

\begin{table}[p]
\begin{tabular}{c|c c c}
type & $x$ & $y$ & $z$\\ 
\hline
C & 4.5382 & -1.4106 & -67.678\\ 
C & 6.5253 & 0.4012 & -66.8023\\ 
C & 5.8359 & -1.4572 & -68.2401\\ 
C & 6.83 & -0.5512 & -67.8019\\ 
H & -2.6346 & 1.9461 & -61.776\\ 
H & -1.0378 & 0.9239 & -63.3376\\ 
H & 0.9204 & 5.1419 & -60.246\\
\end{tabular}
\caption{}
\end{table}
\begin{table}[p]
\begin{tabular}{c|c c c}
type & $x$ & $y$ & $z$\\ 
\hline
C & 8.0029 & 7.0958 & -59.9803\\ 
C & 8.7318 & 9.6683 & -60.9019\\ 
C & 8.9836 & 7.2337 & -60.9893\\ 
C & 9.3484 & 8.5205 & -61.4502\\ 
C & 9.6036 & 6.0863 & -61.535\\ 
C & 10.3294 & 8.6585 & -62.4589\\ 
C & 10.5876 & 6.2248 & -62.5408\\
\end{tabular}
\caption{}
\end{table}
\clearpage

\begin{table}[p]
\begin{tabular}{c|c c c}
type & $x$ & $y$ & $z$\\ 
\hline
C & 7.3889 & 10.8079 & -54.5442\\ 
C & 8.9994 & 9.8096 & -56.0947\\ 
C & 9.3151 & 12.7318 & -53.7798\\ 
C & 10.9289 & 11.7317 & -55.3318\\ 
C & 8.0266 & 12.7498 & -53.1956\\ 
C & 7.0627 & 11.7875 & -53.578\\ 
C & 10.2866 & 9.7884 & -56.6785\\
\end{tabular}
\caption{}
\end{table}
\begin{table}[p]
\begin{tabular}{c|c c c}
type & $x$ & $y$ & $z$\\ 
\hline
H & 9.0755 & 1.7807 & -60.3315\\ 
H & 12.6918 & 4.7593 & -58.5231\\ 
H & 10.667 & 0.8107 & -61.9177\\ 
H & 14.2818 & 3.7941 & -60.1174\\ 
H & 12.974 & 0.7101 & -62.8948\\ 
H & 14.7862 & 2.2032 & -61.9938\\ 
\hline\hline
C242H154\\
\end{tabular}
\caption{}
\end{table}
\clearpage

\begin{table}[p]
\begin{tabular}{c|c c c}
type & $x$ & $y$ & $z$\\ 
\hline
C & 12.0263 & 10.6854 & 3.516\\ 
H & 2.0778 & 6.6235 & 7.5473\\ 
H & 4.4248 & 6.8756 & 6.9105\\ 
H & 1.0676 & 11.1494 & 5.6288\\ 
H & 3.4169 & 11.402 & 4.9875\\ 
H & -0.7704 & 9.7928 & 6.673\\ 
H & -0.2617 & 7.5214 & 7.6335\\
\end{tabular}
\caption{}
\end{table}
\begin{table}[p]
\begin{tabular}{c|c c c}
type & $x$ & $y$ & $z$\\ 
\hline
C & 7.1974 & 19.7272 & 5.6058\\ 
C & 8.8031 & 17.9237 & 6.0214\\ 
C & 8.2269 & 20.5107 & 5.0326\\ 
C & 9.8319 & 18.7069 & 5.4502\\ 
C & 9.5447 & 20.001 & 4.9554\\ 
C & 11.1484 & 18.1965 & 5.3759\\ 
C & 10.5773 & 20.7838 & 4.3869\\
\end{tabular}
\caption{}
\end{table}
\clearpage

\begin{table}[p]
\begin{tabular}{c|c c c}
type & $x$ & $y$ & $z$\\ 
\hline
C & 4.3471 & 10.0458 & 9.477\\ 
C & 1.0289 & 9.4425 & 11.0996\\ 
C & 1.435 & 9.0067 & 12.3828\\ 
C & 6.1282 & 9.5375 & 11.0803\\ 
C & 5.7224 & 9.9686 & 9.7954\\ 
C & 7.5033 & 9.466 & 11.4007\\ 
C & 6.6927 & 10.3227 & 8.8301\\
\end{tabular}
\caption{}
\end{table}
\begin{table}[p]
\begin{tabular}{c|c c c}
type & $x$ & $y$ & $z$\\ 
\hline
H & 11.0667 & 17.5641 & 8.3433\\ 
H & 13.9291 & 14.3752 & 9.0891\\ 
H & 13.4091 & 16.683 & 8.2207\\ 
C & 3.8964 & 18.9761 & 12.675\\ 
C & 3.4889 & 19.3609 & 11.3758\\ 
C & 2.9272 & 18.6437 & 13.6501\\ 
C & 5.2726 & 18.9235 & 12.9979\\
\end{tabular}
\caption{}
\end{table}
\clearpage

\begin{table}[p]
\begin{tabular}{c|c c c}
type & $x$ & $y$ & $z$\\ 
\hline
H & 4.8527 & 6.3535 & 12.9686\\ 
H & 0.7814 & 5.3561 & 15.2737\\ 
H & 1.4929 & 4.3826 & 17.4801\\ 
H & 8.6388 & 4.6842 & 16.7225\\ 
H & 7.1953 & 6.6405 & 12.3296\\ 
H & 10.9893 & 4.9695 & 16.0753\\ 
H & 9.5367 & 6.9316 & 11.6882\\
\end{tabular}
\caption{}
\end{table}
\begin{table}[p]
\begin{tabular}{c|c c c}
type & $x$ & $y$ & $z$\\ 
\hline
C & 10.9681 & 15.9642 & 13.5299\\ 
C & 12.7495 & 15.2662 & 15.059\\ 
C & 11.9389 & 16.3552 & 12.5793\\ 
C & 13.7204 & 15.658 & 14.1072\\ 
C & 13.3148 & 16.2018 & 12.8659\\ 
H & 3.6536 & 13.7642 & 18.5891\\ 
H & 6.0077 & 14.0408 & 17.9567\\
\end{tabular}
\caption{}
\end{table}
\clearpage

\begin{table}[p]
\begin{tabular}{c|c c c}
type & $x$ & $y$ & $z$\\ 
\hline
C & 0.9428 & -9.7348 & -29.628\\ 
C & 0.6943 & -8.3925 & -29.2557\\ 
C & 2.2463 & -10.2718 & -29.5047\\ 
C & 1.7458 & -7.5893 & -28.7581\\ 
C & 3.2994 & -9.4668 & -29.0094\\ 
C & 3.0477 & -8.1255 & -28.6353\\ 
C & 4.6045 & -10.0004 & -28.889\\
\end{tabular}
\caption{}
\end{table}
\begin{table}[p]
\begin{tabular}{c|c c c}
type & $x$ & $y$ & $z$\\ 
\hline
C & -2.5945 & 2.6411 & -38.5379\\ 
C & -2.2832 & 1.3751 & -39.0885\\ 
C & -3.9282 & 3.1117 & -38.5572\\ 
C & -1.5742 & 3.4317 & -37.96\\ 
C & -3.3067 & 0.5804 & -39.6564\\ 
C & -0.9499 & 0.903 & -39.0663\\ 
C & -4.6417 & 1.048 & -39.6678\\
\end{tabular}
\caption{}
\end{table}
\clearpage

\begin{table}[p]
\begin{tabular}{c|c c c}
type & $x$ & $y$ & $z$\\ 
\hline
H & 0.514 & -7.9706 & -24.9835\\ 
H & 0.5029 & -3.3726 & -26.9981\\ 
H & 2.871 & -7.6903 & -24.3617\\ 
H & 2.8507 & -3.0925 & -26.3763\\ 
H & 5.2287 & -7.4065 & -23.7443\\ 
H & 5.1958 & -2.8102 & -25.7603\\ 
H & 7.3099 & -6.0045 & -23.6978\\
\end{tabular}
\caption{}
\end{table}
\begin{table}[p]
\begin{tabular}{c|c c c}
type & $x$ & $y$ & $z$\\ 
\hline
C & 8.2932 & -4.8994 & -31.9527\\ 
C & 8.2673 & -6.1419 & -31.2767\\ 
H & -2.2267 & -4.634 & -35.6684\\ 
H & 0.1193 & -4.3508 & -35.0452\\ 
H & -2.3385 & -9.0376 & -33.2635\\ 
H & 0.0141 & -8.7578 & -32.6429\\ 
H & -4.4007 & -8.1886 & -34.4184\\
\end{tabular}
\caption{}
\end{table}
\clearpage

\begin{table}[p]
\begin{tabular}{c|c c c}
type & $x$ & $y$ & $z$\\ 
\hline
C & 4.107 & -3.8292 & -36.9719\\ 
C & 3.8341 & -2.4888 & -36.6096\\ 
C & 5.4168 & -4.3458 & -36.8327\\ 
C & 4.8694 & -1.667 & -36.1084\\ 
C & 6.4538 & -3.5215 & -36.3358\\ 
C & 6.1792 & -2.1818 & -35.9732\\ 
C & 7.7637 & -4.0367 & -36.1929\\
\end{tabular}
\caption{}
\end{table}
\begin{table}[p]
\begin{tabular}{c|c c c}
type & $x$ & $y$ & $z$\\ 
\hline
C & -7.457 & 3.5883 & -27.6816\\ 
C & -5.1263 & 3.8734 & -26.9809\\ 
C & -8.5946 & 4.0635 & -28.375\\ 
C & -8.5461 & 5.3095 & -29.0431\\ 
C & -3.8845 & 5.8791 & -27.6449\\ 
C & -3.933 & 4.6327 & -26.9764\\ 
C & -2.6894 & 6.6353 & -27.6458\\
\end{tabular}
\caption{}
\end{table}
\clearpage

\begin{table}[p]
\begin{tabular}{c|c c c}
type & $x$ & $y$ & $z$\\ 
\hline
H & 5.6219 & -2.068 & -31.512\\ 
H & 5.6682 & 2.4418 & -33.7098\\ 
H & 7.7043 & -0.6765 & -31.4556\\ 
H & 7.7303 & 1.59 & -32.5559\\ 
\hline\hline
C286H182\\ 
\hline\hline
C & -4.254 & -0.3699 & -57.766\\ 
C & -3.5613 & 0.8537 & -57.9219\\
\end{tabular}
\caption{}
\end{table}
\begin{table}[p]
\begin{tabular}{c|c c c}
type & $x$ & $y$ & $z$\\ 
\hline
H & -1.783 & 1.8173 & -60.6194\\ 
H & -1.6001 & 7.368 & -61.2143\\ 
H & 0.4329 & 6.3156 & -60.3594\\ 
H & -3.9675 & 7.1914 & -62.0185\\ 
H & -5.0832 & 4.9392 & -62.1516\\ 
H & 0.2494 & 0.7662 & -59.7397\\ 
H & 2.4591 & 5.2675 & -59.4757\\
\end{tabular}
\caption{}
\end{table}
\clearpage

\begin{table}[p]
\begin{tabular}{c|c c c}
type & $x$ & $y$ & $z$\\ 
\hline
C & 0.1263 & -2.048 & -52.1504\\ 
C & -0.0303 & -3.5827 & -50.2471\\ 
C & 0.7495 & -2.9 & -51.2103\\ 
C & 0.5971 & -4.4324 & -49.3056\\ 
C & 2.153 & -3.0684 & -51.2335\\ 
C & 2.0023 & -4.5979 & -49.3271\\ 
C & 2.7805 & -3.9161 & -50.2918\\
\end{tabular}
\caption{}
\end{table}
\begin{table}[p]
\begin{tabular}{c|c c c}
type & $x$ & $y$ & $z$\\ 
\hline
C & -2.8956 & 9.8093 & -59.4324\\ 
C & 1.1705 & 7.7469 & -57.6381\\ 
C & 1.9126 & 8.2591 & -58.728\\ 
C & 1.8142 & 6.97 & -56.6482\\ 
C & 3.2965 & 7.9878 & -58.8307\\ 
C & 3.1985 & 6.7021 & -56.7488\\ 
C & 3.9402 & 7.2101 & -57.8409\\
\end{tabular}
\caption{}
\end{table}
\clearpage

\begin{table}[p]
\begin{tabular}{c|c c c}
type & $x$ & $y$ & $z$\\ 
\hline
H & 11.6153 & 9.3054 & -59.3651\\ 
C & -5.0599 & 3.5933 & -51.5734\\ 
C & -4.398 & 4.8411 & -51.6505\\ 
C & -6.4062 & 3.4816 & -51.9923\\ 
C & -4.3736 & 2.4587 & -51.0809\\ 
C & -5.0829 & 5.9748 & -52.1444\\ 
C & -3.0511 & 4.9541 & -51.2358\\
\end{tabular}
\caption{}
\end{table}
\begin{table}[p]
\begin{tabular}{c|c c c}
type & $x$ & $y$ & $z$\\ 
\hline
H & -5.6032 & 9.3322 & -55.6149\\ 
H & -0.3168 & 5.2577 & -52.967\\ 
H & 2.0876 & 9.6435 & -53.388\\ 
H & 1.7165 & 4.2292 & -52.0714\\ 
H & 4.1256 & 8.611 & -52.5068\\ 
H & 3.7536 & 3.1973 & -51.1953\\ 
H & 6.1654 & 7.5742 & -51.6423\\
\end{tabular}
\caption{}
\end{table}
\clearpage

\begin{table}[p]
\begin{tabular}{c|c c c}
type & $x$ & $y$ & $z$\\ 
\hline
C & 4.3804 & 5.7929 & -47.8411\\ 
C & 4.2589 & 4.2268 & -45.9624\\ 
C & 5.0237 & 4.9378 & -46.9168\\ 
H & -5.9353 & 7.8158 & -48.8224\\ 
H & -3.8889 & 6.8222 & -47.9034\\ 
H & -3.1912 & 10.3687 & -52.1601\\ 
H & -1.1537 & 9.3735 & -51.252\\
\end{tabular}
\caption{}
\end{table}
\begin{table}[p]
\begin{tabular}{c|c c c}
type & $x$ & $y$ & $z$\\ 
\hline
C & 1.0741 & 12.7712 & -47.5362\\ 
C & 1.0901 & 10.4669 & -46.702\\ 
C & 1.7722 & 11.6767 & -46.974\\ 
C & 1.789 & 9.3703 & -46.1459\\ 
C & 3.1542 & 11.7875 & -46.6922\\ 
C & 3.1708 & 9.4806 & -45.866\\ 
C & 3.8539 & 10.6894 & -46.139\\
\end{tabular}
\caption{}
\end{table}
\clearpage

\begin{table}[p]
\begin{tabular}{c|c c c}
type & $x$ & $y$ & $z$\\ 
\hline
C & -0.7816 & 8.5736 & -63.9349\\ 
C & 1.2849 & 7.5503 & -63.1068\\ 
C & 0.556 & 11.0467 & -64.2166\\ 
C & 2.6226 & 10.0258 & -63.3896\\ 
C & -0.8139 & 10.9436 & -64.5515\\ 
C & -1.4837 & 9.7056 & -64.4108\\ 
C & 2.65 & 7.6603 & -62.7557\\
\end{tabular}
\caption{}
\end{table}
\begin{table}[p]
\begin{tabular}{c|c c c}
type & $x$ & $y$ & $z$\\ 
\hline
H & 7.3882 & 9.066 & -64.4345\\ 
H & 9.7603 & 13.4617 & -64.9305\\ 
H & 9.4514 & 8.0471 & -63.6081\\ 
H & 11.8226 & 12.4408 & -64.1018\\ 
H & 11.8786 & 8.2265 & -63.0116\\ 
H & 13.0723 & 10.4323 & -63.2588\\ 
C & 5.6656 & 17.7293 & -69.9275\\
\end{tabular}
\caption{}
\end{table}
\clearpage

\begin{table}[p]
\begin{tabular}{c|c c c}
type & $x$ & $y$ & $z$\\ 
\hline
H & -2.7087 & 8.4661 & -57.7043\\ 
H & -0.6431 & 7.4575 & -56.8445\\ 
H & -0.0882 & 10.7163 & -61.345\\ 
H & 1.9686 & 9.7168 & -60.4883\\ 
H & -2.5684 & 11.0441 & -61.2087\\ 
H & -3.8846 & 9.9179 & -59.3815\\ 
H & 1.4232 & 6.4422 & -55.9961\\
\end{tabular}
\caption{}
\end{table}
\begin{table}[p]
\begin{tabular}{c|c c c}
type & $x$ & $y$ & $z$\\ 
\hline
C & 9.0118 & 15.2229 & -62.2189\\ 
C & 10.4487 & 16.3744 & -64.3662\\ 
C & 10.4037 & 15.0039 & -62.3361\\ 
C & 11.1229 & 15.58 & -63.4095\\ 
C & 11.0766 & 14.2078 & -61.3812\\ 
C & 12.515 & 15.3583 & -63.5258\\ 
C & 12.4688 & 13.9881 & -61.4953\\
\end{tabular}
\caption{}
\end{table}
\clearpage

\begin{table}[p]
\begin{tabular}{c|c c c}
type & $x$ & $y$ & $z$\\ 
\hline
C & 0.6124 & 10.5993 & -69.6561\\ 
C & -0.1055 & 10.0328 & -68.5768\\ 
C & 4.0312 & 8.0287 & -66.88\\ 
C & 4.75 & 8.5981 & -67.9573\\ 
C & 4.7045 & 7.2373 & -65.9217\\ 
C & 6.141 & 8.3745 & -68.076\\ 
C & 6.0959 & 7.0167 & -66.0382\\
\end{tabular}
\caption{}
\end{table}
\begin{table}[p]
\begin{tabular}{c|c c c}
type & $x$ & $y$ & $z$\\ 
\hline
H & 13.0966 & 9.0074 & -66.7228\\ 
H & 14.4128 & 10.1337 & -68.5501\\ 
C & 0.4515 & 13.0162 & -57.3986\\ 
C & 1.0936 & 14.2752 & -57.47\\ 
C & -0.9024 & 12.8897 & -57.7873\\ 
C & 1.1641 & 11.8836 & -56.9405\\ 
C & 0.3797 & 15.4075 & -57.9274\\
\end{tabular}
\caption{}
\end{table}
\clearpage

\begin{table}[p]
\begin{tabular}{c|c c c}
type & $x$ & $y$ & $z$\\ 
\hline
H & 0.6952 & 20.5769 & -62.21\\ 
H & -0.5205 & 18.4247 & -61.7459\\ 
H & 4.8629 & 14.4833 & -59.1011\\ 
H & 7.3021 & 18.773 & -60.023\\ 
H & 6.936 & 13.4793 & -58.2775\\ 
H & 9.3806 & 17.7664 & -59.1964\\ 
H & 9.01 & 12.467 & -57.4733\\
\end{tabular}
\caption{}
\end{table}
\begin{table}[p]
\begin{tabular}{c|c c c}
type & $x$ & $y$ & $z$\\ 
\hline
C & 10.1479 & 4.644 & -70.0039\\ 
C & 11.4301 & 7.1618 & -70.1442\\ 
C & 11.5019 & 4.7709 & -69.6143\\ 
C & 12.1434 & 6.0299 & -69.6849\\ 
H & 1.3472 & 7.7239 & -73.2137\\ 
H & 3.4316 & 6.7015 & -72.4313\\ 
H & 3.6413 & 12.1806 & -73.5038\\
\end{tabular}
\caption{}
\end{table}
\clearpage

\begin{table}[p]
\begin{tabular}{c|c c c}
type & $x$ & $y$ & $z$\\ 
\hline
C & 0.606 & 4.1298 & -60.1837\\ 
C & 0.6642 & 1.7308 & -59.6868\\ 
C & 1.9513 & 4.2498 & -59.7645\\ 
C & 2.0087 & 1.8514 & -59.2641\\ 
C & 2.6518 & 3.1116 & -59.3017\\ 
C & 2.71 & 0.7132 & -58.802\\ 
C & 3.9938 & 3.2338 & -58.8731\\
\end{tabular}
\caption{}
\end{table}
\begin{table}[p]
\begin{tabular}{c|c c c}
type & $x$ & $y$ & $z$\\ 
\hline
C & -2.7394 & 4.6552 & -56.8667\\ 
C & -4.8847 & 4.2937 & -55.7421\\ 
C & -2.8432 & 3.2822 & -54.8419\\ 
C & -3.3878 & 5.4448 & -57.8437\\ 
C & -1.3447 & 4.4362 & -56.9439\\ 
C & -4.7843 & 5.6558 & -57.7746\\ 
C & -5.5338 & 5.0802 & -56.7221\\
\end{tabular}
\caption{}
\end{table}
\clearpage

\begin{table}[p]
\begin{tabular}{c|c c c}
type & $x$ & $y$ & $z$\\ 
\hline
H & 3.7512 & 8.6746 & -59.7754\\ 
H & 3.0858 & 5.3719 & -55.32\\ 
H & 5.7938 & 7.6523 & -58.8794\\ 
H & 5.1188 & 4.3656 & -54.4156\\ 
H & 7.8354 & 6.6369 & -57.9719\\ 
H & 7.6015 & 4.037 & -54.4887\\ 
H & 8.9661 & 5.1739 & -56.2735\\
\end{tabular}
\caption{}
\end{table}
\begin{table}[p]
\begin{tabular}{c|c c c}
type & $x$ & $y$ & $z$\\ 
\hline
C & 3.7941 & 0.8004 & -48.2872\\ 
H & -6.8939 & 2.5858 & -52.0747\\ 
H & -4.857 & 1.5577 & -51.1932\\ 
H & -4.5329 & 7.002 & -52.3842\\ 
H & -2.4948 & 5.9755 & -51.5075\\ 
H & -6.936 & 6.823 & -53.0729\\ 
H & -8.1243 & 4.606 & -52.9184\\
\end{tabular}
\caption{}
\end{table}
\clearpage

\begin{table}[p]
\begin{tabular}{c|c c c}
type & $x$ & $y$ & $z$\\ 
\hline
C & 5.0364 & 11.9332 & -56.3309\\ 
C & 4.9819 & 9.6236 & -55.5166\\ 
C & 6.4122 & 12.0138 & -56.0091\\ 
C & 6.3561 & 9.704 & -55.195\\ 
C & 7.073 & 10.8984 & -55.4418\\ 
C & 7.0137 & 8.5889 & -54.6277\\ 
C & 8.4491 & 10.9743 & -55.1207\\
\end{tabular}
\caption{}
\end{table}
\begin{table}[p]
\begin{tabular}{c|c c c}
type & $x$ & $y$ & $z$\\ 
\hline
C & -1.8182 & 8.7837 & -50.825\\ 
C & -5.2612 & 9.986 & -51.6633\\ 
C & -6.0112 & 9.402 & -50.6156\\ 
C & -1.9205 & 7.4029 & -48.8049\\ 
C & -1.1714 & 7.9855 & -49.8541\\ 
C & -1.2735 & 6.6043 & -47.8341\\ 
C & 0.2229 & 7.7663 & -49.9337\\
\end{tabular}
\caption{}
\end{table}
\clearpage

\begin{table}[p]
\begin{tabular}{c|c c c}
type & $x$ & $y$ & $z$\\ 
\hline
H & 7.5113 & 10.8482 & -52.1342\\ 
H & 7.3225 & 8.3282 & -48.5896\\ 
H & 8.6619 & 9.4052 & -50.4306\\ 
C & -5.5872 & 7.7526 & -45.5574\\ 
C & -4.9172 & 8.9874 & -45.7235\\ 
C & -6.9483 & 7.6332 & -45.9227\\ 
C & -4.8944 & 6.6383 & -45.0289\\
\end{tabular}
\caption{}
\end{table}
\begin{table}[p]
\begin{tabular}{c|c c c}
type & $x$ & $y$ & $z$\\ 
\hline
H & -0.3952 & 14.7702 & -48.9613\\ 
H & -4.9018 & 15.6204 & -50.3562\\ 
H & -6.1506 & 13.49 & -49.8766\\ 
H & -0.8505 & 9.5167 & -47.119\\ 
H & 1.66 & 13.7547 & -48.0892\\ 
H & 1.196 & 8.5026 & -46.2453\\ 
H & 3.7148 & 12.734 & -47.2221\\
\end{tabular}
\caption{}
\end{table}
\clearpage

\begin{table}[p]
\begin{tabular}{c|c c c}
type & $x$ & $y$ & $z$\\ 
\hline
C & 5.1378 & 0.8249 & -62.4016\\ 
C & 4.5007 & 0.9879 & -61.1495\\ 
C & 6.3468 & 0.0967 & -62.4894\\ 
C & 5.0755 & 0.4245 & -59.9874\\ 
C & 6.925 & -0.4615 & -61.3256\\ 
C & 6.2884 & -0.2972 & -60.073\\ 
H & -0.2817 & 5.141 & -67.7506\\
\end{tabular}
\caption{}
\end{table}
\begin{table}[p]
\begin{tabular}{c|c c c}
type & $x$ & $y$ & $z$\\ 
\hline
C & 0.6788 & 5.2925 & -59.2201\\ 
C & 1.6554 & 6.1479 & -59.7826\\ 
C & 1.074 & 4.2059 & -58.4065\\ 
C & 3.028 & 5.9144 & -59.532\\ 
C & 2.4456 & 3.9714 & -58.1578\\ 
C & 3.4236 & 4.8242 & -58.722\\ 
C & 2.8397 & 2.8799 & -57.3505\\
\end{tabular}
\caption{}
\end{table}
\clearpage

\begin{table}[p]
\begin{tabular}{c|c c c}
type & $x$ & $y$ & $z$\\ 
\hline
C & -3.5189 & 0.3589 & -64.0702\\ 
C & -2.5461 & 1.1252 & -64.7535\\ 
C & -4.893 & 0.6402 & -64.2497\\ 
C & -3.1161 & -0.6863 & -63.2074\\ 
C & -2.9494 & 2.1708 & -65.6152\\ 
C & -1.1717 & 0.8504 & -64.5691\\ 
C & -4.3231 & 2.4564 & -65.7908\\
\end{tabular}
\caption{}
\end{table}
\begin{table}[p]
\begin{tabular}{c|c c c}
type & $x$ & $y$ & $z$\\ 
\hline
H & 1.8105 & -1.1338 & -69.7899\\ 
H & 4.229 & -1.4135 & -74.1803\\ 
H & 3.5901 & -2.4287 & -68.725\\ 
H & 6.0145 & -2.7136 & -73.1117\\ 
H & 5.3729 & -3.7094 & -67.6548\\ 
H & 7.8034 & -4.0037 & -72.0354\\ 
H & 7.5638 & -4.9175 & -67.7864\\
\end{tabular}
\caption{}
\end{table}
\clearpage

\begin{table}[p]
\begin{tabular}{c|c c c}
type & $x$ & $y$ & $z$\\ 
\hline
C & 3.1786 & -0.4911 & -51.309\\ 
C & 4.1651 & 0.2828 & -51.9642\\ 
H & -5.9758 & 4.5886 & -55.1944\\ 
H & -4.2221 & 3.2595 & -54.1238\\ 
H & -2.476 & 7.3448 & -57.5109\\ 
H & -0.7204 & 6.016 & -56.4389\\ 
H & -4.9073 & 7.8341 & -57.889\\
\end{tabular}
\caption{}
\end{table}
\begin{table}[p]
\begin{tabular}{c|c c c}
type & $x$ & $y$ & $z$\\ 
\hline
C & -2.1081 & -8.2642 & -65.5706\\ 
C & -1.1547 & -7.5661 & -66.3489\\ 
C & -1.6837 & -9.2613 & -64.6609\\ 
C & 0.2207 & -7.8644 & -66.2172\\ 
C & -0.3073 & -9.5632 & -64.5329\\ 
C & 0.6448 & -8.8636 & -65.3113\\ 
C & 0.1199 & -10.5577 & -63.622\\
\end{tabular}
\caption{}
\end{table}
\clearpage

\begin{table}[p]
\begin{tabular}{c|c c c}
type & $x$ & $y$ & $z$\\ 
\hline
C & 3.2192 & 5.7909 & -75.4702\\ 
C & 4.9988 & 4.4788 & -74.4085\\ 
C & 1.8457 & 6.0953 & -75.6211\\ 
C & 0.8776 & 5.3907 & -74.8679\\ 
C & 4.4315 & 2.7738 & -72.743\\ 
C & 5.4025 & 3.4741 & -73.4974\\ 
C & 4.8326 & 1.7725 & -71.8297\\
\end{tabular}
\caption{}
\end{table}
\begin{table}[p]
\begin{tabular}{c|c c c}
type & $x$ & $y$ & $z$\\ 
\hline
H & 7.7723 & 2.3628 & -63.8942\\ 
H & 10.0497 & 1.8165 & -68.3333\\ 
H & 9.9168 & 1.0733 & -64.0458\\ 
H & 11.0616 & 0.7979 & -66.2722\\ 
C & -6.1115 & 1.2559 & -55.8206\\ 
C & -6.7075 & 1.4835 & -54.5579\\ 
C & -6.6919 & 1.8057 & -56.9863\\
\end{tabular}
\caption{}
\end{table}
\clearpage

\begin{table}[p]
\begin{tabular}{c|c c c}
type & $x$ & $y$ & $z$\\ 
\hline
H & -2.6341 & 1.8947 & -68.6013\\ 
H & -0.8597 & 0.5812 & -67.5518\\ 
H & -3.3551 & 3.7714 & -70.1071\\ 
H & -1.6209 & 5.1721 & -71.2818\\ 
H & 4.3701 & 2.0811 & -68.8142\\ 
H & 0.9171 & -0.7247 & -66.4902\\ 
H & 6.1475 & 0.7804 & -67.7434\\
\end{tabular}
\caption{}
\end{table}
\begin{table}[p]
\begin{tabular}{c|c c c}
type & $x$ & $y$ & $z$\\ 
\hline
C & -1.375 & -8.5705 & -72.1028\\ 
C & 0.8317 & -9.5799 & -72.4394\\ 
C & -0.2075 & -9.1664 & -71.5733\\ 
C & 2.0035 & -10.1665 & -71.9059\\ 
C & -0.0748 & -9.3395 & -70.1763\\ 
C & 2.1384 & -10.3335 & -70.5075\\ 
C & 1.0988 & -9.9195 & -69.6421\\
\end{tabular}
\caption{}
\end{table}
\clearpage

\begin{table}[p]
\begin{tabular}{c|c c c}
type & $x$ & $y$ & $z$\\ 
\hline
C & -7.9663 & 3.1485 & -73.1376\\ 
C & -5.3466 & 1.5576 & -69.3133\\ 
C & -4.2483 & 2.3403 & -69.7417\\ 
C & -5.2382 & 0.7676 & -68.145\\ 
C & -3.0431 & 2.3323 & -69.0028\\ 
C & -4.0326 & 0.7599 & -67.4057\\ 
C & -2.9349 & 1.5428 & -67.8352\\
\end{tabular}
\caption{}
\end{table}
\begin{table}[p]
\begin{tabular}{c|c c c}
type & $x$ & $y$ & $z$\\ 
\hline
H & 0.444 & -0.5661 & -83.9456\\ 
C & -2.0268 & -2.8683 & -77.9337\\ 
C & -0.9198 & -2.0871 & -78.3394\\ 
C & -3.228 & -2.832 & -78.6793\\ 
C & -1.9316 & -3.6831 & -76.7821\\ 
C & -1.0171 & -1.2686 & -79.4877\\ 
C & 0.2803 & -2.1188 & -77.5929\\
\end{tabular}
\caption{}
\end{table}
\clearpage

\begin{table}[p]
\begin{tabular}{c|c c c}
type & $x$ & $y$ & $z$\\ 
\hline
H & -1.7408 & 9.8023 & -77.4658\\ 
H & 1.1707 & 7.7182 & -71.2443\\ 
H & 5.0479 & 6.9187 & -74.3321\\ 
H & 2.502 & 6.8999 & -69.36\\ 
H & 6.3789 & 6.0958 & -72.4465\\ 
H & 3.8302 & 6.0807 & -67.478\\ 
H & 7.7059 & 5.2706 & -70.5618\\
\end{tabular}
\caption{}
\end{table}
\begin{table}[p]
\begin{tabular}{c|c c c}
type & $x$ & $y$ & $z$\\ 
\hline
C & -0.3685 & -6.007 & -68.2418\\ 
C & -1.3288 & -7.6831 & -66.7368\\ 
C & -0.253 & -6.842 & -67.1061\\ 
H & -8.5053 & -5.1129 & -74.8985\\ 
H & -7.214 & -5.9328 & -72.9888\\ 
H & -4.7113 & -2.0859 & -76.1659\\ 
H & -3.415 & -2.9121 & -74.2654\\
\end{tabular}
\caption{}
\end{table}
\clearpage

\begin{table}[p]
\begin{tabular}{c|c c c}
type & $x$ & $y$ & $z$\\ 
\hline
C & -0.5628 & 9.1499 & -68.9341\\ 
C & 1.805 & 9.7651 & -68.7981\\ 
C & 0.6819 & 9.0881 & -68.266\\ 
C & 3.0525 & 9.6946 & -68.1346\\ 
C & 0.8061 & 8.3434 & -67.0701\\ 
C & 3.1773 & 8.9471 & -66.9399\\ 
C & 2.0536 & 8.2712 & -66.408\\
\end{tabular}
\caption{}
\end{table}
\begin{table}[p]
\begin{tabular}{c|c c c}
type & $x$ & $y$ & $z$\\ 
\hline
C & -1.8488 & -0.4597 & -74.6644\\ 
C & -5.3493 & 0.8423 & -74.8781\\ 
C & -4.0457 & 0.0531 & -72.9594\\ 
C & -5.4545 & 1.1047 & -76.2641\\ 
C & -4.3532 & 0.8527 & -77.1155\\ 
C & -1.7492 & -0.736 & -73.2817\\ 
C & -2.8478 & -0.4781 & -72.4284\\
\end{tabular}
\caption{}
\end{table}
\clearpage

\begin{table}[p]
\begin{tabular}{c|c c c}
type & $x$ & $y$ & $z$\\ 
\hline
H & 4.4799 & 2.5291 & -73.3452\\ 
H & 9.8626 & 4.0111 & -73.1962\\ 
H & 5.7722 & 1.6839 & -71.4535\\ 
H & 9.9909 & 2.6674 & -71.0788\\ 
H & 7.9373 & 1.5 & -70.2069\\ 
C & 5.1181 & 3.2337 & -81.6242\\ 
C & 6.2383 & 3.0428 & -82.4671\\
\end{tabular}
\caption{}
\end{table}
\begin{table}[p]
\begin{tabular}{c|c c c}
type & $x$ & $y$ & $z$\\ 
\hline
H & 3.5187 & 0.6875 & -78.9809\\ 
H & -1.587 & 2.6263 & -77.8238\\ 
H & -0.291 & 1.8193 & -75.9149\\ 
H & -1.7968 & 3.0536 & -80.2899\\ 
H & 0.1192 & 2.4906 & -81.8267\\ 
H & 4.8083 & -0.1436 & -77.0714\\ 
H & 1.0019 & 1.0001 & -74.0055\\
\end{tabular}
\caption{}
\end{table}
\clearpage

\begin{table}[p]
\begin{tabular}{c|c c c}
type & $x$ & $y$ & $z$\\ 
\hline
C & 5.0525 & -7.4054 & -77.402\\ 
C & 4.0176 & -7.5364 & -75.1852\\ 
C & 5.1414 & -7.7188 & -76.0251\\ 
C & 4.1087 & -7.8372 & -73.8068\\ 
C & 6.3553 & -8.2034 & -75.4833\\ 
C & 5.3219 & -8.3195 & -73.2644\\ 
C & 6.4457 & -8.5037 & -74.1035\\
\end{tabular}
\caption{}
\end{table}
\begin{table}[p]
\begin{tabular}{c|c c c}
type & $x$ & $y$ & $z$\\ 
\hline
C & -3.3633 & 6.4186 & -61.7625\\ 
C & -4.0081 & 5.1594 & -61.7546\\ 
C & 0.089 & 3.0734 & -60.0582\\ 
C & 0.7318 & 4.3342 & -60.0623\\ 
C & 0.7869 & 1.9252 & -59.6161\\ 
C & 2.0699 & 4.4474 & -59.6192\\ 
C & 2.1242 & 2.039 & -59.1694\\
\end{tabular}
\caption{}
\end{table}
\clearpage

\begin{table}[p]
\begin{tabular}{c|c c c}
type & $x$ & $y$ & $z$\\ 
\hline
H & 2.4744 & -5.2788 & -48.4955\\ 
H & 3.8359 & -4.1769 & -50.3039\\ 
C & -3.4432 & 4.1442 & -55.7712\\ 
C & -2.6761 & 4.7521 & -56.7921\\ 
C & -4.8403 & 4.3548 & -55.7171\\ 
C & -2.8126 & 3.3241 & -54.8077\\ 
C & -3.3089 & 5.5663 & -57.7591\\
\end{tabular}
\caption{}
\end{table}
\begin{table}[p]
\begin{tabular}{c|c c c}
type & $x$ & $y$ & $z$\\ 
\hline
H & -2.686 & 11.2686 & -61.2404\\ 
H & -4.0622 & 10.0509 & -59.5223\\ 
H & 1.1109 & 6.4786 & -56.0656\\ 
H & 3.8722 & 8.8995 & -59.4889\\ 
H & 3.134 & 5.446 & -55.1611\\ 
H & 5.9023 & 7.8631 & -58.5813\\ 
H & 5.1546 & 4.4255 & -54.2445\\
\end{tabular}
\caption{}
\end{table}
\clearpage

\begin{table}[p]
\begin{tabular}{c|c c c}
type & $x$ & $y$ & $z$\\ 
\hline
C & 1.7076 & -0.6599 & -48.5542\\ 
C & 3.0461 & 1.826 & -48.6748\\ 
C & 3.0585 & -0.5653 & -48.1468\\ 
C & 3.7288 & 0.679 & -48.2072\\ 
H & -6.9026 & 2.5594 & -52.1066\\ 
H & -4.8801 & 1.5104 & -51.2168\\ 
H & -4.5281 & 6.9746 & -52.3029\\
\end{tabular}
\caption{}
\end{table}
\begin{table}[p]
\begin{tabular}{c|c c c}
type & $x$ & $y$ & $z$\\ 
\hline
C & 3.019 & 10.7974 & -56.0929\\ 
C & 4.3832 & 13.2659 & -56.3094\\ 
C & 4.3729 & 10.8936 & -55.6976\\ 
C & 5.0562 & 12.1279 & -55.8065\\ 
C & 5.0429 & 9.7571 & -55.1901\\ 
C & 6.4101 & 12.2235 & -55.4087\\ 
C & 6.3969 & 9.8521 & -54.7945\\
\end{tabular}
\caption{}
\end{table}
\clearpage

\begin{table}[p]
\begin{tabular}{c|c c c}
type & $x$ & $y$ & $z$\\ 
\hline
C & -5.4038 & 8.5473 & -49.5193\\ 
C & -3.3767 & 7.5224 & -48.6009\\ 
C & -3.8729 & 9.7284 & -51.5802\\ 
C & -1.8462 & 8.7021 & -50.6656\\ 
C & -5.2701 & 9.9401 & -51.5295\\ 
C & -6.0367 & 9.3488 & -50.498\\ 
C & -1.9822 & 7.2979 & -48.6636\\
\end{tabular}
\caption{}
\end{table}
\begin{table}[p]
\begin{tabular}{c|c c c}
type & $x$ & $y$ & $z$\\ 
\hline
H & 2.6774 & 9.5843 & -48.9011\\ 
H & 5.3962 & 11.6837 & -52.5612\\ 
H & 4.6996 & 8.5547 & -47.9909\\ 
H & 7.4175 & 10.6506 & -51.6504\\ 
H & 7.1742 & 8.1719 & -48.0797\\ 
H & 8.5422 & 9.2234 & -49.9168\\ 
C & 1.312 & 17.6584 & -55.3308\\
\end{tabular}
\caption{}
\end{table}
\clearpage

\begin{table}[p]
\begin{tabular}{c|c c c}
type & $x$ & $y$ & $z$\\ 
\hline
H & -7.5028 & 6.6004 & -45.7881\\ 
H & -5.4757 & 5.569 & -44.8813\\ 
H & -5.1096 & 10.9986 & -46.1333\\ 
H & -3.0847 & 9.9697 & -45.2307\\ 
H & -7.5064 & 10.8231 & -46.8477\\ 
H & -8.7093 & 8.6171 & -46.6759\\ 
H & -3.4445 & 4.5379 & -43.9773\\
\end{tabular}
\caption{}
\end{table}
\begin{table}[p]
\begin{tabular}{c|c c c}
type & $x$ & $y$ & $z$\\ 
\hline
C & 4.2697 & 13.8396 & -49.0527\\ 
C & 5.6713 & 16.275 & -49.3778\\ 
C & 5.6328 & 13.9245 & -48.6883\\ 
C & 6.3347 & 15.1421 & -48.8502\\ 
C & 6.2968 & 12.7888 & -48.1714\\ 
C & 7.7014 & 15.221 & -48.4928\\ 
C & 7.6626 & 12.8664 & -47.8149\\
\end{tabular}
\caption{}
\end{table}
\clearpage

\begin{table}[p]
\begin{tabular}{c|c c c}
type & $x$ & $y$ & $z$\\ 
\hline
C & -3.4795 & 18.76 & -48.2753\\ 
C & -4.2274 & 18.2159 & -47.2048\\ 
C & -0.1407 & 16.2458 & -45.3523\\ 
C & 0.606 & 16.7863 & -46.4262\\ 
C & 0.5054 & 15.4697 & -44.3618\\ 
C & 1.9968 & 16.5455 & -46.5127\\ 
C & 1.896 & 15.2262 & -44.45\\
\end{tabular}
\caption{}
\end{table}
\begin{table}[p]
\begin{tabular}{c|c c c}
type & $x$ & $y$ & $z$\\ 
\hline
H & 6.6008 & 3.2245 & -57.8616\\ 
H & 6.5358 & -0.9472 & -56.6335\\ 
H & 7.7851 & 1.2133 & -56.9395\\ 
C & 0.6117 & 8.6876 & -63.6132\\ 
C & 1.2821 & 9.931 & -63.6812\\ 
C & -0.7411 & 8.5923 & -64.0136\\ 
C & 1.2944 & 7.542 & -63.1435\\
\end{tabular}
\caption{}
\end{table}
\clearpage

\begin{table}[p]
\begin{tabular}{c|c c c}
type & $x$ & $y$ & $z$\\ 
\hline
H & 5.6776 & 15.5602 & -66.2639\\ 
H & 1.2321 & 16.4393 & -67.8003\\ 
H & 0.0333 & 14.2271 & -67.652\\ 
H & 5.3306 & 10.1123 & -65.0883\\ 
H & 7.7125 & 14.5205 & -65.3845\\ 
H & 7.3655 & 9.0726 & -64.209\\ 
H & 9.747 & 13.4807 & -64.5056\\
\end{tabular}
\caption{}
\end{table}
\begin{table}[p]
\begin{tabular}{c|c c c}
type & $x$ & $y$ & $z$\\ 
\hline
C & 15.1349 & 20.4661 & -69.9176\\ 
C & 15.074 & 18.12 & -69.2157\\ 
C & 16.5026 & 20.5351 & -69.5613\\ 
C & 16.4405 & 18.1878 & -68.8602\\ 
C & 17.1555 & 19.3961 & -69.0331\\ 
H & 6.442 & 21.4081 & -72.611\\ 
H & 8.4638 & 20.3573 & -71.7367\\
\end{tabular}
\caption{}
\end{table}
\clearpage

\begin{table}[p]
\begin{tabular}{c|c c c}
type & $x$ & $y$ & $z$\\ 
\hline
C & 4.3897 & 12.984 & -61.6324\\ 
C & 4.2849 & 11.5983 & -59.6162\\ 
C & 5.7814 & 12.7521 & -61.7191\\ 
C & 5.6763 & 11.3639 & -59.7046\\ 
C & 6.4253 & 11.9429 & -60.7556\\ 
C & 6.3212 & 10.554 & -58.7418\\ 
C & 7.8171 & 11.7119 & -60.8426\\
\end{tabular}
\caption{}
\end{table}
\begin{table}[p]
\begin{tabular}{c|c c c}
type & $x$ & $y$ & $z$\\ 
\hline
C & 7.4589 & 23.2046 & -68.3875\\ 
C & 5.3075 & 22.8165 & -67.2817\\ 
C & 7.3461 & 21.8009 & -66.3829\\ 
C & 6.8135 & 24.0061 & -69.3579\\ 
C & 8.8554 & 22.9917 & -68.4582\\ 
C & 5.4151 & 24.2088 & -69.2935\\ 
C & 4.6616 & 23.6149 & -68.2539\\
\end{tabular}
\caption{}
\end{table}
\clearpage

\begin{table}[p]
\begin{tabular}{c|c c c}
type & $x$ & $y$ & $z$\\ 
\hline
H & 4.2487 & 4.3254 & -65.7812\\ 
H & 3.6038 & 1.183 & -61.2078\\ 
H & 6.2812 & 3.293 & -64.8938\\ 
H & 5.6452 & 0.169 & -60.308\\ 
H & 8.3141 & 2.2775 & -63.9991\\ 
H & 8.124 & -0.1938 & -60.4221\\ 
H & 9.4638 & 0.8606 & -62.2743\\
\end{tabular}
\caption{}
\end{table}
\begin{table}[p]
\begin{tabular}{c|c c c}
type & $x$ & $y$ & $z$\\ 
\hline
C & 13.4827 & 10.1834 & -68.1046\\ 
H & 2.6123 & 13.5009 & -69.6969\\ 
H & 4.6472 & 12.4732 & -68.8147\\ 
H & 5.2629 & 15.5575 & -73.4259\\ 
H & 7.3006 & 14.5251 & -72.547\\ 
H & 2.792 & 15.9482 & -73.2922\\ 
H & 1.4575 & 14.9169 & -71.4197\\
\end{tabular}
\caption{}
\end{table}
\clearpage

\begin{table}[p]
\begin{tabular}{c|c c c}
type & $x$ & $y$ & $z$\\ 
\hline
C & 5.0203 & 12.0731 & -55.567\\ 
C & 5.0395 & 9.6713 & -55.0806\\ 
C & 6.372 & 12.1667 & -55.1608\\ 
C & 6.3933 & 9.7635 & -54.682\\ 
C & 7.0605 & 11.0109 & -54.7238\\ 
C & 7.0838 & 8.6069 & -54.2508\\ 
C & 8.4173 & 11.1003 & -54.3334\\
\end{tabular}
\caption{}
\end{table}
\begin{table}[p]
\begin{tabular}{c|c c c}
type & $x$ & $y$ & $z$\\ 
\hline
C & 7.0392 & 23.3189 & -63.1136\\ 
C & 3.626 & 24.2502 & -64.3408\\ 
C & 2.9249 & 23.0407 & -64.1243\\ 
C & 7.004 & 20.9934 & -62.3434\\ 
C & 7.7076 & 22.2018 & -62.5594\\ 
C & 7.6718 & 19.8739 & -61.7964\\ 
C & 9.0805 & 22.2884 & -62.2285\\
\end{tabular}
\caption{}
\end{table}
\clearpage

\begin{table}[p]
\begin{tabular}{c|c c c}
type & $x$ & $y$ & $z$\\ 
\hline
H & 9.9828 & 3.6485 & -67.0647\\ 
H & 9.9328 & -0.4815 & -65.6978\\ 
H & 11.1785 & 1.669 & -66.0838\\ 
C & 3.5943 & 8.9624 & -72.9521\\ 
C & 4.2349 & 10.2237 & -72.9291\\ 
C & 2.2495 & 8.8615 & -73.3798\\ 
C & 4.2972 & 7.8053 & -72.5417\\
\end{tabular}
\caption{}
\end{table}
\begin{table}[p]
\begin{tabular}{c|c c c}
type & $x$ & $y$ & $z$\\ 
\hline
H & 8.5176 & 15.9288 & -75.4209\\ 
H & 4.0673 & 16.8378 & -76.9327\\ 
H & 2.8814 & 14.6208 & -76.8631\\ 
H & 8.1914 & 10.4394 & -74.4287\\ 
H & 10.5516 & 14.8708 & -74.5591\\ 
H & 10.2224 & 9.3834 & -73.5569\\ 
H & 12.5802 & 13.8163 & -73.6812\\
\end{tabular}
\caption{}
\end{table}
\clearpage

\begin{table}[p]
\begin{tabular}{c|c c c}
type & $x$ & $y$ & $z$\\ 
\hline
C & 4.1273 & 0.8261 & -58.432\\ 
C & 4.7558 & 2.0929 & -58.3913\\ 
C & 4.8307 & -0.3246 & -58.0054\\ 
C & 6.085 & 2.209 & -57.9233\\ 
C & 6.1604 & -0.2083 & -57.5369\\ 
C & 6.7885 & 1.0588 & -57.4966\\ 
H & -3.8186 & 2.8713 & -61.5149\\
\end{tabular}
\caption{}
\end{table}
\begin{table}[p]
\begin{tabular}{c|c c c}
type & $x$ & $y$ & $z$\\ 
\hline
C & -4.0466 & -1.52 & -52.0124\\ 
C & -3.2842 & -0.9049 & -53.0335\\ 
C & -3.4108 & -2.3457 & -51.0559\\ 
C & -1.8879 & -1.1163 & -53.0985\\ 
C & -2.0137 & -2.5561 & -51.1202\\ 
C & -1.2525 & -1.941 & -52.1423\\ 
C & -1.3756 & -3.376 & -50.1603\\
\end{tabular}
\caption{}
\end{table}
\clearpage

\begin{table}[p]
\begin{tabular}{c|c c c}
type & $x$ & $y$ & $z$\\ 
\hline
C & -0.9903 & 8.7337 & -58.6703\\ 
C & -0.2438 & 9.3004 & -59.7292\\ 
C & -2.3828 & 8.9636 & -58.5854\\ 
C & -0.3435 & 7.9363 & -57.6986\\ 
C & -0.8924 & 10.0954 & -60.7017\\ 
C & 1.1477 & 9.066 & -59.8181\\ 
C & -2.2861 & 10.3204 & -60.6214\\
\end{tabular}
\caption{}
\end{table}
\begin{table}[p]
\begin{tabular}{c|c c c}
type & $x$ & $y$ & $z$\\ 
\hline
H & 3.5356 & 10.9933 & -59.0519\\ 
H & 6.2566 & 13.2844 & -62.5946\\ 
H & 5.5622 & 9.9505 & -58.1672\\ 
H & 8.2895 & 12.2328 & -61.7105\\ 
H & 7.586 & 8.9174 & -57.2726\\ 
H & 10.3215 & 11.1879 & -60.8147\\ 
H & 10.0649 & 8.5628 & -57.3521\\
\end{tabular}
\caption{}
\end{table}
\clearpage

\begin{table}[p]
\begin{tabular}{c|c c c}
type & $x$ & $y$ & $z$\\ 
\hline
C & 3.1345 & -0.4857 & -48.1999\\ 
C & 3.7989 & 0.7622 & -48.2508\\ 
H & -6.8489 & 2.6274 & -52.1118\\ 
H & -4.8202 & 1.5803 & -51.2346\\ 
H & -4.5046 & 7.0609 & -52.2469\\ 
H & -2.4724 & 6.0141 & -51.38\\ 
H & -6.8973 & 6.8943 & -52.9737\\
\end{tabular}
\caption{}
\end{table}
\begin{table}[p]
\begin{tabular}{c|c c c}
type & $x$ & $y$ & $z$\\ 
\hline
C & 4.159 & 10.7623 & -55.7814\\ 
C & 4.8367 & 12.0005 & -55.8759\\ 
C & 4.8372 & 9.6214 & -55.295\\ 
C & 6.1907 & 12.0972 & -55.4806\\ 
C & 6.1928 & 9.7168 & -54.9058\\ 
C & 6.8704 & 10.9547 & -54.999\\ 
C & 6.8743 & 8.5738 & -54.4291\\
\end{tabular}
\caption{}
\end{table}
\clearpage

\begin{table}[p]
\begin{tabular}{c|c c c}
type & $x$ & $y$ & $z$\\ 
\hline
C & -6.3132 & 5.156 & -48.6393\\ 
C & -4.2896 & 4.1202 & -47.7329\\ 
C & -7.7174 & 5.3193 & -48.6289\\ 
C & -8.5026 & 4.6141 & -47.6869\\ 
C & -4.4502 & 2.5407 & -45.8674\\ 
C & -3.6682 & 3.2472 & -46.8115\\ 
C & -3.8267 & 1.6626 & -44.9499\\
\end{tabular}
\caption{}
\end{table}
\begin{table}[p]
\begin{tabular}{c|c c c}
type & $x$ & $y$ & $z$\\ 
\hline
H & 2.2196 & 4.0501 & -45.1173\\ 
H & 4.8875 & 6.1485 & -48.8116\\ 
H & 4.6935 & 3.6696 & -45.2385\\ 
H & 6.0363 & 4.7219 & -47.093\\ 
C & -1.5082 & 13.0304 & -52.5232\\ 
C & -0.7544 & 13.6151 & -53.5676\\ 
C & -2.9044 & 13.2443 & -52.4545\\
\end{tabular}
\caption{}
\end{table}
\clearpage

\begin{table}[p]
\begin{tabular}{c|c c c}
type & $x$ & $y$ & $z$\\ 
\hline
H & 1.865 & 19.4706 & -58.373\\ 
H & 3.9089 & 18.4584 & -57.477\\ 
H & -0.6125 & 19.8385 & -58.2492\\ 
H & -1.9454 & 18.7967 & -56.3847\\ 
H & 3.2803 & 15.3412 & -52.8843\\ 
H & 5.9486 & 17.4271 & -56.5905\\ 
H & 5.3133 & 14.3018 & -52.0043\\
\end{tabular}
\caption{}
\end{table}
\begin{table}[p]
\begin{tabular}{c|c c c}
type & $x$ & $y$ & $z$\\ 
\hline
C & 3.2425 & 9.4644 & -45.7179\\ 
C & 3.9122 & 10.7099 & -45.7647\\ 
C & 3.9281 & 8.3094 & -45.2771\\ 
C & 5.2686 & 10.7975 & -45.3722\\ 
C & 5.2851 & 8.3951 & -44.8899\\ 
C & 5.9563 & 9.6395 & -44.9386\\ 
H & -4.7026 & 11.552 & -48.727\\
\end{tabular}
\caption{}
\end{table}
\clearpage

\begin{table}[p]
\begin{tabular}{c|c c c}
type & $x$ & $y$ & $z$\\ 
\hline
C & -2.4349 & 11.2499 & -42.5995\\ 
C & -1.6672 & 12.0574 & -43.4719\\ 
C & -1.7912 & 10.3379 & -41.7303\\ 
C & -0.257 & 11.9534 & -43.4741\\ 
C & -0.3802 & 10.2345 & -41.732\\ 
C & 0.3867 & 11.0418 & -42.6053\\ 
C & 0.2641 & 9.319 & -40.867\\ 
C & 1.7965 & 10.9333 & -42.6132\\ 
C & 1.6749 & 9.2124 & -40.8732\\ 
C & 2.4408 & 10.0188 & -41.7479\\ 
C & 2.3204 & 8.2946 & -40.0117\\ 
C & 3.8506 & 9.9071 & -41.7604\\ 
C & 3.7308 & 8.1847 & -40.0228\\ 
C & 4.4962 & 8.9905 & -40.8982\\ 
H & -6.4955 & 11.7141 & -42.8226\\ 
H & -4.441 & 10.716 & -41.9351\\ 
H & -3.7685 & 14.5793 & -45.9134\\ 
H & -1.7162 & 13.586 & -45.0248\\ 
H & -6.2704 & 14.7402 & -45.9412\\ 
H & -7.6383 & 13.3086 & -44.3896\\ 
H & -2.386 & 9.7057 & -41.0615\\ 
H & 0.3367 & 12.5762 & -44.1525\\ 
H & -0.3306 & 8.6873 & -40.1974\\ 
H & 2.389 & 11.5555 & -43.2932\\ 
H & 1.7255 & 7.6639 & -39.3412\\ 
H & 4.4395 & 10.5301 & -42.4428\\ 
H & 4.228 & 7.4716 & -39.3559\\ 
H & 5.5883 & 8.9053 & -40.9107\\ 
\end{tabular}
\caption{}
\end{table}
\clearpage

\end{document}